\documentclass[iop,numberedappendix]{emulateapj}
\pdfoutput=1
\usepackage{color}
\usepackage{amsmath}
\usepackage{mathtools}
\usepackage{multirow}
\usepackage{subfigure}

\newcommand{\eref}[1]{(\ref{#1})}
\newcommand{\dnds}{$\mathrm{d}N/\mathrm{d}S$}
\newcommand{\mpd}{{(p)}}
\newcommand{\pmh}{\mbox{MH11}}
\newcommand{\opdf}{1pPDF}
\newcommand{\shy}{\mathrm{h}}

\slugcomment{Manuscript accepted for publication in ApJS (2016 June 1)}

\shortauthors{Zechlin et al.}

\begin{document}

\title{UNVEILING THE GAMMA-RAY SOURCE COUNT DISTRIBUTION BELOW THE \emph{FERMI} DETECTION LIMIT WITH PHOTON STATISTICS}

\author{Hannes-S. Zechlin\altaffilmark{1,2}, Alessandro Cuoco\altaffilmark{2,3},\\
        Fiorenza Donato\altaffilmark{1,2}, Nicolao Fornengo\altaffilmark{1,2}, and
	Andrea Vittino\altaffilmark{4}}

\altaffiltext{1}{Dipartimento di Fisica, Universit\`a di Torino, via P. Giuria, 1, I-10125 Torino, Italy; zechlin@to.infn.it}
\altaffiltext{2}{Istituto Nazionale di Fisica Nucleare, Sezione di Torino, via P. Giuria, 1, I-10125 Torino, Italy}
\altaffiltext{3}{Institute for Theoretical Particle Physics and Cosmology (TTK), RWTH Aachen University, D-52056 Aachen, Germany}
\altaffiltext{4}{Physik-Department T30d, Technische Universit\"at M\"unchen, James-Franck-Stra\ss e, D-85748 Garching, Germany}

\begin{abstract}
The source-count distribution as a function of their flux, \dnds, is
one of the main quantities characterizing gamma-ray source
populations.  We employ statistical properties of the \emph{Fermi}
Large Area Telescope (LAT) photon counts map to measure the
composition of the extragalactic gamma-ray sky at high latitudes ($|b|
\geq 30^\circ$) between 1\,GeV and 10\,GeV.  We present a new method,
generalizing the use of standard pixel-count statistics, to decompose
the total observed gamma-ray emission into (a) point-source
contributions, (b) the Galactic foreground contribution, and (c) a
truly diffuse isotropic background contribution. Using the 6-year
\emph{Fermi}-LAT data set (\texttt{P7REP}), we show that the
\dnds\ distribution in the regime of so far undetected point sources
can be consistently described with a power law of index between 1.9
and 2.0. We measure \dnds\ down to an integral flux of $\sim\! 2\times
10^{-11}\,\mathrm{cm}^{-2}\,\mathrm{s}^{-1}$, improving beyond the
3FGL catalog detection limit by about one order of magnitude.  The
overall \dnds\ distribution is consistent with a broken power law,
with a break at $2.1^{+1.0}_{-1.3} \times
10^{-8}\,\mathrm{cm}^{-2}\,\mathrm{s}^{-1}$.  The power-law index $n_1
= 3.1^{+0.7}_{-0.5}$ for bright sources above the break hardens to
$n_2 = 1.97\pm 0.03$ for fainter sources below the break.  A possible
second break of the \dnds\ distribution is constrained to be at fluxes
below $6.4\times 10^{-11}\,\mathrm{cm}^{-2}\,\mathrm{s}^{-1}$ at 95\%
confidence level. The high-latitude gamma-ray sky between 1\,GeV and
10\,GeV is shown to be composed of $\sim$25\% point sources,
$\sim$69.3\% diffuse Galactic foreground emission, and $\sim$6\%
isotropic diffuse background.
\end{abstract}
\keywords{methods: statistical ---  gamma rays: general --- gamma rays: diffuse background}

\section{INTRODUCTION}\label{sec:intro}
The decomposition of the extragalactic gamma-ray background (EGB; see
\citet{2015PhR...598....1F} for a recent review) is pivotal for
unveiling the origin of the nonthermal cosmic radiation field. The EGB
comprises the emission from all individual and diffuse gamma-ray
sources of extragalactic origin, and thus it originates from different
mechanisms of gamma-ray production in the Universe. The EGB can be
dissected by resolving the various point-source contributions,
characterized by their differential source-count distribution
\dnds\ as a function of the integral source flux $S$ \citep[see,
  e.g.,][]{2010ApJ...720..435A,2015MNRAS.454..115S}.  Conventionally,
the EGB emission that is left after subtracting the resolved gamma-ray
sources is referred to as the isotropic diffuse gamma-ray background
\citep[IGRB;][]{2015ApJ...799...86A}.  The Large Area Telescope (LAT)
on board the \emph{Fermi} satellite \citep{2012ApJS..203....4A} has
allowed the discovery of more than 3,000 gamma-ray point sources,
collected in the 3FGL catalog \citep{2015ApJS..218...23A}.  Resolved
sources amount to about 30\% of the EGB \citep{2015ApJ...799...86A}
below $\sim$100 GeV (while above $\sim$100 GeV this percentage can
rise to about 50\%).

For resolved point sources listed in catalogs the \dnds\ distributions
of different source classes can be characterized.  Among these,
blazars represent the brightest and most numerous population, and,
consequently, their \dnds\ is the best-determined one. Blazars exhibit
two different subclasses: flat-spectrum radio quasars \mbox{(FSRQs)},
with a typically soft gamma-ray spectrum characterized by an average
power-law photon index of $\sim$2.4, and BL Lacertae (BL Lac) objects,
with a harder photon index of $\sim$2.1. The \dnds\ distribution of
blazars has been studied in detail in several works
\citep{Ajello:2011zi,Ajello:2013lka,Chang:2013yia,DiMauro:2013zfa,Harding:2012gk,Inoue:2008pk,Stecker:2010di,Stecker:1996ma}.
Besides blazars, the EGB includes fainter sources like misaligned
active galactic nuclei \citep[mAGN;][]{DiMauro:2013xta,Inoue:2011bm}
and star-forming galaxies
\citep[SFGs;][]{Ackermann:2012vca,Fields:2010bw,Lacki:2012si,Tamborra:2014xia,Thompson:2006qd}.
A contribution from Galactic sources such as millisecond pulsars
(MSPs) located at high Galactic latitude is possible, although it has
been constrained to be subdominant
\citep{Calore:2014oga,Gregoire:2013yta}.  Finally, pure diffuse (not
point-like) components can contribute, for instance caused by pair
halo emission from AGN, clusters of galaxies, or cascades of
ultra-high-energy cosmic rays (UHECRs) on the CMB (see
\citet{2015PhR...598....1F} and references therein).

In the usual approach, the \dnds\ distributions of different
populations (inferred from resolved sources) are extrapolated to the
unresolved regime and used to investigate the composition of the IGRB
(i.e., the unresolved EGB). This approach has revealed that the
above-mentioned three main components well explain the observed IGRB
spectrum, constraining further contributions to be subdominant,
including a possible exotic contribution from dark matter (DM)
annihilation or decay
\citep{Ajello:2015mfa,Cholis:2013ena,DiMauro:2015tfa}.  While the
above-mentioned approach is very useful, a clear drawback is caused by
the fact that it relies on the extrapolation of
\dnds\ distributions. In this work, we will focus on a method to
overcome this problem by conducting a direct measurement of the
\dnds\ in the unresolved regime.

Detection capabilities for individual point sources are intrinsically
limited by detector angular resolution and backgrounds. This makes in
particular the IGRB a quantity that depends on the actual observation
\citep[][]{2015ApJ...799...86A}.  The common approach of detecting
individual sources \citep{2015ApJS..218...23A,2016ApJS..222....5A} can
be complemented by decomposing gamma-ray skymaps by statistical means,
using photon-count or intensity maps.  One of the simplest ways of
defining such a statistic is to consider the probablity distribution
function (PDF) of photon counts or fluxes in pixels, commonly known as
$P(D)$ distribution in the radio \citep[e.g.,][and references
  therein]{1974ApJ...188..279C,1957PCPS...53..764S,2014MNRAS.440.2791V,2015MNRAS.447.2243V}
and X-ray \citep[e.g.,][and references
  therein]{1993A&A...275....1H,2011A&A...532A..19S} bands.  Recently,
this technique has been adapted to photon-count measurements in the
gamma-ray band; see \citet{2011ApJ...738..181M}, henceforth \pmh, for
details.  Various theoretical studies have also been performed
\citep{2010PhRvD..82l3511B,Dodelson:2009ih,Feyereisen:2015cea,Lee:2008fm}.
In addition, this method has been used to probe unresolved gamma-ray
sources in the region of the Galactic Center
\citep{2016PhRvL.116e1102B,2015JCAP...05..056L,2016PhRvL.116e1103L},
as well as to constrain the source-count distribution above 50\,GeV
\citep{2015arXiv151100693T}.

As argued above, this method has the advantage of directly measuring
the \dnds\ in the unresolved regime, thus not relying on any
extrapolation.  A difference with respect to the use of resolved
sources is that in the PDF approach only the global \dnds, i.e., the
sum of all components, can be directly measured: since no individual
source can be identified with this method, counterpart association and
the separation of \dnds\ into different source components become
impossible.  The PDF approach nonetheless offers another important
advantage with respect to the standard method: the use of the
\dnds\ built from cataloged sources close to the detection threshold
of the catalog is hampered by the fact that the threshold is not sharp
but rather characterized by a detection efficiency as a function of
flux \citep{2010ApJ...720..435A,2015arXiv151100693T}.  The \dnds\ thus
needs to be corrected for the catalog detection efficiency, which, in
turn, is a nontrivial quantity to determine
\citep{2010ApJ...720..435A}.  On the contrary, the PDF approach treats
all the sources in the same way, resolved and unresolved, and can thus
determine the \dnds\ in a significantly larger flux range, without
requiring the use of any efficiency function.

In the following, we will measure the high-latitude \dnds\ with the
PDF methodology using 6 years of gamma-ray data collected with the
{\it Fermi}-LAT.  We will show that for the 1\,GeV to 10\,GeV energy
band we can measure the \dnds\ down to an integral flux of $\sim
10^{-11}\,\mathrm{cm}^{-2}\,\mathrm{s}^{-1}$, which is a factor of
$\sim 20$ lower than the nominal threshold of the 3FGL catalog.

This article is structured as follows: In Section~\ref{sec:theory} we
introduce the mathematical framework of the analysis method,
supplemented by a detailed description of our extensions to previous
approaches, the modeling of source and background components, and the
fitting procedure.  The gamma-ray data analysis is addressed in
Section~\ref{sec:Fermi_data}.  Section~\ref{sec:analysis_routine} is
dedicated to details of the statistical analysis approach and the
fitting technique.  The resulting global source-count distribution and
the composition of the gamma-ray sky are considered in
Section~\ref{sec:application}.  Section~\ref{sec:anisotropy} addresses
the angular power of unresolved sources detected with this
analysis. Possible systematic and modeling uncertainties are discussed
in Section~\ref{sec:systematics}.  Eventually, final results are
summarized in Section~\ref{sec:conclusions}.

\section{THE STATISTICS OF GAMMA-RAY PHOTON COUNTS}\label{sec:theory}
In the present analysis, we assume the gamma-ray sky at high Galactic
latitudes to be composed of three different contributions:
\begin{itemize}
\item A population of gamma-ray point sources.  Given that the
  analysis is restricted to high Galactic latitudes, this source
  population is considered to be dominantly of extragalactic origin.
  Sources can thus be assumed to be distributed homogeneously across
  the sky.
\item Diffuse gamma-ray emission from our Galaxy, mostly bright along
  the Galactic plane but extending also to the highest Galactic
  latitudes. We will refer to this component as Galactic foreground
  emission.  The photon flux in map pixel $p$ from this component will
  be denoted as $F^\mpd_\mathrm{gal}$.
\item Gamma-ray emission from all contributions that are
  indistinguishable from diffuse isotropic emission, such as extremely
  faint sources. We will include in this component possible truly
  diffuse emission of extragalactic or Galactic origin, such as, for
  example, gamma rays from cosmological cascades from UHECRs, or
  possible isotropic subcomponents of the Galactic foreground
  emission.  In addition, the component comprises the residual
  cosmic-ray background.  All together this emission will be denoted
  as $F_\mathrm{iso}$.
\end{itemize}
A more detailed account of the individual components is given in
Section~\ref{sec:intro} and later in this section.

Following the method of \pmh, we considered the celestial region of
interest (ROI) to be partitioned into $N_\mathrm{pix}$ pixels of equal
area \mbox{$\Omega_\mathrm{pix} = 4\pi
  f_\mathrm{ROI}/N_\mathrm{pix}\,\mathrm{sr}$}, where $f_\mathrm{ROI}$
is the fraction of sky covered by the ROI.  The probability $p_k$ of
finding $k$ photons in a given pixel is by definition the 1-point PDF
(\opdf).  In the simplest scenario of purely isotropic emission, $p_k$
follows a Poisson distribution with an expectation value equal to the
mean photon rate.  The imprints of more complex diffuse components and
a distribution of point sources alter the shape of the \opdf, in turn
allowing us to investigate these components by measuring the \opdf\ of
the data.

The usual way in which the \opdf\ is used requires us to bin the
photon counts of each pixel into a histogram of the number of pixels,
$n_k$, containing $k$ photon counts, and to compare the $p_k$
predicted by the model with the estimator $n_k/N_\mathrm{pix}$. This
method is the one adopted by \pmh.  By definition, this technique does
not preserve any spatial information of the measurement or its
components (for example, the uneven morphology of the Galactic
foreground emission), resulting in an undesired loss of information.
We will instead use the \opdf\ in a more general form, including
pixel-dependent variations in order to fully exploit all the available
information.

\subsection{Generating Functions}\label{ssec:gen_funcs}
An elegant way of deriving the \opdf\ including all the desired
components exploits the framework of probability generating functions
(see \pmh\ and references therein for details).  The generating
function $\mathcal{P}^\mpd (t)$ of a discrete probability distribution
$p^\mpd_k$, which may depend on the pixel $p$ and where
$k=0,1,2,\dots$ is a discrete random variable, is defined as a power
series in an auxiliary variable $t$ by
\begin{equation}\label{eq:gf}
\mathcal{P}^\mpd (t) = \sum_{k=0}^{\infty} p^\mpd_k t^k . 
\end{equation}
The series coefficients $p^\mpd_k$ can be derived from a given
$\mathcal{P}^\mpd (t)$ by differentiating with respect to $t$ and
evaluating them at $t=0$,
\begin{equation}\label{eq:pkcalc}
p^\mpd_k = \frac{1}{k!} \left. \frac{\mathrm{d}^k \mathcal{P}^\mpd (t)}
{\mathrm{d}t^k}\right|_{t=0} .
\end{equation}
The method of combining individual components into a single
$\mathcal{P}^\mpd (t)$ makes use of the summation property of
generating functions, i.e., the fact that the generating function for
the sum of two independent random variables is given by the product of
the generating functions for each random variable itself.

In our case, the general representation of $\mathcal{P}^\mpd (t)$ for
photon-count maps can be derived from considering a superposition of
Poisson processes; see Appendix~\ref{app:genfunc_poisson} and
\pmh\ for a more detailed explanation. The generating function is
therefore given by
\begin{equation}\label{eq:gfgen1}
\mathcal{P}^\mpd (t) = \exp \left[ \sum_{m=1}^{\infty} x^\mpd_m 
\left( t^m -1 \right) \right] ,
\end{equation}
where the coefficients $x^\mpd_m$ are the expected number of point
sources per pixel $p$ that contribute exactly $m$ photons to the total
photon count of the pixel, and $m$ is a positive integer.  In the
derivation of Equation~\eqref{eq:gfgen1}, it has been assumed that the
$x^\mpd_m$ are mean values of underlying Poisson PDFs.  The quantities
$x^\mpd_m$ are related to the differential source-count distribution
\dnds, where $S$ denotes the integral photon flux of a source in a
given energy range $[E_\mathrm{min}, E_\mathrm{max}]$, by
\begin{equation}\label{eq:xm}
x^\mpd_m = \Omega_\mathrm{pix} \int_0^\infty \mathrm{d}S 
\frac{\mathrm{d}N}{\mathrm{d}S} \frac{(\mathcal{C}^\mpd\!(S) )^m}{m!} 
e^{-\mathcal{C}^\mpd\!(S)} .
\end{equation}
The number of counts $\mathcal{C}^\mpd\!(S)$ expected in pixel $p$ is
given as a function of $S$ by
\begin{equation}\label{eq:counts}
\mathcal{C}^\mpd\!(S) = S\,\frac{ \int_{E_\mathrm{min}}^{E_\mathrm{max}}
\mathrm{d}E\,E^{-\Gamma} \mathcal{E}^\mpd(E) }
{  \int_{E_\mathrm{min}}^{E_\mathrm{max}} \mathrm{d}E\,E^{-\Gamma} }
\end{equation}
for sources with a power-law-type energy spectrum $\propto
E^{-\Gamma}$, where $\Gamma$ denotes the photon index and the
pixel-dependent exposure\footnote{The experiment exposure, which
  depends on energy and position, is discussed in
  Section~\ref{sec:Fermi_data}.} as a function of energy is denoted by
$\mathcal{E}^\mpd(E)$. In Equation~\eref{eq:xm}, we have assumed that
the PDF for a source to contribute $m$ photons to a pixel $p$ follows
a Poisson distribution with mean $\mathcal{C}^\mpd\!(S)$.  Gamma-ray
sources have been assumed to be isotropically distributed across the
sky, i.e., \dnds\ is pixel independent, while, in principle,
Equation~\eref{eq:xm} allows for an extension of the method to
spatially dependent \dnds\ distributions.

The generating functions for diffuse background components correspond
to $1$-photon source terms, with $x^\mpd_m = 0$ for all $m$ except
$m=1$:
\begin{equation}\label{eq:Dgen}
\mathcal{D}^\mpd (t) = \exp \left[ x_\mathrm{diff}^\mpd\,(t-1) \right] ,
\end{equation}
where $x^\mpd_\mathrm{diff}$ denotes the number of diffuse photon
counts expected in pixel $p$ for a given
observation.\footnote{Equation~\eref{eq:Dgen} can be derived from
  Equation~\eref{eq:gf} by taking $p^\mpd_k$ as a Poissonian with mean
  $x_\mathrm{diff}^\mpd$.}  This quantity is given by
\begin{equation}\label{eq:xdiff}
x^\mpd_\mathrm{diff} = \int_{\Omega_\mathrm{pix}} \,\mathrm{d}\Omega
\int_{E_\mathrm{min}}^{E_\mathrm{max}} 
\mathrm{d}E\,f^\mpd_\mathrm{diff}(E)\,\mathcal{E}^\mpd (E)\, ,
\end{equation}
with $f^\mpd_\mathrm{diff}(E)$ being the differential flux of the
diffuse component as a function of energy.

The relation in Equation~\eref{eq:xm} allows measuring the
source-count distribution \dnds\ from pixel-count
statistics. Furthermore, we can observe that the \opdf\ approach may
allow the detection of point-source populations below catalog
detection thresholds: if the source-count distribution implies a large
number of faint emitters, pixels containing photon counts originating
from these sources will be stacked in an $n_k$-histogram, increasing
the statistical significance of corresponding $k$-bins.  The average
number of photons required from individual sources for the statistical
detection of the entire population will therefore be significantly
smaller than the photon contribution required for individual source
detection.

The simple \opdf\ approach refers to a measurement of $p_k$ which is
averaged over the considered ROI.  The generating function for the
\opdf\ measurement therefore reduces to a pixel average,
\begin{equation}\label{eq:gf1ppdf}
\mathcal{P}(t) = \frac{1}{N_\mathrm{pix}} \sum_{p=1}^{N_\mathrm{pix}}
\mathcal{P}_\mathrm{S}^\mpd (t) \, \mathcal{D}^\mpd (t),
\end{equation}
where we made use of the fact that the total generating function
factorizes in the point-source component and the diffuse component,
$\mathcal{P}^\mpd (t) = \mathcal{P}_\mathrm{S}^\mpd
(t)\,\mathcal{D}^\mpd (t)$ (see Equations~\eref{eq:gfgen1} and
\eref{eq:Dgen}).

The numerical implementation of Equation~\eref{eq:gf1ppdf} in its most
general form is computationally complex \citep[see
  \pmh;][]{2015JCAP...05..056L}. In the ideal situation of an
isotropic point-source distribution and homogeneous exposure,
$\mathcal{P}_\mathrm{S} (t) \equiv \mathcal{P}_\mathrm{S}^\mpd (t)$
factorizes out of the sum, reducing the pixel-dependent part of
Equation~\eref{eq:gf1ppdf} to the diffuse component, which is easy to
handle.  The exposure of \emph{Fermi}-LAT data is, however, not
uniformly distributed in the ROI (see Section~\ref{sec:Fermi_data})
and requires appropriate consideration.

To correct the point-source component for exposure inhomogeneities, we
divided the exposure map into $N_\mathrm{exp}$ regions, separated by
contours of constant exposure such that the entire exposure range is
subdivided into $N_\mathrm{exp}$ equally spaced bins.  In each region,
the exposure values were replaced with the region averages, yielding
$N_\mathrm{exp}$ regions of constant exposure. The approximation
accuracy is thus related to the choice of $N_\mathrm{exp}$.  In this
case, Equation~\eref{eq:gf1ppdf} reads
\begin{equation}\label{eq:Pexpinhomo}
\mathcal{P}(t) = \frac{1}{N_\mathrm{pix}} \sum_{i=1}^{N_\mathrm{exp}}
\sum_{\mathrm{P}_i} \mathcal{P}_\mathrm{S}^\mpd (t) \, \mathcal{D}^\mpd (t),
\end{equation}
where $\mathrm{P}_i = \{ p | p \in \mathrm{R}_i\}$ denotes the subset
of pixels belonging to region $\mathrm{R}_i$.  In this way,
$\mathcal{P}_\mathrm{S}^\mpd (t)$ becomes independent of the inner sum
and factorizes, significantly reducing the required amount of
computation time.

The probability distributions $p_k$ or $p^\mpd_k$ can eventually be
calculated from $\mathcal{P}(t)$ or $\mathcal{P}^\mpd (t)$,
respectively, by using Equation~\eref{eq:pkcalc}.

\subsection{Model Description}
\subsubsection{Source-count Distribution}\label{sssec:dnds_model}
The source-count distribution \dnds\ characterizes the number of point
sources $N$ in the flux interval $(S, S+dS)$, where $S$ is the
integral flux of a source in a given energy range. The quantity $N$
actually denotes the areal source density per solid angle element
$\mathrm{d}\Omega$, which is omitted in our notation for
simplicity. In this analysis, we parameterized the source-count
distribution with a power law with \textit{multiple breaks}, referred
to as multiply broken power law (MBPL) in the remainder. An MBPL with
$N_\mathrm{b}$ breaks located at $S_{\mathrm{b}j}$,
$j=1,2,\dots,N_\mathrm{b}$, is defined as
\begin{equation}\label{eq:mbpl}
\frac{\mathrm{d}N}{\mathrm{d}S} \propto
\begin{cases} 
\left( \frac{S}{S_0} \right)^{-n_1} \qquad\qquad\qquad\qquad\qquad , 
S > S_{\mathrm{b}1} &  \\
\left( \frac{S_{\mathrm{b}1}}{S_0} \right)^{-n_1+n_2} 
\left( \frac{S}{S_0} \right)^{-n_2}  \qquad\qquad , 
S_{\mathrm{b}2} < S \leq S_{\mathrm{b}1} \\
\vdotswithin{\left(\frac{S_{\mathrm{b}1}}{S_0}\right)} 
\qquad\qquad\qquad\qquad\qquad \vdotswithin{S_{\mathrm{b}2} < S} & \\
\left( \frac{S_{\mathrm{b}1}}{S_0} \right)^{-n_1+n_2} 
\left( \frac{S_{\mathrm{b}2}}{S_0} \right)^{-n_2+n_3} 
\cdots \ \left( \frac{S}{S_0} \right)^{-n_{N_\mathrm{b}+1}} & \\
\hspace{14em} , S \leq S_{\mathrm{b}N_\mathrm{b}} & \\
\end{cases}
\end{equation}
where $S_0$ is a normalization constant. The $n_j$ denote the indices
of the power-law components.  The \dnds\ distribution is normalized
with an overall factor $A_\mathrm{S}$, which is given by $A_\mathrm{S}
= \mathrm{d}N/\mathrm{d}S\,(S_0)$ if $S_0 > S_{\mathrm{b}1}$.  We
required a finite total flux, i.e., we imposed $n_1 > 2$ and
$n_{N_\mathrm{b}+1} < 2$.

\subsubsection{Source Spectra} \label{sec:source_spectra}
The whole population of gamma-ray sources is disseminated by a variety
of different source classes (see Section~\ref{sec:intro} for details).
In particular, FSRQs and BL Lac objects contribute to the overall
\dnds\ at high Galactic latitudes. The spectral index distribution of
all resolved sources in the energy band between 100\,MeV and 100\,GeV
(assuming power-law spectra) is compatible with a Gaussian centered on
$\Gamma = 2.40\pm 0.02$, with a half-width of $\sigma_\Gamma = 0.24
\pm 0.02$ \citep{2010ApJ...720..435A}.  We thus used an index of
$\Gamma=2.4$ in Equation~\eref{eq:counts}.

\subsubsection{Galactic Foreground and Isotropic Background}\label{ssec:bckgs}
The Galactic foreground and the diffuse isotropic background were
implemented as described in Equation~\eref{eq:Dgen}. The total diffuse
contribution was modeled by
\begin{equation}
x^\mpd_\mathrm{diff} = A_\mathrm{gal}\,x^\mpd_\mathrm{gal} + 
\frac{x^\mpd_\mathrm{iso} }{F_\mathrm{iso}}\,F_\mathrm{iso} \,,
\end{equation} 
with $A_\mathrm{gal}$ being a normalization parameter of the Galactic
foreground component $x^\mpd_\mathrm{gal}$.  For the isotropic
component $x^\mpd_\mathrm{iso}$ the integral flux $F_\mathrm{iso}$ was
directly used as a sampling parameter, in order to have physical units
of flux.

\paragraph{Galactic Foreground}
The Galactic foreground was modeled using a template
(\texttt{gll\_iem\_v05\_rev1.fit}) developed by the Fermi-LAT
collaboration to compile the 3FGL catalog
\citep{2015ApJS..218...23A}\footnote{See also
  http://fermi.gsfc.nasa.gov/ssc/data/access/lat/
  \\BackgroundModels.html for details.}.  The Galactic foreground
model is based on a fit of multiple templates to the gamma-ray data.
The templates used are radio-derived gas maps splitted into various
galactocentric annuli, a further dust-derived gas map, an inverse
Compton emission template derived with the GALPROP
code,\footnote{http://galprop.stanford.edu/} and some patches designed
to describe observed residual emission not well represented by the
pervious templates, such as the Fermi bubbles and Galactic Loop\,I.

The Galactic foreground template comprises predictions of the
differential intensity at 30 logarithmically spaced energies in the
interval between 50\,MeV and 600\,GeV.  The spatial map resolution is
$0.125^\circ$, which was resampled to match the pixelization scheme
and spatial resolutions used in our analysis.  The predicted number of
counts per pixel $x^\mpd_\mathrm{gal}$ was obtained from integration
in the energy range $[E_\mathrm{min}, E_\mathrm{max}]$ as described in
Section~\ref{ssec:gen_funcs}.

In order to include the effects caused by the point spread function
(PSF) of the detector, we smoothed the final template map with a
Gaussian kernel of $0.5^\circ$.  We checked that systematics of this
coarse PSF approximation (see Section~\ref{sec:Fermi_data}) were
negligible, by comparing kernels with half-widths between $0^\circ$
and $1^\circ$.

\begin{figure}[t]
\epsscale{1.0}
\plotone{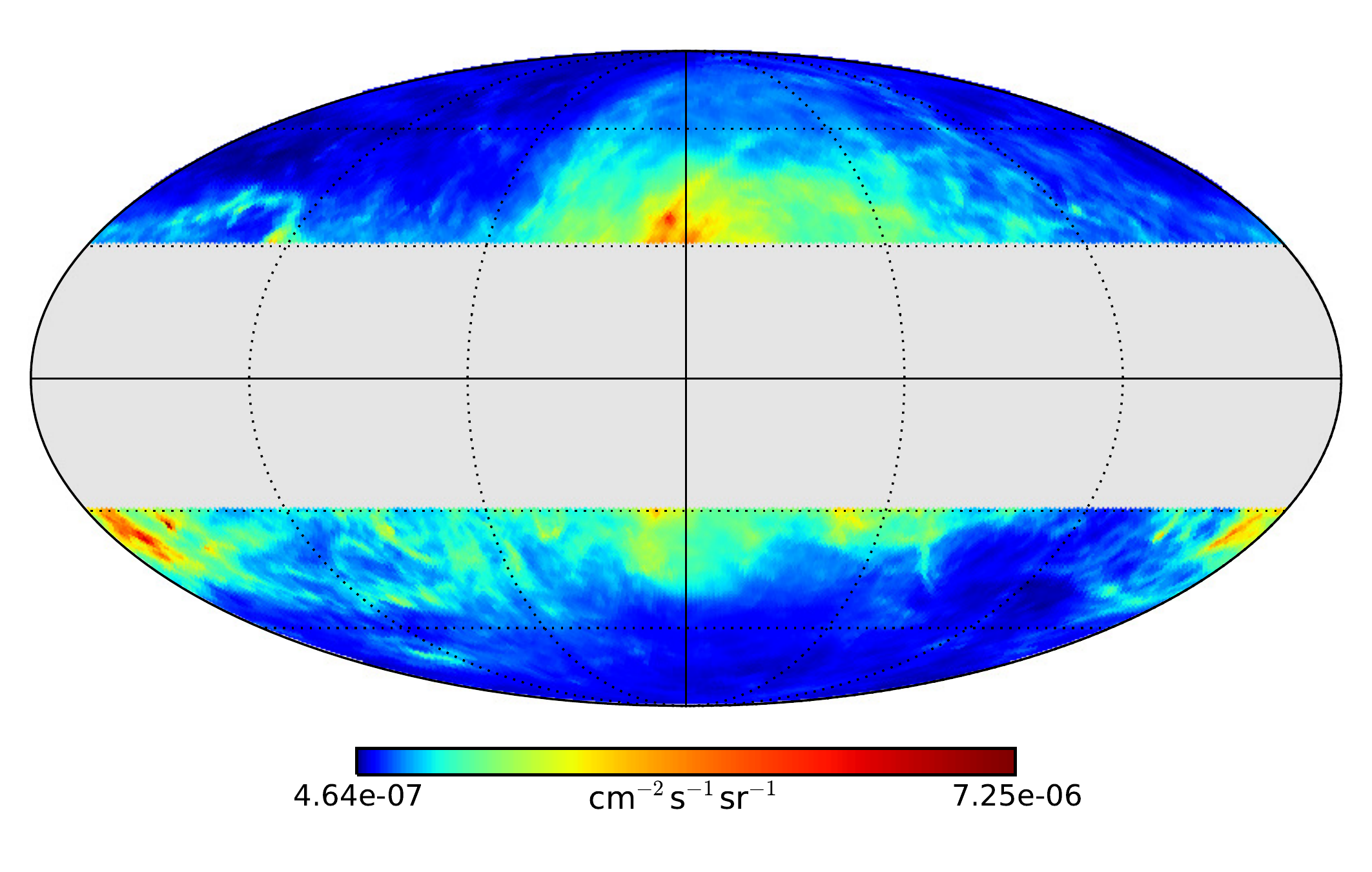}
\caption{Diffuse Galactic foreground emission between 1\,GeV and
  10\,GeV as predicted by the model template (see text for
  details). The integral flux $F_\mathrm{gal}^\mpd$ is plotted in
  Galactic coordinates $(l,b)$ using a Mollweide projection of the
  sphere. The Galactic Center is in the middle of the map. The
  Galactic plane has been masked for latitudes $|b| < 30^\circ$ (in
  gray). The color mapping is log-linear.
\label{fig:f1}}
\end{figure}

Figure \ref{fig:f1} shows the model prediction for the diffuse
Galactic foreground flux between $1\,\mathrm{GeV}$ and
$10\,\mathrm{GeV}$ and Galactic latitudes $|b| \geq 30^\circ$. The
complex spatial morphology of the Galactic foreground emission is
evident. The intensity of Galactic foreground emission significantly
decreases with increasing latitude.  The integral flux predicted by
the model in the energy range $\Delta E$ between 1\,GeV and 10\,GeV is
$F_\mathrm{gal}(\Delta E) \simeq 4.69\times
10^{-5}\,\mathrm{cm}^{-2}\,\mathrm{s}^{-1}$ for the full sky and
$F_\mathrm{gal}(\Delta E; |b| \geq 30^\circ) \simeq 6.42\times
10^{-6}\,\mathrm{cm}^{-2}\,\mathrm{s}^{-1}$ for high Galactic
latitudes $|b| \geq 30^\circ$.

Since the model reported in \texttt{gll\_iem\_v05\_rev1.fit} was
originally normalized to best reproduce the whole gamma-ray sky, we
allowed for an overall different normalization parameter
$A_\mathrm{gal}$ in our analysis, given that we explored different
ROIs. Nonetheless, $A_\mathrm{gal}$ is expected to be of order unity
when considered a free fit parameter.

\paragraph{Isotropic Background}
The expected counts for the diffuse isotropic background component
$F_\mathrm{iso}$ were derived assuming a power-law spectrum with
spectral index $\Gamma_\mathrm{iso} = 2.3$
\citep{2015ApJ...799...86A}. We verified that using the specific
energy spectrum template provided by the \emph{Fermi}-LAT
Collaboration (\texttt{iso\_clean\_front\_v05.txt}) had no impact on
our results.

\subsection{PSF Smearing}\label{sec:psf}
The detected photon flux from point sources is distributed over a
certain area of the sky as caused by the finite PSF of the instrument.
Photon contributions from individual point sources are therefore
spread over several adjacent pixels, each containing a fraction $f$ of
the total photon flux from the source. Apart from being a function of
the pixel position, the fractions $f$ depend on the location of a
source within its central pixel. A smaller pixel size, i.e., a
higher-resolution map, decreases the values of $f$, corresponding to a
relatively larger PSF smoothing.

Equation~\eqref{eq:xm} must therefore be corrected for PSF
effects. Following \pmh, the PSF correction was incorporated by
statistical means, considering the average distribution of fractions
$\rho(f)$ among pixels for a given pixel size. To determine $\rho(f)$,
we used Monte Carlo simulations distributing a number of $N$ fiducial
point sources at random positions on the sky. The sources were
convolved with the detector PSF, and the fractions $f_i$,
$i=1,\dots,N_\mathrm{pix}$, were evaluated for each source. The sums
of the fractions $f_i$ were normalized to 1.  We used the effective
detector PSF derived from the data set analyzed below, corresponding
to the specific event selection cuts used in our analysis. The
effective detector PSF was obtained by averaging the detector PSF over
energy and spectral index distribution.  This is further explained in
Section~\ref{sec:Fermi_data}.

The average distribution function $\rho(f)$ is then given by
\begin{equation}\label{eq:rho_f}
\rho(f) = \left. \frac{\Delta N(f)}{N \Delta f} 
\right|_{\Delta f \rightarrow 0,\, N \rightarrow \infty}\,,
\end{equation}
where $\Delta N(f)$ denotes the number of fractions in the interval
$(f, f+\Delta f)$.  The distribution obeys the normalization condition
\begin{equation}\label{eq:rho_f_norm}
\int \mathrm{d}f\,f \rho(f) = 1\,.
\end{equation}
The expected number of $m$-photon sources in a given pixel corrected
for PSF effects is given by
\begin{equation}\label{eq:xm_corr}
x^\mpd_m = \Omega_\mathrm{pix} \int_0^\infty \mathrm{d}S 
\frac{\mathrm{d}N}{\mathrm{d}S} \int \mathrm{d}f \rho(f)
\frac{(f\,\mathcal{C}^\mpd\!(S) )^m}{m!} e^{-f\,\mathcal{C}^\mpd\!(S)}\,.
\end{equation}

\begin{figure}[t]
\epsscale{1.15}
\plotone{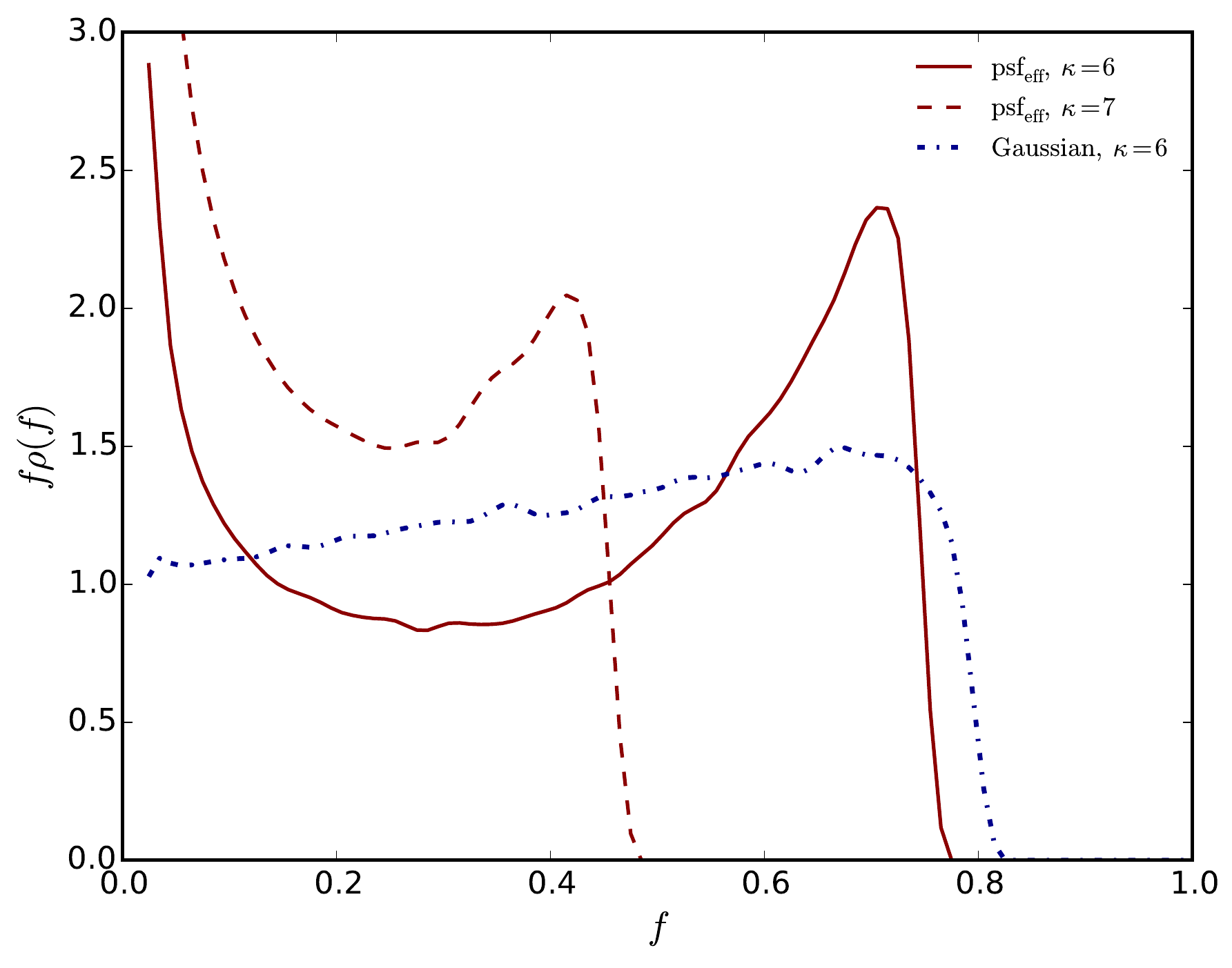}
\caption{Average distribution function $\rho(f)$ of the fractional
  photon flux $f$ from a point source in a given pixel. The solid and
  dashed red lines depict the distribution function for the effective
  detector PSF for two different pixel sizes: a HEALPix grid with
  resolution parameter $\kappa=6$ (solid) and $\kappa=7$ (dashed); see
  Section~\ref{sec:Fermi_data} for details. The dot-dashed blue line
  depicts the $\kappa = 6$ distribution function for a Gaussian PSF
  with a 68\% containment radius resembling the one of the actual
  effective PSF. The average distributions have been derived from
  Monte Carlo simulations of \mbox{$5\times 10^4$} fiducial point
  sources at random positions on the sky.  The numerical resolution is
  $\Delta f = 0.01$.
\label{fig:psf_corr}}
\end{figure}

Figure~\ref{fig:psf_corr} depicts the distribution function $\rho(f)$
derived for the effective PSF of the data set for two different pixel
sizes. The function $\rho(f)$ is also shown assuming a Gaussian PSF
with a 68\% containment radius resembling the one of the actual PSF.
Compared to the Gaussian case, the more pronounced peak of the
detector PSF reflects in a strongly peaked $\rho(f)$ at large flux
fractions. Reducing the pixel size, i.e., effectively increasing PSF
smoothing (in the sense of this analysis), shifts the peak of
$\rho(f)$ to smaller $f$. The impact of the large tails of the
detector PSF becomes evident at small fractions.

\subsection{Data Fitting}
To fit the model ($H$) to a given data set (D), we used the method of
maximum likelihood (see, e.g., \citet{2014ChPhC..38i0001O} for a
review). We defined the likelihood $\mathcal{L}({\bf \Theta}) \equiv
P(\mathrm{D}|{\bf \Theta}, H)$ in two different ways, which we refer
to as L1 and L2 in the following. The likelihood function describes
the probability distribution function $P$ of obtaining the data set
$\mathrm{D}$, under the assumption of the model (hypothesis) $H$ with
a given parameter set ${\bf \Theta}$.

For a source-count distribution following an MBPL with $N_\mathrm{b}$
breaks and the previously defined background contributions, the
parameter vector is given by
\begin{equation}
{\bf \Theta} = (A_\mathrm{S},S_{\mathrm{b}1},\dots,S_{\mathrm{b}N_\mathrm{b}},
n_1,\dots,n_{N_\mathrm{b}+1},A_\mathrm{gal},F_\mathrm{iso}),
\end{equation}
containing $N_{\bf \Theta} = 2N_\mathrm{b} + 4$ free parameters.

\subsubsection{Likelihood L1}
The L1 approach resembles the method of the simple \opdf (see MH11).
Given the probability distribution $p_k$ for a given ${\bf \Theta}$,
the expected number of pixels containing $k$ photons is $\nu_k({\bf
  \Theta}) = N_\mathrm{pix}\,p_k({\bf \Theta})$. The probability of
finding $n_k$ pixels with $k$ photons follows a Poissonian (if pixels
are considered statistically independent), resulting in the total
likelihood function
\begin{equation}\label{eq:L1}
\mathcal{L}_1({\bf \Theta}) = \prod_{k=0}^{k_\mathrm{max}} 
\frac{\nu_k({\bf \Theta})^{n_k}}{n_k!} e^{-\nu_k({\bf \Theta})}\,,
\end{equation}
where $k_\mathrm{max}$ denotes the maximum value of $k$ considered in
the analysis.

\subsubsection{Likelihood L2}
The simple \opdf\ approach can be improved by including morphological
information provided by templates.  The L2 approach defines a
likelihood function that depends on the location of the pixel.  The
probability of finding $k$ photons in a pixel $p$ is given by
$p_k^\mpd$ for a given parameter vector ${\bf \Theta}$. We emphasize
that now the data set comprises the measured number of photons $k_p$
in each pixel $p$, instead of the $n_k$-histogram considered in
L1. For clarity, the function $p_k^\mpd$ is therefore denoted by
$P(k_p) \equiv p_k^\mpd$ in the following. The likelihood function for
the entire ROI is then given by
\begin{equation}
\mathcal{L}_2 ({\bf \Theta}) = \prod_{p=1}^{N_\mathrm{pix}} P(k_p)\,.
\end{equation}

It should be noted that the L2 approach is a direct generalization of
the L1 approach.  The \opdf\ approach already provides the PDF for
each pixel, and it is thus natural to use the appropriate PDF for each
pixel instead of using the average one and comparing it with the
$n_k$-histogram.  The L2 approach can then be seen as building a
different $n_k$-histogram for each pixel, comparing it with the
appropriate $p_k$ distribution and then joining the likelihoods of all
the pixels together in the global L2 one.  The fact that for each
pixel the $n_k$-histogram actually reduces to a single count does not
pose a matter-of-principle problem.

\subsubsection{Bayesian Parameter Estimation}\label{sssec:par_est}
The sampling of the likelihood functions $\mathcal{L}_1({\bf \Theta})$
and $\mathcal{L}_2({\bf \Theta})$ is numerically demanding and
requires advanced Markov Chain Monte Carlo (MCMC) methods to account
for multimodal behavior and multiparameter degeneracies. We used the
multimodal nested sampling algorithm
\texttt{MultiNest}\footnote{Version v3.8, 2014 October}
\citep{2008MNRAS.384..449F,2009MNRAS.398.1601F,2013arXiv1306.2144F} to
sample the posterior distribution $P({\bf \Theta}|\mathrm{D},H)$. The
posterior is defined by Bayes's theorem as $P({\bf
  \Theta}|\mathrm{D},H) = \mathcal{L}({\bf \Theta}) \pi({\bf \Theta})
/ \mathcal{Z}$, where $\mathcal{Z} \equiv P(\mathrm{D}|H)$ is the
Bayesian evidence given by
\begin{equation}
\mathcal{Z} = \int \mathcal{L}({\bf \Theta}) 
\pi({\bf \Theta})\,\mathrm{d}^{N_{\bf \Theta}} {\bf \Theta}\,,
\end{equation}
and $\pi ({\bf \Theta})$ is the prior. \texttt{MultiNest} was used in
its recommended configuration regarding sampling efficiency. For our
analysis setups, we checked that sufficient sampling accuracy was
reached using 1,500 live points with a tolerance setting of 0.2.
Final acceptance rates typically resulted in values between 5\% and
10\%, while the final samples of approximately equal-weight parameter
space points consisted of about $10^4$ points.

From the marginalized one-dimensional posterior distributions, for
each parameter we quote the median, and the lower and upper
statistical uncertainties were derived from the 15.85\% and 84.15\%
quantiles, respectively. In the case of log-flat priors (see below),
we assumed the marginalized posterior distribution to be Gaussian for
deriving single-parameter uncertainty estimates in linear space.  The
derivation of uncertainty bands of the \dnds\ fit exploited the same
method but using the full posterior.

Priors were chosen to be flat or log flat, depending on the numerical
range required for a parameter.  Details are discussed in
Section~\ref{sssec:priors}.

\subsubsection{Frequentist Parameter Estimation}\label{sssec:par_est_freq}
Bayesian parameter estimates from the posterior distributions are
compared to parameter estimates employing the frequentist
approach. The MCMC method intrinsically provides samples of a
posterior distribution that depends on the prior. Nonetheless, if the
number of samples is sufficiently high such that also the tails of the
posterior are well explored, it can be assumed that the final sample
reasonably explored the likelihood function. Profile likelihood
functions \citep[see, e.g.,][]{2005NIMPA.551..493R} can be built from
the posterior sample.  In particular, we built the profile likelihood
of the \dnds\ fit and one-dimensional profile likelihoods for each
parameter.  We quote the maximum likelihood parameter values and 68\%
confidence level (CL) intervals derived under the assumption that the
profiled $-2\ln \mathcal{L}$ follows a chi-squared distribution with
one degree of freedom, i.e., we quote the values of the parameters for
which $-2\Delta \ln \mathcal{L}=1$.\footnote{We defined $\Delta \ln
  \mathcal{L} = \ln \left(\mathcal{L}/\mathcal{L}_\mathrm{max}
  \right)$, where $\mathcal{L}_\mathrm{max} = \max(\mathcal{L})$.}
The advantage of profile likelihood parameter estimates is that they
are prior independent.

\section{\emph{FERMI}-LAT DATA}\label{sec:Fermi_data}
The analysis is based on all-sky gamma-ray data that were recorded
with the \emph{Fermi}-LAT\footnote{\emph{Fermi}-LAT data are publicly
  available at
  http://heasarc.gsfc.nasa.gov/FTP/fermi/data/lat/weekly/p7v6d/}
within the first 6 years of the mission.\footnote{The data set covers
  the time period between 2008 August 4 (239,557,417 MET) and 2014
  August 4 (428,859,819 MET).}  Event selection and processing were
performed with the public version of the Fermi Science Tools (v9r33p0,
release date 2014 May 20).\footnote{See
  http://fermi.gsfc.nasa.gov/ssc/data/analysis/software/} We used
\texttt{Pass 7 Reprocessed} (\texttt{P7REP}) data along with
\texttt{P7REP\_V15} instrument response functions.

The application of the analysis method presented here is restricted to
the energy bin between $E_\mathrm{min}=1\,\mathrm{GeV}$ and
$E_\mathrm{max}=10\,\mathrm{GeV}$.  The lower bound in energy was
motivated by the size of the PSF, which increases significantly to
values larger than $1^\circ$ for energies below 1\,GeV
\citep{2012ApJS..203....4A}.  The significant smoothing of point
sources caused by a larger PSF may lead to large uncertainties in this
analysis (see Section~\ref{sec:psf}).  Effects of a possible energy
dependence of \dnds\ are mitigated by selecting an upper bound of
10\,GeV.

Data selection was restricted to events passing \texttt{CLEAN} event
classification, as recommended for diffuse gamma-ray analyses. We
furthermore required \texttt{FRONT}-converting events, in order to
select events with a better PSF and to avoid a significant broadening
of the effective PSF.  Contamination from the Earth's limb was
suppressed by allowing a maximum zenith angle of $90^\circ$. We used
standard quality selection criteria, i.e., \texttt{DATA\_QUAL==1} and
\texttt{LAT\_CONFIG==1}, and the rocking angle of the satellite was
constrained to values smaller than $52^\circ$. The data selection
tasks were carried out with the tools \texttt{gtselect} and
\texttt{gtmktime}.

\begin{figure}[t]
\epsscale{1.00}
\plotone{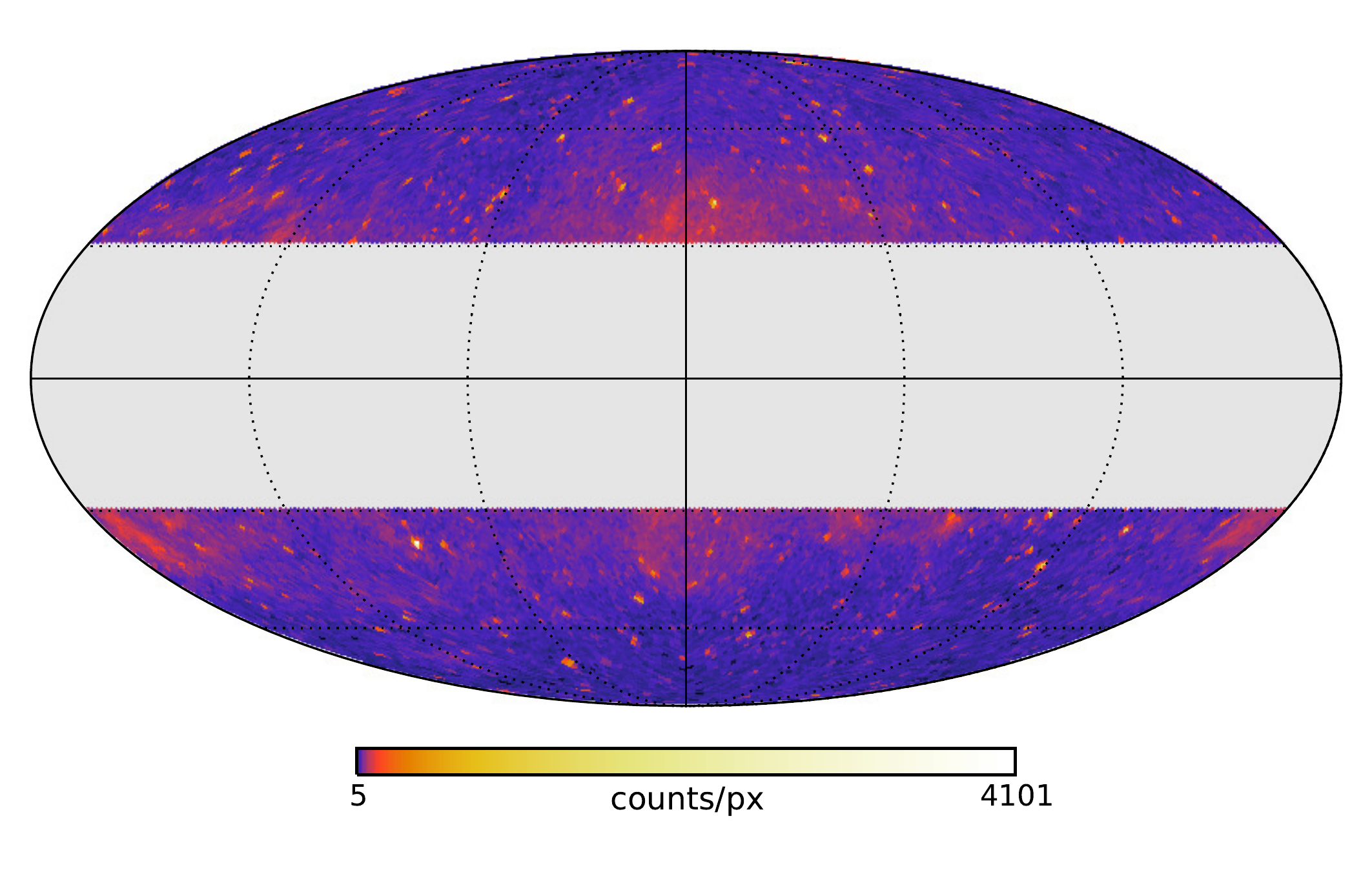}
\caption{Photon counts map between 1\,GeV and 10\,GeV as derived from
  the \emph{Fermi}-LAT data covering the time period of 6 years. The
  Mollweide projection of the celestial sphere is shown in Galactic
  coordinates $(l,b)$, centered on the position of the Galactic
  Center. The Galactic plane has been omitted within $|b| < 30^\circ$
  (gray area).  The color scale is log-linear.
\label{fig:cnts}}
\end{figure}

The resulting counts map was pixelized with \texttt{gtbin} using the
equal-area HEALPix pixelization scheme
\citep{2005ApJ...622..759G}. The resolution of the discretized map is
given by the pixel size, $\theta_\mathrm{pix} =
\sqrt{\Omega_\mathrm{pix}}$.  For the statistical analysis employed
here, the optimum resolution is expected to be of the order of the
PSF: while undersampling the PSF leads to information loss on
small-scale structures such as faint point sources, oversampling
increases the statistical uncertainty on the number of counts per
pixel.  We thus compared two choices for the map resolution, where
$\kappa$ denotes the HEALPix resolution parameter: $\kappa=6$
($N_\mathrm{side} = 64$),\footnote{The number of pixels of the all-sky
  map is given by $N_\mathrm{pix} = 12 N_\mathrm{side}^2$;
  $N_\mathrm{side}$ can be obtained from the resolution parameter by
  $N_\mathrm{side} = 2^\kappa$.} corresponding to a resolution of
\mbox{$\sim\!0.92^\circ$}, and $\kappa=7$ ($N_\mathrm{side} = 128$),
corresponding to a resolution of \mbox{$\sim\!0.46^\circ$}.  These
choices slightly undersample or oversample the actual PSF,
respectively.

We used \texttt{gtltcube} and \texttt{gtexpcube2} to derive the
exposure map as a function of energy.  The lifetime cube was
calculated on a spatial grid with a $1^\circ$ spacing. The exposure
map imposed a spatial grating of $0.125^\circ$ (in Cartesian
projection) and the same energy binning as used in the Galactic
foreground template. The map was projected into HEALPix afterwards.

The statistical analysis requires a careful correction for effects
imposed by the PSF; see Section~\ref{sec:psf} for details.  The PSF of
the data set was calculated with \texttt{gtpsf} for a fiducial
Galactic position $(l,b)=(45^\circ,45^\circ)$ as a function of the
displacement angle $\theta$ and the energy $E$.  We checked that
changes of the PSF at other celestial positions were negligible. Given
that the PSF strongly depends on energy, analyzing data in a single
energy bin requires appropriate averaging. The effective PSF of the
data set was calculated by weighting with the energy-dependent
exposure and power-law type energy spectra $\propto E^{-\Gamma}$,
\begin{equation}
\mathrm{psf} (\theta,\Delta E) = \frac{\int_{E_\mathrm{min}}^{E_\mathrm{max}}
\mathrm{d}E\,E^{-\Gamma}\,\mathcal{E}(E)\,\mathrm{psf} (\theta,E) }
{ \int_{E_\mathrm{min}}^{E_\mathrm{max}} \mathrm{d}E\,E^{-\Gamma}\,\mathcal{E}(E) } ,
\end{equation}
where $\mathcal{E}(E) = \langle \mathcal{E}^\mpd (E)
\rangle_\mathrm{ROI}$ denotes the exposure averaged over the ROI. An
average spectral index $\Gamma = 2.4$ was assumed.

The analysis presented in this article was carried out for high
Galactic latitudes $|b|\geq 30^\circ$, aiming at measuring the
source-count distribution and compositon of the extragalactic
gamma-ray sky. For $|b|\geq 30^\circ$ (corresponding to
$f_\mathrm{ROI}=0.5$), the photon counts map comprises 862,459 events
distributed in 24,576 pixels ($\kappa = 6$). The counts map, with a
minimum of 5 events per pixel and a maximum of 4,101 events, is shown
in Figure \ref{fig:cnts}.

\begin{figure}[t]
\epsscale{1.00}
\plotone{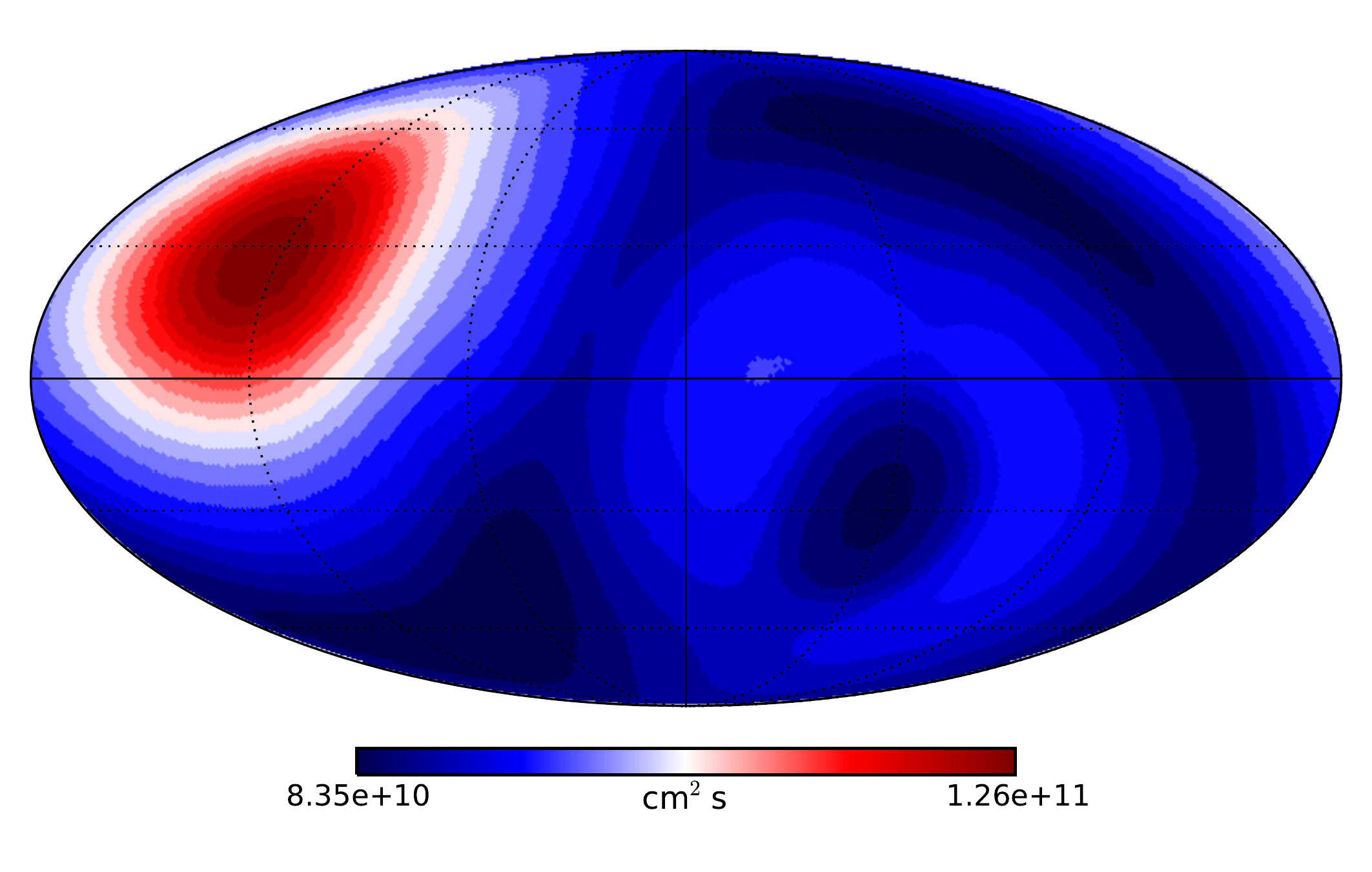}
\caption{\emph{Fermi}-LAT exposure map of the 6-year data set,
  averaged over the energy band between 1\,GeV and 10\,GeV.  The
  exposure map is divided into 20 equally spaced regions. For each
  region, the mean exposure is plotted.  The coordinate system matches
  the one used in Figure \ref{fig:cnts}.  The color mapping is
  linear. \label{fig:exp}}
\end{figure}

The energy-averaged exposure map of the data set is shown in Figure
\ref{fig:exp} for the full sky, divided into 20 equal-exposure regions
(see Equation~\eref{eq:Pexpinhomo}).  The full-sky (unbinned) exposure
varies from $8.22\times 10^{10}\,\mathrm{cm}^2\,\mathrm{s}$ to
$1.27\times 10^{11}\,\mathrm{cm}^2\,\mathrm{s}$.  The mean of the
energy-averaged exposure is $9.18\times
10^{10}\,\mathrm{cm}^2\,\mathrm{s}$ for $|b|\geq 30^\circ$.

The effective PSF width (68\% containment radius) is
$\sigma_\mathrm{psf} = 0.43^\circ$.

\section{ANALYSIS ROUTINE}\label{sec:analysis_routine}
The following section is dedicated to details of the analysis method
and to the analysis strategy developed in this article. The analysis
aims at measuring (i) the contribution from resolved and unresolved
gamma-ray point sources to the EGB, (ii) the shape of their
source-count distribution \dnds, and (iii) the resulting total
composition of the gamma-ray sky, in the energy band between 1\,GeV
and 10\,GeV.  The restriction to Galactic latitudes $|b| \geq
30^\circ$ provides a reasonable choice for ensuring that the dominant
source contributions are of extragalactic origin.

\subsection{Expected Sensitivity}\label{ssec:sensitivity}
The source-population sensitivity of the method can be estimated from
the theoretical framework discussed in Section \ref{sec:theory}. By
definition, the total PDF incorporates background components as
populations of $1$-photon sources (see
Equation~\ref{eq:Dgen}). Sources contributing on average two photons
per pixel should be clearly distinguishable from background
contributions.  The limiting sensitivity on the point-source flux is
thus given by the inverse of the average exposure, yielding a value of
$S_\mathrm{sens} \simeq 2.31\times
10^{-11}\,\mathrm{cm}^{-2}\,\mathrm{s}^{-1}$ for a pixel size
corresponding to resolution $\kappa=6$.  This value gives a
back-of-the-envelope estimate of the sensitivity to the point-source
population, while the actual sensitivity additionally depends on
quantities such as the unknown shape of the source-count distribution,
the relative contribution from foreground and background components,
and the number of evaluated pixels $N_\mathrm{pix}$ (i.e., the
Galactic latitude cut).  The actual sensitivity will be determined
from a data-driven approach in Section~\ref{sec:application}, as well
as from simulations in Appendix~\ref{app:sims}.

In comparison, the sensitivity of the 3FGL catalog drops at a flux of
$\sim\! 2.2 \times 10^{-10}\,\mathrm{cm}^{-2}\,\mathrm{s}^{-1}$ for
the energy band between 1\,GeV and 10\,GeV.\footnote{See Section~4.2
  in \citet{2015ApJS..218...23A}. The catalog threshold has been
  rescaled to the 1\,GeV to 10\,GeV energy band assuming an average
  photon index of 2.4.}  Additional sensitivity can be achieved to
lower fluxes by correcting for point-source detection
efficiency. However, determining the point-source detection efficiency
is nontrivial. The catalog detection procedure needs to be accurately
reproduced with Monte Carlo simulations and the method is not
completely free from assumptions regarding the properties of the
unresolved sources.  A clear advantage of the method employed here is,
instead, that no detection efficiency is involved.  As indicated by
the value of $S_\mathrm{sens}$, we will see that this analysis
increases the sensitivity to faint point-source populations by about
one order of magnitude with respect to the 3FGL catalog.

\subsection{Analysis Setup}
The L2 approach emerged to provide significantly higher sensitivity
than the L1 approach, as a consequence of the inclusion of spatial
information. We will thus use the second method $\mathcal{L}_2({\bf
  \Theta})$ as our reference analysis in the remainder.  We will
nonetheless present in the main text a comparison of the two
approaches, showing that they lead to consistent results.

All pixels in the ROI were considered in the calculation of the
likelihood. The upper bound on the number of photon counts per pixel,
$k_\mathrm{max}$, as used in Equation~\eref{eq:L1} was always chosen
to be slightly larger than the maximum number of counts per map pixel.

\subsection{Source-count Distribution Fit}
The source-count distribution \dnds\ was parameterized with the MBPL
defined in Equation~\eref{eq:mbpl}.  For readability, the following
terminology will be used in the remainder: the source-count
distribution is subdivided into three different regimes, defined by
splitting the covered flux range $S$ into three disjoint intervals,
\begin{equation*}
\begin{aligned}
 \left[\,0, 10^{-10} \right) \,\mathrm{cm}^{-2}\,\mathrm{s}^{-1}: 
& \quad \text{faint-source region,} \\
 \left[ 10^{-10}, 10^{-8} \right) \,\mathrm{cm}^{-2}\,\mathrm{s}^{-1}: 
& \quad \text{intermediate region,} \\
 \left[ 10^{-8}, S_\mathrm{cut} \right]\,\mathrm{cm}^{-2}\,\mathrm{s}^{-1}: 
& \quad \text{bright-source region.} \\
\end{aligned}
\end{equation*}
The quantity $S_\mathrm{cut}$ corresponds to a high cutoff flux of the
source-count distribution.  The observational determination of
$S_\mathrm{cut}$ is limited by cosmic variance, and a precise value is
therefore lacking. Unless stated otherwise, we chose a cutoff value
$S_\mathrm{cut} = 10^{-6}\,\mathrm{cm}^{-2}\,\mathrm{s}^{-1}$, which
is almost one order of magnitude higher than the flux of the brightest
source listed in the 3FGL catalog within the ROIs considered in this
work (see Section~\ref{ssec:data_mbpl}).  The stability of this choice
was checked by comparing with $S_\mathrm{cut} =
10^{-5}\,\mathrm{cm}^{-2}\,\mathrm{s}^{-1}$.

In the following, we describe our strategy to fit the
\dnds\ distribution to the data.  A validation of the analysis method
with Monte Carlo simulations is described in Appendix~\ref{app:sims}.

\subsubsection{Parameters of \dnds}\label{sssec:dnds_params}
\paragraph{Normalization}
The reference normalization flux $S_0$ was kept fixed during the fit.
A natural choice for $S_0$ would be the flux where the uncertainty
band of the \dnds\ reaches its minimum (pivot flux).  In this way,
undesired correlations among the fit parameters are minimized.  We
refrained from a systematic determination of the pivot point, but we
instead fixed $S_0$ to a value of $S_0 = 3\times
10^{-8}\,\mathrm{cm}^{-2}\,\mathrm{s}^{-1}$ after optimization
checks. We checked for robustness by varying $S_0$ within the range
$[0.1\,S_0, S_0]$, obtaining stable results.\footnote{Given that the
  choice of $S_0$ turns out to be larger than the position of the
  first break, we note that increasing the interval to larger fluxes
  is not required.}  Remaining parameter degeneracies were handled
well by the sampling.

\paragraph{Number of Breaks}
Previous works investigating the gamma-ray \dnds\ distribution with
cataloged sources concluded that the \dnds\ distribution above
$|b|>10^\circ$ is well described by a broken power law down to a flux
of $\sim 5\times 10^{-10}\,\mathrm{cm}^{-2}\,\mathrm{s}^{-1}$, with a
break at $(2.3\pm 0.6)\times
10^{-9}\,\mathrm{cm}^{-2}\,\mathrm{s}^{-1}$
\citep{2010ApJ...720..435A}.  The following analysis increases the
sensitivity to resolving point sources with a flux above $\sim 2\times
10^{-11}\,\mathrm{cm}^{-2}\,\mathrm{s}^{-1}$ and provides a
significantly smaller statistical uncertainty. We therefore
parameterized \dnds\ with up to three free breaks ($N_\mathrm{b} \leq
3$), in order to find the minimum number of breaks required to
properly fit the data. In the case of $N_\mathrm{b} = 3$, one break
was placed in the bright-source region, a second in the intermediate
region, and the last one in the faint-source region; see Section
\ref{sssec:priors} for details. We compared these results with setups
reducing $N_\mathrm{b}$ to one or two free breaks, to investigate
stability and potential shortcomings in the different approaches.

\subsubsection{Fitting Techniques}\label{sssec:fit_approach}
We employed three different techniques of fitting the
\dnds\ distribution to the data, in order to investigate the stability
of the analysis and to study the sensitivity limit. The third
technique, which we refer to as the \emph{hybrid approach}, is a
combination of the two other techniques. This hybrid approach proved
to provide the most robust results.

\paragraph{MBPL Approach}
The MBPL approach comprises fitting a pure MBPL with a number of
$N_\mathrm{b}$ free break positions. The total number of free
parameters is given by $N_{\bf \Theta} = 2N_\mathrm{b} + 4$ (including
free parameters of the background components). The parameters of the
MBPL are sampled directly.

\paragraph{Node-based Approach}
The complexity of the parameter space, including degeneracies between
breaks and power-law indices, can be reduced by imposing a grid of
$N_\mathrm{nd}$ fixed flux positions, which we refer to as nodes
$S_{\mathrm{nd}j}$, where $j=0,1,\dots,N_\mathrm{nd}-1$. Nodes are
counted starting from the one with the highest flux, in order to
maintain compatibility with the numbering of breaks in the MBPL
described in Equation~\eref{eq:mbpl}. The free parameters of the
source-count distribution correspond to the values of \dnds\ at the
positions of the nodes, i.e., $A_{\mathrm{nd} j } =
\mathrm{d}N/\mathrm{d}S\,(S_{\mathrm{nd}j})$. The index of the
power-law component below the last node, $n_\mathrm{f}$, is kept fixed
in this approach.

The parameter set $\{ A_{\mathrm{nd} j }, S_{\mathrm{nd} j },
n_\mathrm{f} \}$ can then be mapped to the MBPL parameters using
Equation~\eref{eq:mbpl}, i.e., the \dnds\ distribution between
adjacent nodes is assumed to follow power laws. Technically, it should
be noted that $S_\mathrm{cut} \equiv S_{\mathrm{nd} 0 }$ in this case.
A choice of $N_\mathrm{nd}$ nodes therefore corresponds to choosing an
MBPL with $N_\mathrm{nd} - 1$ fixed breaks. The quantity
$A_\mathrm{S}$ is to be calculated at a value close to the
decorrelation flux to ensure a stable fit.  The total number of free
parameters is given by $N_{\bf \Theta} = N_\mathrm{nd} + 2$.

While this technique comes with the advantage of reducing the
complexity of the parameter space, the choice of the node positions is
arbitrary. This can introduce biases between nodes and can thus bias
the overall \dnds\ fit.  The node-based approach is further considered
in Appendix~\ref{app:node_based}.

We note that a similar approach has been recently used by
\cite{2014MNRAS.440.2791V} for measuring the source-count distribution
of radio sources.

\paragraph{Hybrid Approach}
The hybrid approach combines the MBPL approach and the node-based
approach. Free break positions as used in the MBPL approach are
required to robustly fit the \dnds\ distribution and to determine the
sensitivity; see Section~\ref{sec:application} for
details. \textit{Fitting a pure MBPL, however, was found to
  underestimate the uncertainty band of the fit at the lower end of
  the faint-source region. In addition, the fit obtained from the
  Bayesian posterior can suffer a bias for very faint sources, as
  demonstrated by Monte-Carlo simulations in Appendix~\ref{app:sims}.}
We therefore chose to incorporate a number of nodes around the
sensitivity threshold of the analysis, resolving the issues of the
MBPL approach.

The hybrid approach is characterized by choosing a number
$N^\shy_\mathrm{b}$ of free breaks, a number $N^\shy_\mathrm{nd}$ of
nodes, and the index of the power-law component below the last node,
$n_\mathrm{f}$. We note that the lower limit of the prior of the last
free break $S_{\mathrm{b}N^\shy_\mathrm{b}}$ technically imposes a
fixed node $S_\mathrm{nd0}$, given that the first free node
$S_\mathrm{nd1}$ is continuously connected with a power law to the
MBPL component at higher fluxes.  The setup corresponds to choosing an
MBPL with $N^\shy_\mathrm{b} + N^\shy_\mathrm{nd} + 1$ breaks, with
the last ones at fixed positions.  The total number of free parameters
in the hybrid approach is $N_{\bf \Theta} = 2N^\shy_\mathrm{b} +
N^\shy_\mathrm{nd} +4$.

\begin{deluxetable*}{lllccc}
\tablecaption{Prior Ranges\label{tab:priors}}
\tablewidth{0pt}
\tablehead{
\colhead{} & \colhead{} & \colhead{} & \multicolumn{3}{c}{Prior Range} \\ \cline{4-6}
\colhead{Method} & \colhead{Parameter\tablenotemark{a}} & \colhead{Prior} & 
\colhead{$N_\mathrm{b} = 1$} & \colhead{$N_\mathrm{b} = 2$} & \colhead{$N_\mathrm{b} = 3$}
}
\startdata
\multirow{3}{*}{Generic}  & $A_\mathrm{S}$ & log-flat & [1,\,30] & [1,\,30] & [1,\,30] \\
  & $A_\mathrm{gal}$ & flat & [0.95,\,1.1] & [0.95,\,1.1] & [0.95,\,1.1] \\
  & $F_\mathrm{iso}$ & log-flat & [0.5,\,5] & [0.5,\,5] & [0.5,\,5] \\
\tableline
\multirow{7}{*}{MBPL} & $S_\mathrm{b1}$ & log-flat & [3E-13,\,5E-8] & [3E-9,\,5E-8]  & [3E-9,\,5E-8] \\
  & $S_\mathrm{b2}$ & log-flat & \nodata & [3E-13,\,3E-9]  & [2E-11,\,3E-9] \\
  & $S_\mathrm{b3}$ & log-flat & \nodata & \nodata & [3E-13,\,2E-11] \\
  & $n_1$ & flat & [1.0,\,4.3]\tablenotemark{b} & [2.05,\,4.3] & [2.05,\,4.3] \\
  & $n_2$ & flat & [-2.0,\,2.0] & [1.4,\,2.3] & [1.7,\,2.2] \\
  & $n_3$ & flat & \nodata & [-2.0,\,2.0] & [1.4,\,2.3] \\
  & $n_4$ & flat & \nodata & \nodata & [-2.0,\,2.0] \\
\tableline
\multirow{10}{*}{Hybrid} & $S_\mathrm{b1}$ & log-flat & [1E-11,\,5E-8] & [3E-9,\,5E-8]  & [3E-9,\,5E-8]  \\
  & $S_\mathrm{b2}$ & log-flat & \nodata & [1E-11,\,3E-9]  & [2E-10,\,3E-9] \\
  & $S_\mathrm{b3}$ & log-flat & \nodata & \nodata & [1E-11,\,2E-10] \\
  & $n_1$ & flat & [2.05,\,4.3] & [2.05,\,4.3] & [2.05,\,4.3] \\
  & $n_2$ & flat & [1.4,\,2.3] & [1.7,\,2.3] & [1.7,\,2.3] \\
  & $n_3$ & flat & \nodata & [1.3,\,3.0] & [1.4,\,2.2] \\
  & $n_4$ & flat & \nodata & \nodata & [1.3,\,3.0] \\
  & $A_\mathrm{nd1}$ & log-flat & [1,\,300] & [1,\,300] & [1,\,300] \\
  & $S_\mathrm{nd1}$ & fixed & 5E-12 & 5E-12 & 5E-12 \\
  & $n_\mathrm{f}$ & fixed & -10 & -10 & -10 
\enddata
\tablecomments{Prior types and ranges used for the different
  \dnds\ and background parameterizations investigated in this work.
  Either the \dnds\ was parameterized with a pure MBPL, or the MBPL
  was extended with a node (hybrid approach).  For the node-based
  approach see Appendix~\ref{app:node_based}.  The table lists the
  prior ranges used for the MBPL and hybrid approaches, given a number
  of $N_\mathrm{b}$ or $N^\shy_\mathrm{b}$ free breaks,
  respectively. Priors listed in the first panel "Generic" were used
  in both setups identically. Ellipses indicate parameters not present
  in the specific model.}
\tablenotetext{a}{The normalization $A_\mathrm{S}$ is given in units
  of $10^7\,\mathrm{s}\,\mathrm{cm}^2\,\mathrm{sr}^{-1}$, while the
  normalization of the node $A_\mathrm{nd1}$ is given in
  $10^{14}\,\mathrm{s}\,\mathrm{cm}^2\,\mathrm{sr}^{-1}$.  The
  normalizations refer to $S_0 = 3\times
  10^{-8}\,\mathrm{cm}^{-2}\,\mathrm{s}^{-1}$. The breaks
  $S_{\mathrm{b} 1}$, $S_{\mathrm{b} 2}$, and $S_{\mathrm{b} 3}$, as
  well as the node position $S_\mathrm{nd1}$, are given in units of
  $\mathrm{cm}^{-2}\,\mathrm{s}^{-1}$.  The diffuse flux component
  $F_\mathrm{iso}$ is given in units of
  $10^{-7}\,\mathrm{cm}^{-2}\,\mathrm{s}^{-1}\,\mathrm{sr}^{-1}$.  All
  other quantities are dimensionless.}
\tablenotetext{b}{The lower prior limit poses an exception to our
  requirement of $n_1 > 2$ (see Section~\ref{sssec:dnds_model}).
  Because only a single break is allowed in this model, a suitable
  prior coverage must include the cases of a faint break with a hard
  index $n_1 < 2$, and a break in the bright-source region with a
  consequently softer index $n_1 > 2$.}
\end{deluxetable*}

\subsubsection{Priors}\label{sssec:priors}
We used log-flat priors for the normalization $A_\mathrm{S}$, the
nodes $A_{\mathrm{nd} j }$, the breaks $S_{\mathrm{b} j }$, and the
isotropic diffuse background flux $F_\mathrm{iso}$, while the indices
$n_j$ and the normalization of the Galactic foreground map
$A_\mathrm{gal}$ were sampled with flat priors. Prior types and prior
ranges are listed in Table \ref{tab:priors} for the MBPL and hybrid
approaches.  In general, priors were limited to physically reasonable
ranges. Prior ranges were chosen to cover the posterior distributions
well.

In particular, data from the 3FGL catalog motivate that $S^2
\mathrm{d}N/\mathrm{d}S \simeq
10^{-11}\,\mathrm{cm}^{-2}\,\mathrm{s}^{-1}\,\mathrm{deg}^{-2}$ in the
intermediate region; see Section~\ref{sec:application}. The range of
the prior for $A_\mathrm{S}$ was therefore adjusted to cover the
corresponding interval between $3\times
10^{-12}\,\mathrm{cm}^{-2}\,\mathrm{s}^{-1}\,\mathrm{deg}^{-2}$ and
$8\times
10^{-11}\,\mathrm{cm}^{-2}\,\mathrm{s}^{-1}\,\mathrm{deg}^{-2}$ at
least (assuming an index of 2).  The ranges of the priors for the node
normalizations were chosen similarly, but reducing the lower bound to
a value of $\sim
10^{-12}\,\mathrm{cm}^{-2}\,\mathrm{s}^{-1}\,\mathrm{deg}^{-2}$.

The ranges of the priors for the breaks were chosen to connect
continuously and not to overlap, preserving a well-defined order of
the break points.  For both the MBPL and the hybrid approach, the
upper bound of the first break $S_\mathrm{b1}$ approximately matched
the bright end of the 3FGL data points (excluding the brightest
source). It is advantageous to keep the prior range for the first
break sufficiently small, in order to reduce a possible bias of the
intermediate region by bright sources (mediated through the index
$n_2$).  For the MBPL approach, the lower bound of the last break was
chosen almost two orders of magnitude below the sensitivity estimate
of $S_\mathrm{sens} \simeq 2\times
10^{-11}\,\mathrm{cm}^{-2}\,\mathrm{s}^{-1}$, to fully explore the
sensitivity range. In the case of three breaks, the lower bound of the
intermediate break was selected to match the sensitivity estimate. For
the hybrid approach, the lower bound of the last free break was set to
$\sim\! S_\mathrm{sens}/2$\,. We comment on the choice of the nodes in
Section~\ref{ssec:data_hybrid}.

Index ranges were selected according to expectations, allowing enough
freedom to explore the parameter space. The stability of these choices
was checked iteratively. For the MBPL approach, the lower bound of the
last index allowed for a sharp cutoff of the \dnds\ distribution. For
the hybrid approach, the index $n_\mathrm{f}$ was fixed to a value of
$-10$, introducing a sharp cutoff manually. This choice will be
motivated in Section~\ref{ssec:data_mbpl}.

As discussed in Section \ref{ssec:bckgs}, $A_\mathrm{gal}$ is expected
to be of order unity. The selection of the prior boundaries for
$F_\mathrm{iso}$ was based on previous measurements
\citep[see][]{2015ApJ...799...86A} and was further motivated
iteratively.

Prior ranges reported in Table \ref{tab:priors} are further discussed
in Section~\ref{sec:application}.

\subsubsection{Exposure Correction: $N_\mathrm{exp}$}
The results were checked for robustness with respect to variations of
$N_\mathrm{exp}$ (see Section~\ref{ssec:gen_funcs}).  We found that
the choice of this parameter is critical for a correct recovery of the
final result, and it is closely related to the sensitivity of the
analysis.  In particular, small values $\lesssim 5$ were found
insufficient. Results were stabilized by using at least
$N_\mathrm{exp}=15$ contours, and we tested that increasing up to
$N_\mathrm{exp}=40$ did not have further impact.  Insufficient
sampling of the exposure (i.e., small values of $N_\mathrm{exp}$) was
seen to affect the faint end of the \dnds\ by introducing an early
cutoff and attributing a larger flux to the isotropic component.  At
the same time, the best-fit likelihood using small $N_\mathrm{exp}$
values was significantly smaller than the one obtained choosing larger
values, indicating that indeed the sampling was insufficient.  As a
final reference value we chose $N_\mathrm{exp}=20$.

\subsubsection{3FGL Catalog Data}\label{ssec:3FGLpoints}
The results are compared to the differential (\dnds) and integral
($N(>S)$) source-count distributions derived from the 3FGL catalog for
the same energy band and ROI. The method of deriving the source-count
distribution from catalog data is described in
Appendix~\ref{app:dnds_cat}.

\section{APPLICATION TO THE DATA}\label{sec:application}
In this section, a detailed description and discussion of the data
analysis and all setups chosen in this article are given.  Final
results are summarized in Section~\ref{sec:conclusions}.

The data were fit by employing the MBPL approach and the hybrid
approach consecutively.  The use of the hybrid approach was mostly
chosen to inspect the uncertainties in the faint-source region. It
should be emphasized that the prior of the last free break and the
position of the node depend on the results obtained with the MBPL
approach.

All analyses were carried out using two different pixel sizes, i.e.,
HEALPix grids of order $\kappa = 6$ ($N_\mathrm{side}=64$) and $\kappa
= 7$ ($N_\mathrm{side}=128$). Details are discussed in Section
\ref{sec:Fermi_data}. We chose $\kappa=6$ as a reference, due to the
expected sensitivity gain.  All parameters were stable within their
uncertainty bands against changes to $\kappa=7$. Results using
$\kappa=7$ are shown in Section~\ref{ssec:hp7}.

\subsection{MBPL Approach}\label{ssec:data_mbpl}
\begin{figure*}[t]
\begin{centering}
\subfigure[MBPL, $N_\mathrm{b}=2$]{%
  \includegraphics[width=0.49\textwidth]{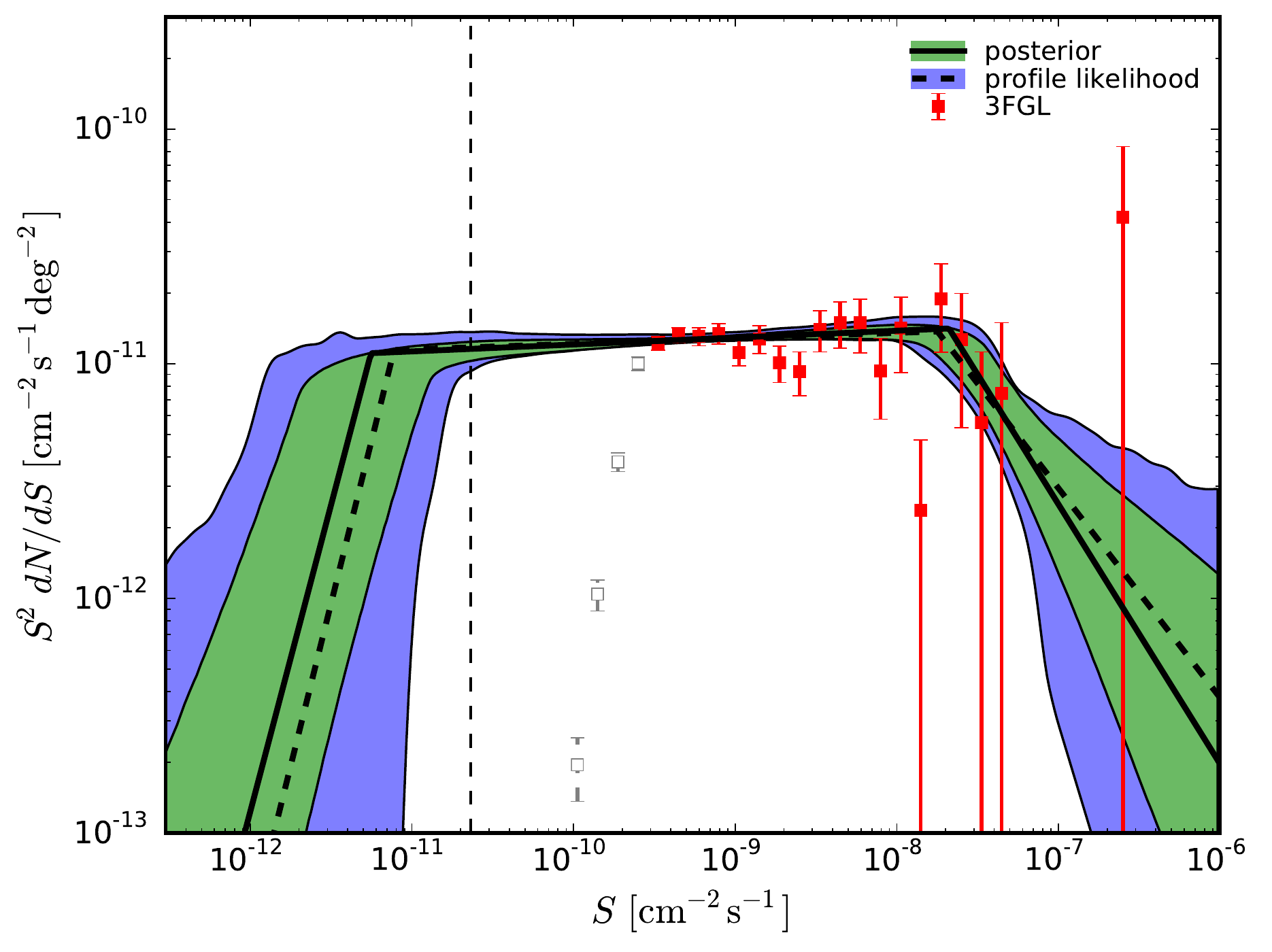}
  \label{sfig:s2dnds_mbpl_Nb2}
}
\subfigure[MBPL, $N_\mathrm{b}=3$]{%
  \includegraphics[width=0.49\textwidth]{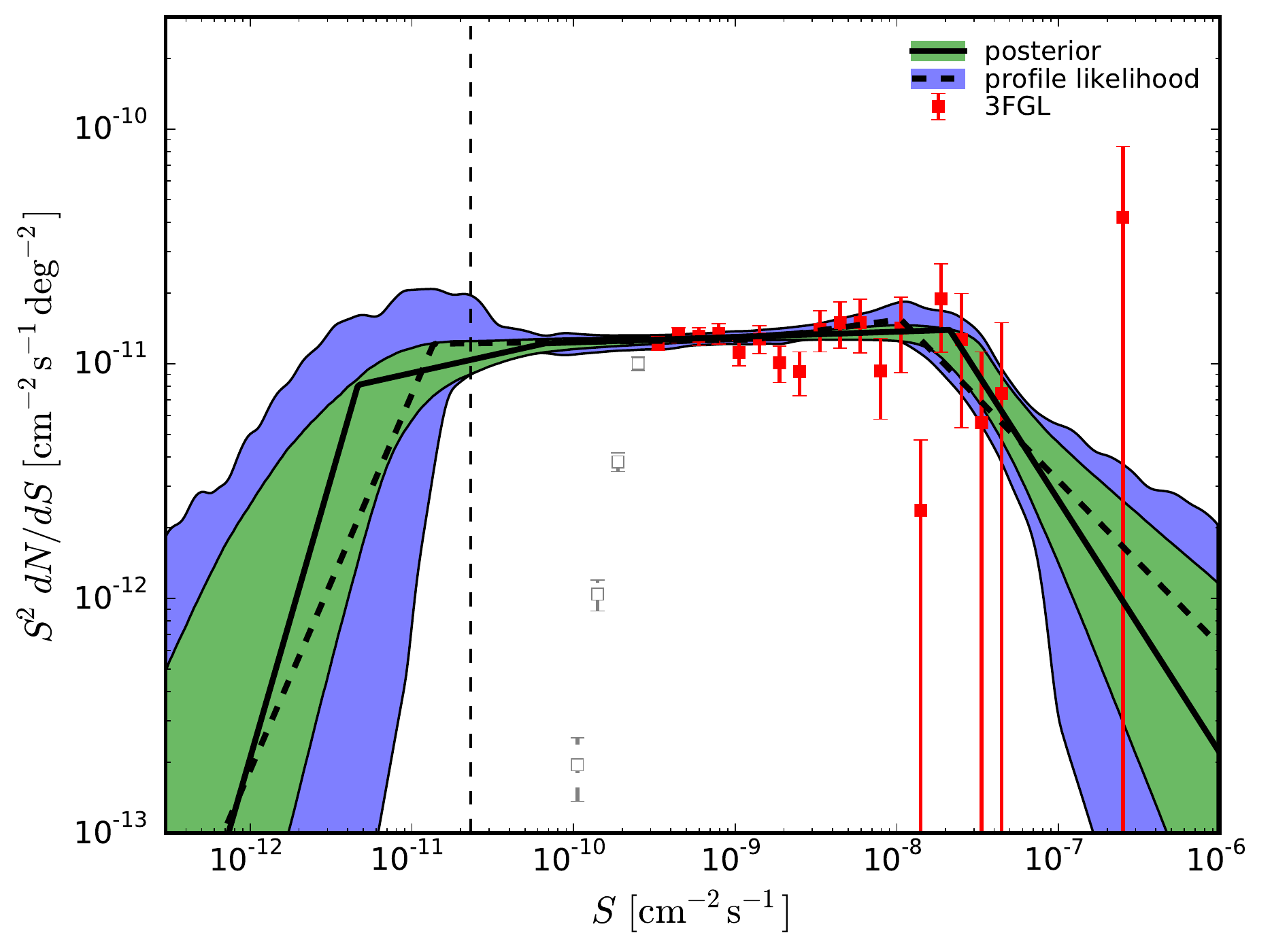}
  \label{sfig:s2dnds_mbpl_Nb3}
}
\caption{Source-count distribution \dnds\ as obtained from the 6-year
  \emph{Fermi}-LAT data set using the MBPL approach.  The
  \dnds\ distribution has been parameterized with a pure MBPL with (a)
  two and (b) three free breaks. The solid black line depicts the
  best-fit \dnds\ given by the Bayesian posterior; the corresponding
  statistical uncertainty is shown by the green band. The dashed black
  line and the blue band show the same quantities as derived from the
  profile likelihood. Red points depict the \dnds\ distribution
  derived from the 3FGL catalog. Poissonian errors $\propto \sqrt{N}$
  have been assumed. Gray points depict the same quantity derived for
  low fluxes, but without any correction for catalog detection
  efficiency applied (see Section \ref{sec:analysis_routine}). These
  points have been included for completeness only, while lacking any
  meaning for comparison. The vertical dashed line depicts the
  sensitivity estimate $S_\mathrm{sens}$ discussed in
  Section~\ref{ssec:sensitivity}.
\label{fig:mbpl_fit_3_2}}
\end{centering}
\end{figure*}

The MBPL fit was employed using the priors as discussed in Section
\ref{sssec:priors}. The results are shown in Table
\ref{tab:mbpl_fit_3_2} and Figure \ref{fig:mbpl_fit_3_2}.

\begin{deluxetable*}{lcccccc}
\tablecaption{MBPL Fit\label{tab:mbpl_fit_3_2}}
\tablewidth{0pt}
\tablehead{
\colhead{} & \multicolumn{2}{c}{$N_\mathrm{b} = 1$} & \multicolumn{2}{c}{$N_\mathrm{b} = 2$} & \multicolumn{2}{c}{$N_\mathrm{b} = 3$} \\
\colhead{Parameter\tablenotemark{a}} & \colhead{Posterior} & \colhead{PL} & \colhead{Posterior} & \colhead{PL} 
& \colhead{Posterior} & \colhead{PL}
}
\startdata
$A_\mathrm{S}$ & $4.1^{+0.3}_{-0.3}$ & $4.1^{+0.4}_{-0.5}$ &  $3.5^{+1.6}_{-1.0}$ & $3.1^{+3.9}_{-1.1}$ &  $3.5^{+1.4}_{-0.9}$ &  $2.7^{+3.1}_{-0.6}$ \\
$S_\mathrm{b1}$ & $1.3^{+1.3}_{-1.3}$\,E-3 & $2.1^{+5.7}_{-1.8}$\,E-3 &  $2.1^{+0.9}_{-1.2}$ &  $1.8^{+2.1}_{-1.1}$ &  $2.1^{+0.8}_{-1.2}$ &  $1.1^{+2.4}_{-0.3}$ \\
$S_\mathrm{b2}$ & \nodata & \nodata & $5.6^{+5.6}_{-5.1}$\,E-2 &  $7.8^{+24.4}_{-6.8}$\,E-2 &  $0.7^{+1.1}_{-0.5}$ &  $12.8^{+17.0}_{-12.6}$ \\
$S_\mathrm{b3}$ & \nodata & \nodata &  \nodata &  \nodata &  $4.6^{+4.1}_{-6.3}$ &  $13.6^{+6.4}_{-13.0}$ \\
$n_1$ & $2.03^{+0.02}_{-0.02}$ & $2.03^{+0.04}_{-0.03}$ &  $3.11^{+0.69}_{-0.55}$ &  $2.89^{+1.41}_{-0.59}$ &  $3.08^{+0.65}_{-0.50}$ &  $2.70^{+1.35}_{-0.35}$ \\
$n_2$ & $-0.49^{+1.20}_{-1.04}$ & $-0.69^{+2.34}_{-1.31}$ &  $1.97^{+0.03}_{-0.03}$ &  $1.98^{+0.03}_{-0.05}$ &  $1.98^{+0.03}_{-0.03}$ &  $1.91^{+0.13}_{-0.19}$ \\
$n_3$ & \nodata & \nodata &  $-0.61^{+1.13}_{-0.89}$ &  $-0.77^{+2.40}_{-1.23}$ &  $1.85^{+0.18}_{-0.25}$ &  $1.99^{+0.31}_{-0.59}$ \\
$n_4$ & \nodata & \nodata & \nodata &  \nodata  &  $-0.38^{+1.06}_{-0.97}$ &  $0.40^{+1.04}_{-2.40}$ \\
$A_\mathrm{gal}$ & $1.071^{+0.005}_{-0.005}$ & $1.072^{+0.005}_{-0.007}$ &  $1.072^{+0.004}_{-0.004}$ & $1.073^{+0.005}_{-0.006}$ &  $1.072^{+0.004}_{-0.004}$ &  $1.072^{+0.005}_{-0.006}$ \\
$F_\mathrm{iso}$ & $1.0^{+0.3}_{-0.4}$ & $1.2^{+0.3}_{-0.7}$ & $0.9^{+0.3}_{-0.3}$ &  $1.0^{+0.4}_{-0.5}$ &  $0.9^{+0.2}_{-0.3}$ &  $1.1^{+0.2}_{-0.6}$ \\
\tableline
$\ln \mathcal{L}_1({\bf \Theta})$ & $-851.9$ & $-855.0$ &  $-850.7$ & $-853.2$  &  $-851.7$ &  $-853.5$ \\
$\ln \mathcal{L}_2({\bf \Theta})$ & $-86793.1$ & $-86789.0$ &  $-86786.8$ & $-86785.3$  &  $-86785.9$ &  $-86785.2$ \\
$\ln \mathcal{Z}$ & \multicolumn{2}{c}{$-86804.10 \pm 0.09$} & \multicolumn{2}{c}{$-86799.17 \pm 0.09$} & \multicolumn{2}{c}{$-86798.34 \pm 0.08$} 
\enddata
\tablecomments{Best-fit values and statistical uncertainties (68.3\%
  CL) obtained for a pure MBPL fit to the data (MBPL approach). The
  table compares \dnds\ fits with one, two, and three free
  breaks. Both the parameter values obtained from the Bayesian
  posterior and the values derived from the profile likelihood (PL)
  are given. The last three rows list the values of the
  $\mathcal{L}_1$ and $\mathcal{L}_2$ likelihoods for the best-fit
  results. The value of the Bayesian evidence $\mathcal{Z}$ is given
  in addition. Ellipses indicate parameters not present in the
  specific model.}
\tablenotetext{a}{$A_\mathrm{S}$ is given in units of
  $10^{7}\,\mathrm{s}\,\mathrm{cm}^2\,\mathrm{sr}^{-1}$.  We remind
  that the values correspond to a flux normalization constant of $S_0
  = 3\times 10^{-8}\,\mathrm{cm}^{-2}\,\mathrm{s}^{-1}$.  The breaks
  $S_{\mathrm{b} 1}$, $S_{\mathrm{b} 2}$, and $S_{\mathrm{b} 3}$ are
  given in units of $10^{-8}$, $10^{-10}$, and
  $10^{-12}\,\mathrm{cm}^{-2}\,\mathrm{s}^{-1}$, respectively. The
  units of the diffuse flux component $F_\mathrm{iso}$ are
  $10^{-7}\,\mathrm{cm}^{-2}\,\mathrm{s}^{-1}\,\mathrm{sr}^{-1}$.  All
  other quantities are dimensionless.}
\end{deluxetable*}

\begin{figure*}[t]
\begin{centering}
\subfigure[MBPL, $N_\mathrm{b}=3$, posterior]{%
  \includegraphics[width=1.0\textwidth]{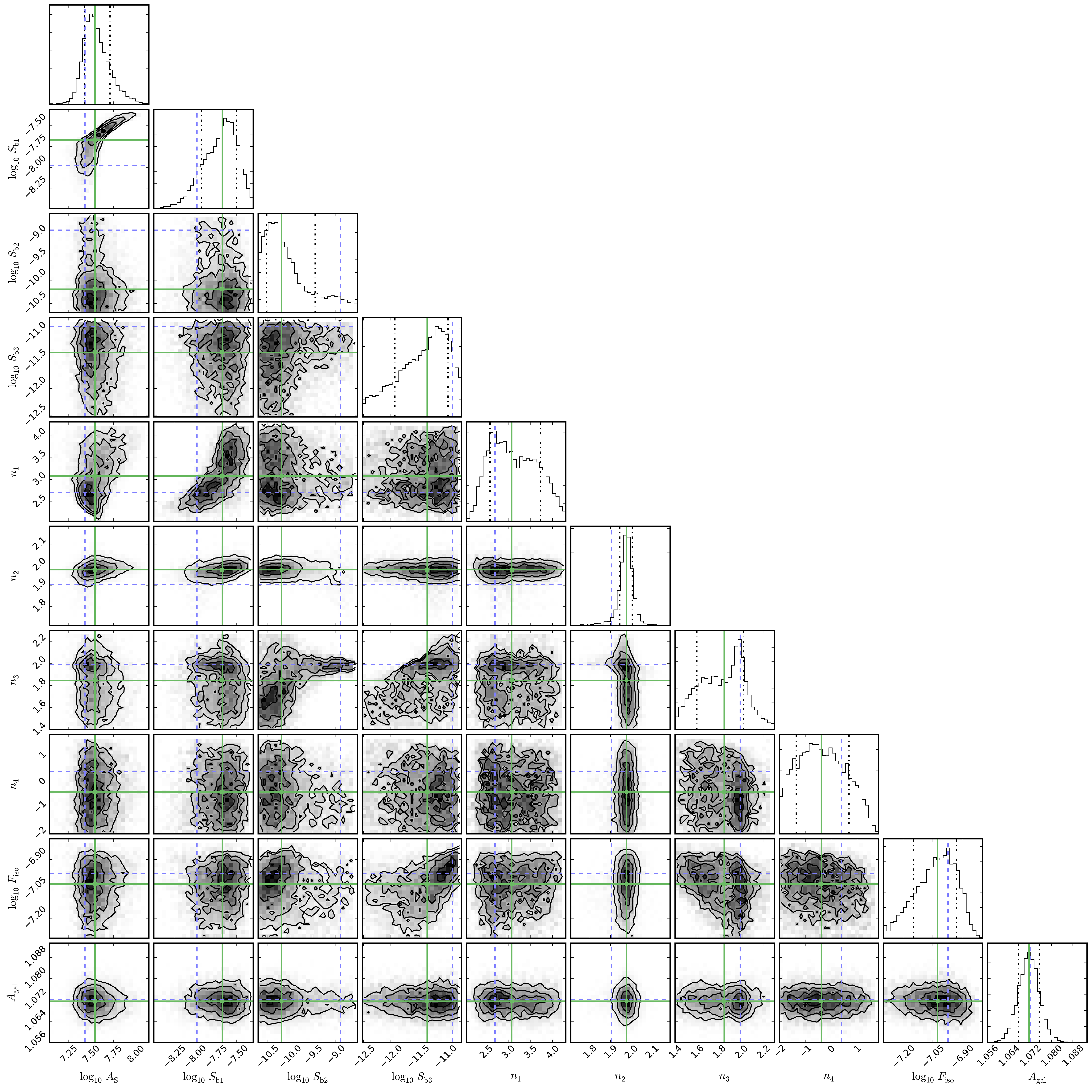}
  \label{sfig:triangle_mbpl_3}
}\\
\subfigure[MBPL, $N_\mathrm{b}=3$, profile likelihood]{%
  \includegraphics[width=1.0\textwidth]{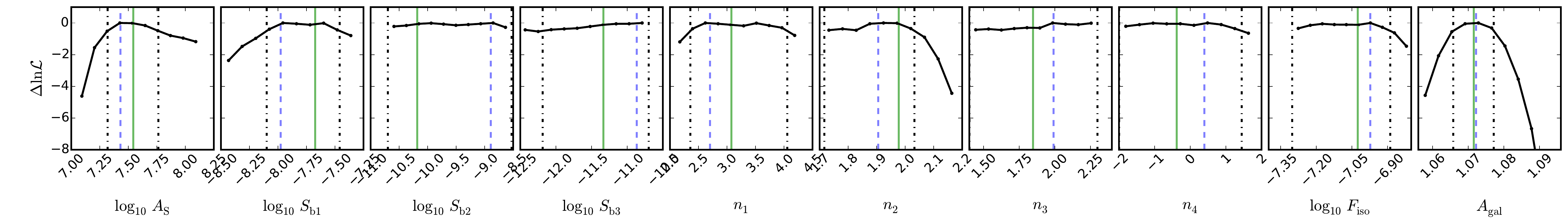}
  \label{sfig:plike_mbpl_3}
}
\caption{(a) Triangle plot of the Bayesian posterior and (b)
  corresponding profile likelihood functions of the sampling
  parameters.  The data have been fit using the MBPL approach with
  three free breaks ($N_\mathrm{b}=3$).  The posterior median is
  depicted by the solid green line. Maximum likelihood parameter
  values are depicted by the dashed blue lines. Dot-dashed black lines
  show the uncertainty estimates. The scalings of the $x$-axes have
  been chosen in accordance with the priors (flat or log-flat). The
  parameter $A_\mathrm{S}$ is given in units of
  $\mathrm{s}\,\mathrm{cm}^2\,\mathrm{sr}^{-1}$, while the units of
  the breaks $S_{\mathrm{b}j }$ and the integral flux $F_\mathrm{iso}$
  are $\mathrm{cm}^{-2}\,\mathrm{s}^{-1}$.  (a) The diagonal shows the
  marginalized posteriors for single parameters, while two-dimensional
  correlations between parameters become evident from the off-diagonal
  plots. Uncertainties have been calculated from the 15.85\% and
  84.15\% quantiles.  Note that contours have been chosen for
  visibility puposes only and do not represent a specific confidence
  interval.  (b) The profile likelihood has been derived from the
  posterior samples.  Black circles depict bin centers. Uncertainty
  estimates are 68.3\% CL.  In case the profile likelihood was not
  sufficiently constraining, uncertainty estimates have been
  approximated with the limits of the sample
  data. \label{fig:triangle_plike_mbpl_3}}
\end{centering}
\end{figure*}

The source-count distribution was parameterized with one, two, and
three free breaks.  Table \ref{tab:mbpl_fit_3_2} lists all best-fit
values and statistical uncertainties obtained for individual fit
parameters, in addition to the corresponding likelihoods of the
best-fit solutions. Single-parameter uncertainties can be large in
general, given that correlations were integrated over.  Comparing
Bayesian (posterior) and frequentist (profile likelihood\footnote{See
  Section~\ref{sssec:par_est_freq}.  Further details on the derivation
  of uncertainties are given in the caption of
  Figure~\ref{sfig:plike_mbpl_3}.}) parameter estimates, best-fit
values match within their uncertainties.

Figure \ref{fig:mbpl_fit_3_2} shows the best-fit results and
corresponding statistical uncertainty bands for the
\dnds\ distributions parameterized with two and three free breaks. We
can see that there is good agreement between the \dnds\ distributions
derived from the Bayesian posterior (solid black line and green band)
and the \dnds\ fits derived from the profile likelihood (dashed black
line and blue band): they match well within their uncertainty bands.
The uncertainty given by the profile likelihood is larger than the
band from the posterior in all cases. The frequentist uncertainty
estimates can therefore be considered more conservative. In common,
the statistical uncertainty bands of the \dnds\ fits obtained here are
small compared to fits employing catalog points only
\citep[see][]{2010ApJ...720..435A}.  This directly reflects the fact
that the method is independent of source-detection or binning
effects. The smallest statistical uncertainty appears to be around a
flux of $\sim 10^{-9}\,\mathrm{cm}^{-2}\,\mathrm{s}^{-1}$.

As shown in Table~\ref{tab:mbpl_fit_3_2}, the fit of the simplest
\dnds\ model with only a single break prefers a break at low fluxes,
i.e., at $\sim 10^{-11}\,\mathrm{cm}^{-2}\,\mathrm{s}^{-1}$. Below
that break, the \dnds\ cuts off steeply. The source-count distribution
in the entire flux range above that break was fit with the single
power-law component, with an index of $n_1 = 2.03 \pm 0.02$. We found
that adding a break at higher fluxes, i.e., parameterizing \dnds\ with
two free breaks, instead improved the fit with a significance of
$\sim$3$\sigma$. Here the bright-source region is resolved with a
break at $\sim 2\times
10^{-8}\,\mathrm{cm}^{-2}\,\mathrm{s}^{-1}$. The region between the
two breaks (faint-source region and intermediate region) is compatible
with an index of $n_2 = 1.97 \pm 0.03$, while the index in the
bright-source region $n_1=3.1^{+0.7}_{-0.6}$ is softer (see
Figure~\ref{fig:mbpl_fit_3_2}).

The intermediate region is populated with numerous sources
contributing a comparably large number of photons. Given the high
statistical impact of these sources, it was found that a fit of the
faint-source and intermediate regions with only a single power-law
component can be significantly driven by brighter sources of the
intermediate region. We therefore extended the \dnds\ model to three
free breaks, properly investigating possible features in the
faint-source region below \mbox{$\sim 3\times
  10^{-9}\,\mathrm{cm}^{-2}\,\mathrm{s}^{-1}$}.  We found that the
model comprising three free breaks is not statistically preferred
against the two-break model (see Table~\ref{tab:mbpl_fit_3_2}).
Furthermore, the three-break \dnds\ distribution is consistent with
the previous scenario within uncertainties (see
Figure~\ref{fig:mbpl_fit_3_2}).  Differences between the best fit from
Bayesian inference and the best fit given by the maximum likelihood
are not statistically significant.

It can be seen in Figure~\ref{fig:mbpl_fit_3_2} that the source-count
distribution as resolved by the 3FGL catalog (red data points; see
Section \ref{ssec:3FGLpoints}) in the intermediate and the
bright-source regions is well reproduced with both the two-break and
the three-break fits.  Again, we emphasize that this analysis is
independent of catalog data, which are shown in the plot for
comparison only.

From the MBPL approach, we therefore conclude that parameterizing
\dnds\ with two free breaks is sufficient to fit the data.  The index
$n_2 = 1.97 \pm 0.03$, characterizing the intermediate region of
\dnds, is determined with exceptionally high precision ($\sim$2\%),
originating from the high statistics of sources populating that
region.  The accuracy of the Galactic foreground normalization
$A_\mathrm{gal}$ fit is at the per mil level.

We found that the fit prefers a source-count distribution that
continues with an almost flat slope (in $S^2\,\mathrm{d}N/\mathrm{d}S$
representation) in the regime of unresolved sources, i.e., faint
sources not detected in the 3FGL catalog.  A strong cutoff was found
at fluxes between $\sim\!5\times
10^{-12}\,\mathrm{cm}^{-2}\,\mathrm{s}^{-1}$ and
$\sim\!10^{-11}\,\mathrm{cm}^{-2}\,\mathrm{s}^{-1}$.  This cutoff,
however, falls well within the flux region where this method is
expected to lose sensitivity and where the uncertainty bands widen. It
should thus be considered with special care.  Indeed, Monte Carlo
simulations were used to demonstrate that such a cutoff can originate
either from the sensitivity limit of the analysis or from an intrinsic
end of the source-count distribution (see Appendix~\ref{app:sims} for
details).  In the former case, possible point-source contributions
below the cutoff are consistent with diffuse isotropic emission, and
the fit therefore attributes them to $F_\mathrm{iso}$.

It was found that the uncertainty band below the cutoff can be
underestimated due to lacking degrees of freedom in the faint-source
end. Moreover, simulations revealed that the fit obtained from the
Bayesian posterior can be biased in the regime of very faint
sources. We therefore chose to improve the fit procedure by using the
hybrid approach in Section~\ref{ssec:data_hybrid}.

\paragraph{Sampling}
The triangle plot of the Bayesian posterior and the corresponding
profile likelihood functions are shown in
Figure~\ref{fig:triangle_plike_mbpl_3} for parameterizing \dnds\ with
three free breaks.

It can be seen that the marginalized posterior distributions are well
defined.  We attenuated strong parameter degeneracies by adapting the
normalization constant $S_0$ to the value quoted in the previous
section.  It becomes evident from the posteriors that the breaks
$S_\mathrm{b2}$ and $S_\mathrm{b3}$ tended to merge to a single break;
this is supplemented by the flatness of their profile likelihoods. It
therefore explains the previous observation that adding a third break
is not required to improve the fit of the data.

\begin{deluxetable*}{lcccccc}
\tablecaption{Hybrid Fit\label{tab:hybrid_fit_3_2_1}}
\tablewidth{0pt}
\tablehead{
\colhead{} & \multicolumn{2}{c}{$N^\shy_\mathrm{b} = 1$} & \multicolumn{2}{c}{$N^\shy_\mathrm{b} = 2$} & \multicolumn{2}{c}{$N^\shy_\mathrm{b} = 3$} \\
\colhead{Parameter\tablenotemark{a}} & \colhead{Posterior} & \colhead{PL} & \colhead{Posterior} & \colhead{PL}
 & \colhead{Posterior} & \colhead{PL}
}
\startdata
$A_\mathrm{S}$ & $3.6^{+1.8}_{-1.1}$ &  $3.2^{+3.7}_{-1.2}$ & $3.5^{+1.7}_{-1.0}$ &  $3.3^{+2.9}_{-1.3}$ & $3.3^{+1.2}_{-0.8}$ &  $3.4^{+2.9}_{-1.3}$ \\
$S_\mathrm{b1}$ &  $2.2^{+1.0}_{-1.3}$ &  $1.9^{+3.1}_{-1.3}$ &  $2.1^{+1.0}_{-1.3}$ &  $2.0^{+1.5}_{-1.3}$ & $1.8^{+0.9}_{-1.0}$ &  $2.1^{+1.5}_{-1.5}$ \\
$S_\mathrm{b2}$ & \nodata &  \nodata &  $0.3^{+0.3}_{-0.2}$ &  $2.4^{+27.2}_{-2.3}$ & $7.6^{+6.8}_{-6.8}$ &  $4.4^{+25.6}_{-2.4}$ \\
$S_\mathrm{b3}$  &  \nodata &  \nodata &  \nodata &  \nodata  &  $27.7^{+25.3}_{-17.3}$ &  $124^{+41}_{-114}$ \\
$n_1$ &  $3.16^{+0.69}_{-0.59}$ &  $2.99^{+1.16}_{-0.66}$ & $3.10^{+0.71}_{-0.54}$ &  $3.20^{+0.95}_{-0.85}$ & $2.99^{+0.67}_{-0.43}$ &  $3.13^{+0.76}_{-0.76}$\\
$n_2$ &  $1.98^{+0.02}_{-0.03}$ &  $1.97^{+0.04}_{-0.06}$ & $1.97^{+0.03}_{-0.03}$ &  $1.95^{+0.07}_{-0.23}$ & $1.96^{+0.06}_{-0.08}$ &  $1.97^{+0.07}_{-0.27}$ \\
$n_3$ & \nodata &  \nodata &  $2.02^{+0.49}_{-0.38}$ &  $2.07^{+0.93}_{-0.77}$ & $1.98^{+0.06}_{-0.06}$ &  $1.87^{+0.33}_{-0.20}$ \\
$n_4$ &  \nodata &  \nodata &  \nodata &  \nodata & $2.02^{+0.46}_{-0.40}$ &  $2.24^{+0.76}_{-0.94}$ \\
$A_\mathrm{nd1}$ & $10.0^{+14.1}_{-15.2}$ &  $21.6^{+90.3}_{-20.6}$ & $8.7^{+12.0}_{-11.9}$ &  $5.0^{+80.9}_{-4.0}$ &  $8.3^{+10.9}_{-10.1}$ &  $2.4^{+84.1}_{-1.4}$ \\
$A_\mathrm{gal}$ & $1.072^{+0.004}_{-0.004}$ &  $1.073^{+0.005}_{-0.007}$ & $1.072^{+0.004}_{-0.004}$ &  $1.072^{+0.005}_{-0.006}$ &  $1.072^{+0.004}_{-0.004}$ &  $1.070^{+0.006}_{-0.003}$ \\
$F_\mathrm{iso}$ & $1.0^{+0.1}_{-0.3}$ & $0.9^{+0.3}_{-0.4}$  &  $0.9^{+0.2}_{-0.2}$ &  $0.9^{+0.3}_{-0.4}$ &  $0.9^{+0.2}_{-0.3}$ &  $0.9^{+0.5}_{-0.4}$ \\
\tableline
$\ln \mathcal{L}_1({\bf \Theta})$ &  $-853.9$ &  $-853.8$ &  $-849.3$ &  $-852.9$ &  $-851.4$ &  $-853.7$ \\
$\ln \mathcal{L}_2({\bf \Theta})$ &  $-86786.4$ &  $-86785.3$  &  $-86788.4$ &  $-86785.1$  &  $-86786.7$ &  $-86785.0$ \\
$\ln \mathcal{Z}$ & \multicolumn{2}{c}{$-86799.16 \pm 0.09$}  & \multicolumn{2}{c}{$-86798.34 \pm 0.09$}  & \multicolumn{2}{c}{$-86798.38 \pm 0.09$} 
\enddata
\tablecomments{Best-fit values and statistical uncertainties (68.3\%
  CL) obtained with the hybrid approach.  The table compares
  \dnds\ fits with one, two, and three free breaks. Both the parameter
  values obtained from the Bayesian posterior and the values derived
  from the profile likelihood (PL) are given.  The last three rows
  list the values of the $\mathcal{L}_1$ and $\mathcal{L}_2$
  likelihoods for the best-fit results.  The value of the Bayesian
  evidence $\mathcal{Z}$ is given in addition.  Ellipses indicate
  parameters not present in the specific model.}
\tablenotetext{a}{$A_\mathrm{S}$ is given in units of
  $10^{7}\,\mathrm{s}\,\mathrm{cm}^2\,\mathrm{sr}^{-1}$.  We remind
  that the values correspond to a flux normalization constant of $S_0
  = 3\times 10^{-8}\,\mathrm{cm}^{-2}\,\mathrm{s}^{-1}$.  The breaks
  $S_{\mathrm{b} 1}$, $S_{\mathrm{b} 2}$, and $S_{\mathrm{b} 3}$ are
  given in units of $10^{-8}$, $10^{-10}$, and
  $10^{-12}\,\mathrm{cm}^{-2}\,\mathrm{s}^{-1}$, respectively.  The
  normalization $A_\mathrm{nd1}$ of the node is given in units of
  $10^{14}\,\mathrm{s}\,\mathrm{cm}^2\,\mathrm{sr}^{-1}$.  The diffuse
  flux component $F_\mathrm{iso}$ is listed in units of
  $10^{-7}\,\mathrm{cm}^{-2}\,\mathrm{s}^{-1}\,\mathrm{sr}^{-1}$.  All
  other quantities are dimensionless.}
\end{deluxetable*}

\begin{figure*}[t]
\begin{centering}
\subfigure[\dnds, hybrid, $N^\shy_\mathrm{b}=1$]{%
  \includegraphics[width=0.49\textwidth]{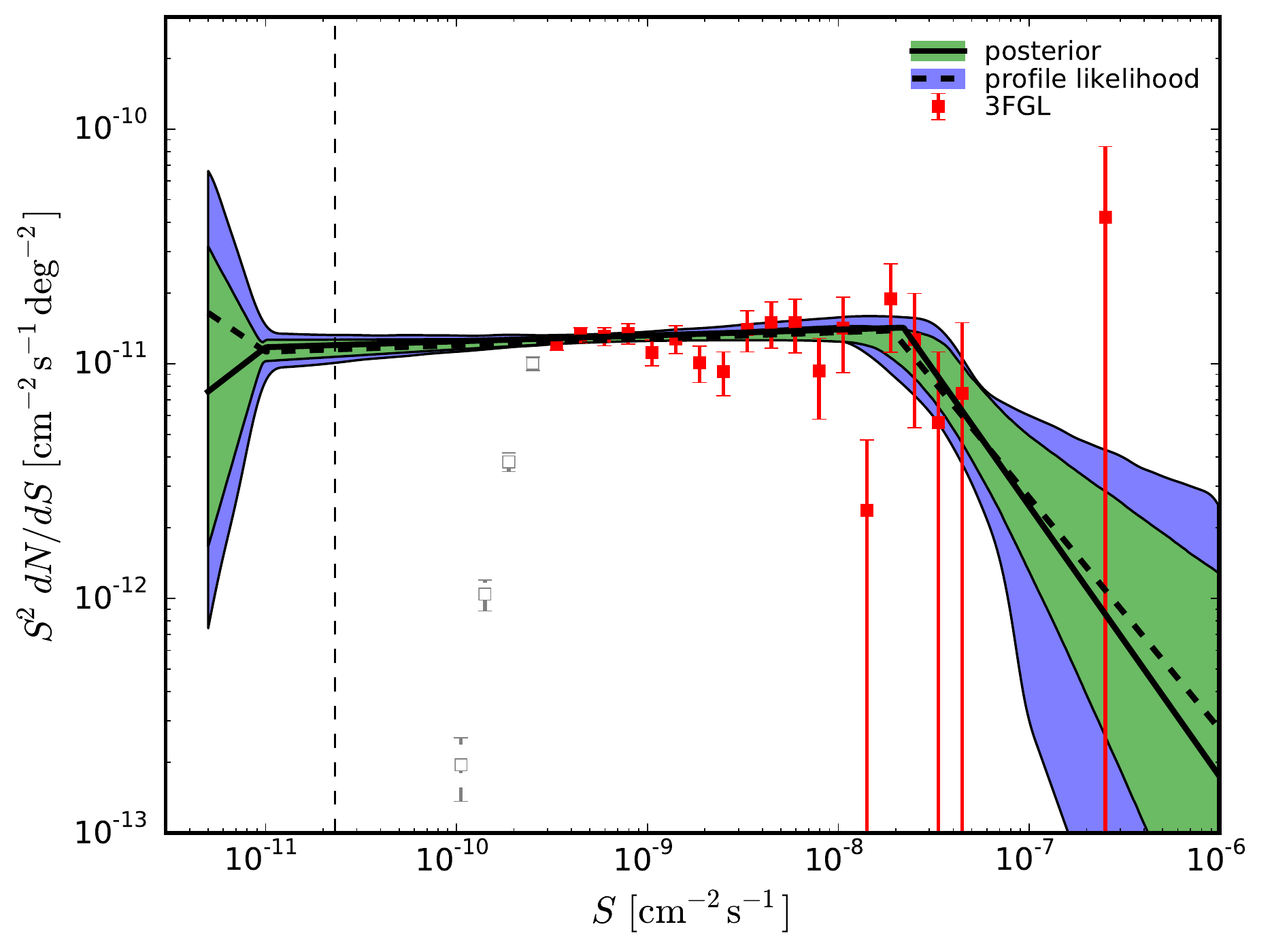}
  \label{sfig:s2dnds_Nb1}
}
\subfigure[$N(>S)$, hybrid, $N^\shy_\mathrm{b}=1$]{%
  \includegraphics[width=0.49\textwidth]{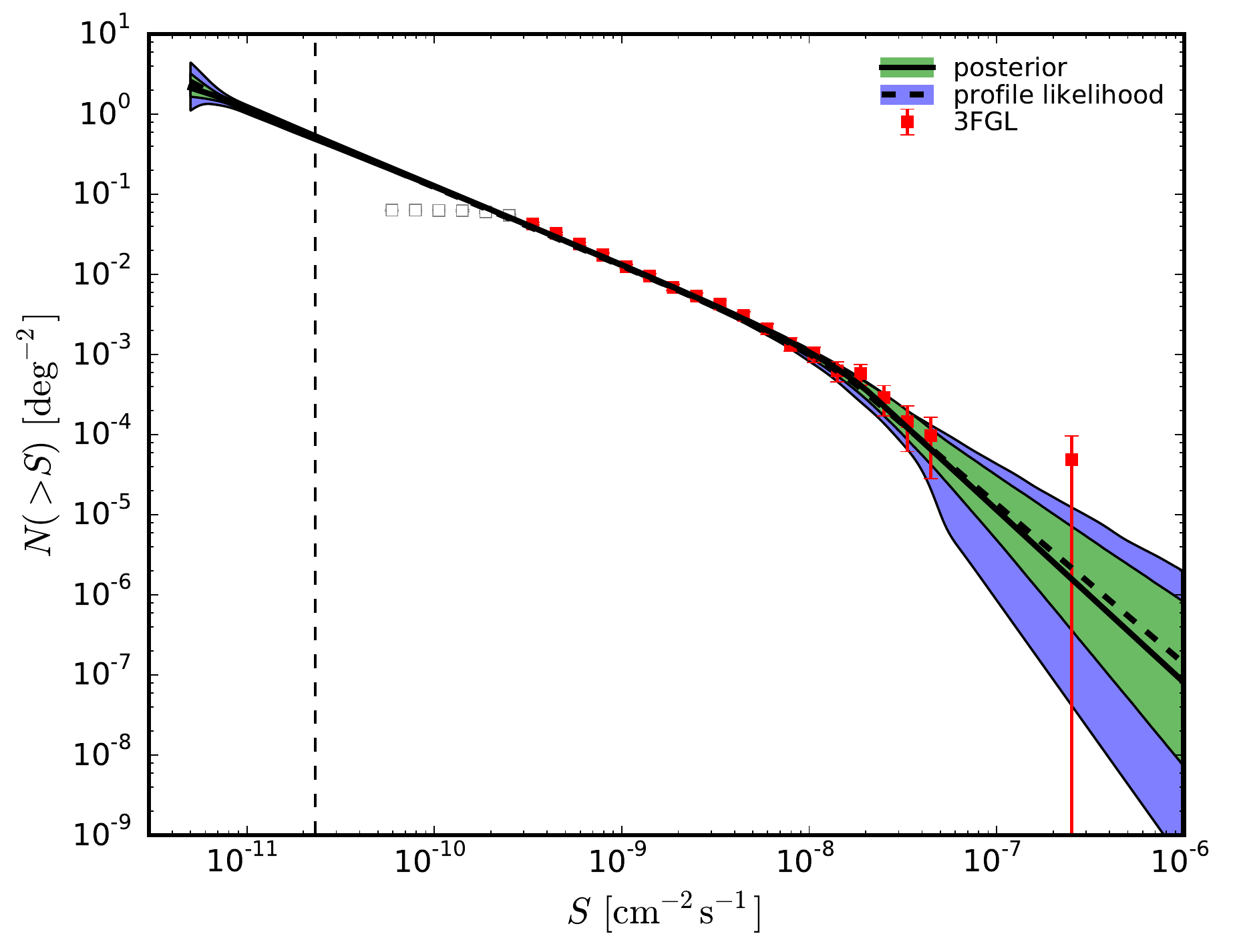}
  \label{sfig:SNS_Nb1}
}\\
\subfigure[\dnds, hybrid, $N^\shy_\mathrm{b}=2$]{%
  \includegraphics[width=0.49\textwidth]{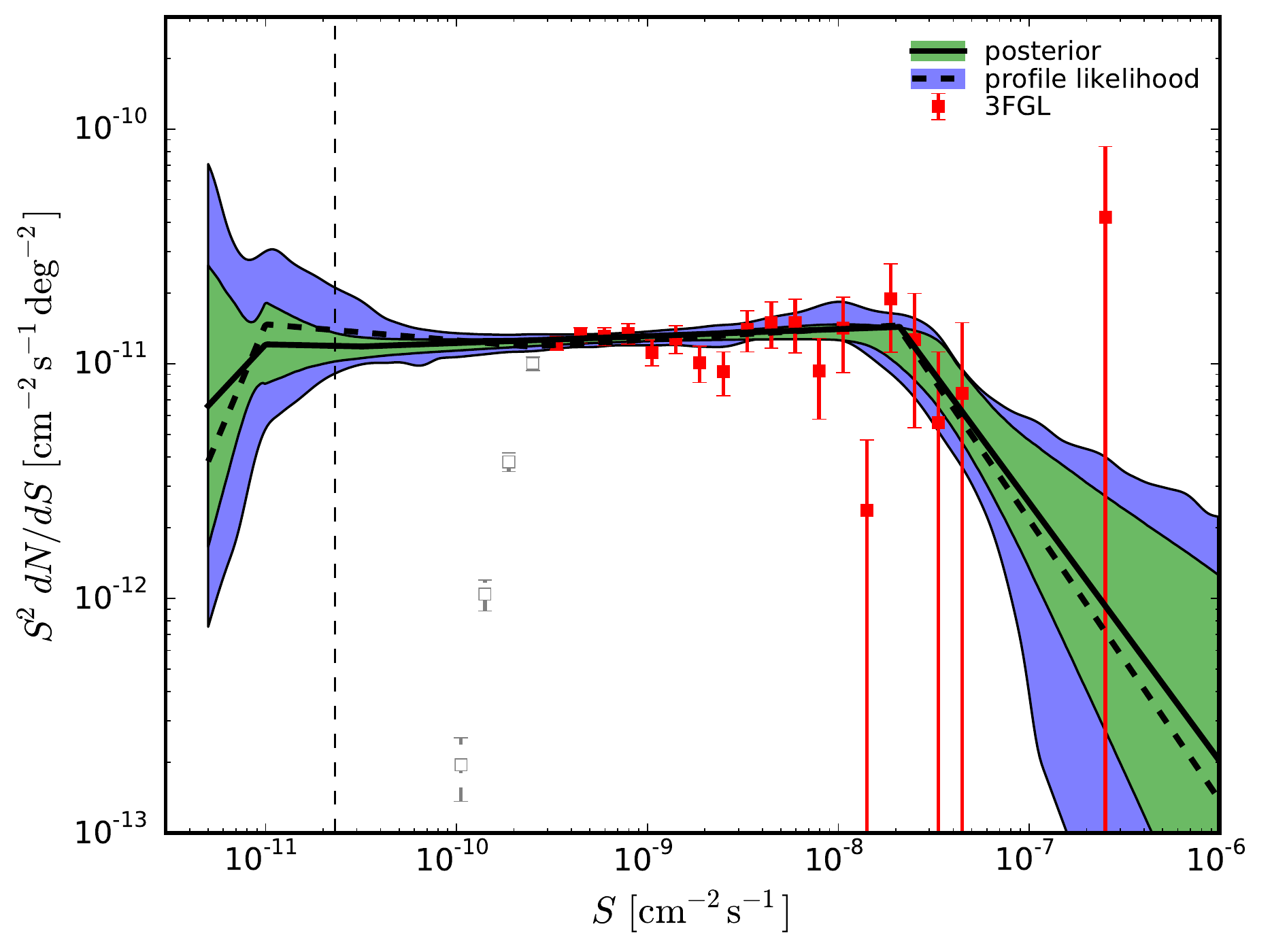}
  \label{sfig:s2dnds_Nb2}
}
\subfigure[$N(>S)$, hybrid, $N^\shy_\mathrm{b}=2$]{%
  \includegraphics[width=0.49\textwidth]{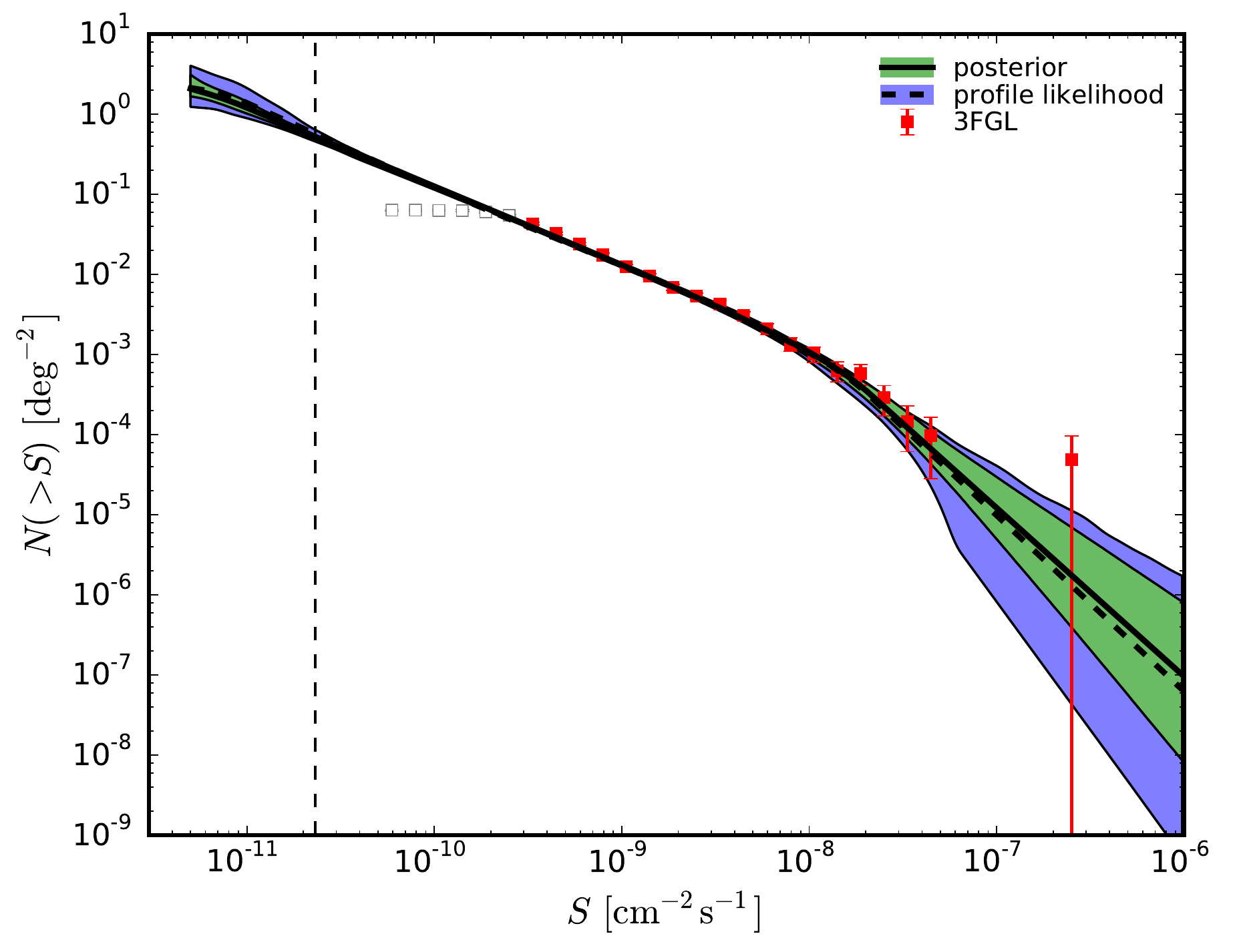}
  \label{sfig:SNS_Nb2}
}
\subfigure[\dnds, hybrid, $N^\shy_\mathrm{b}=3$]{%
  \includegraphics[width=0.49\textwidth]{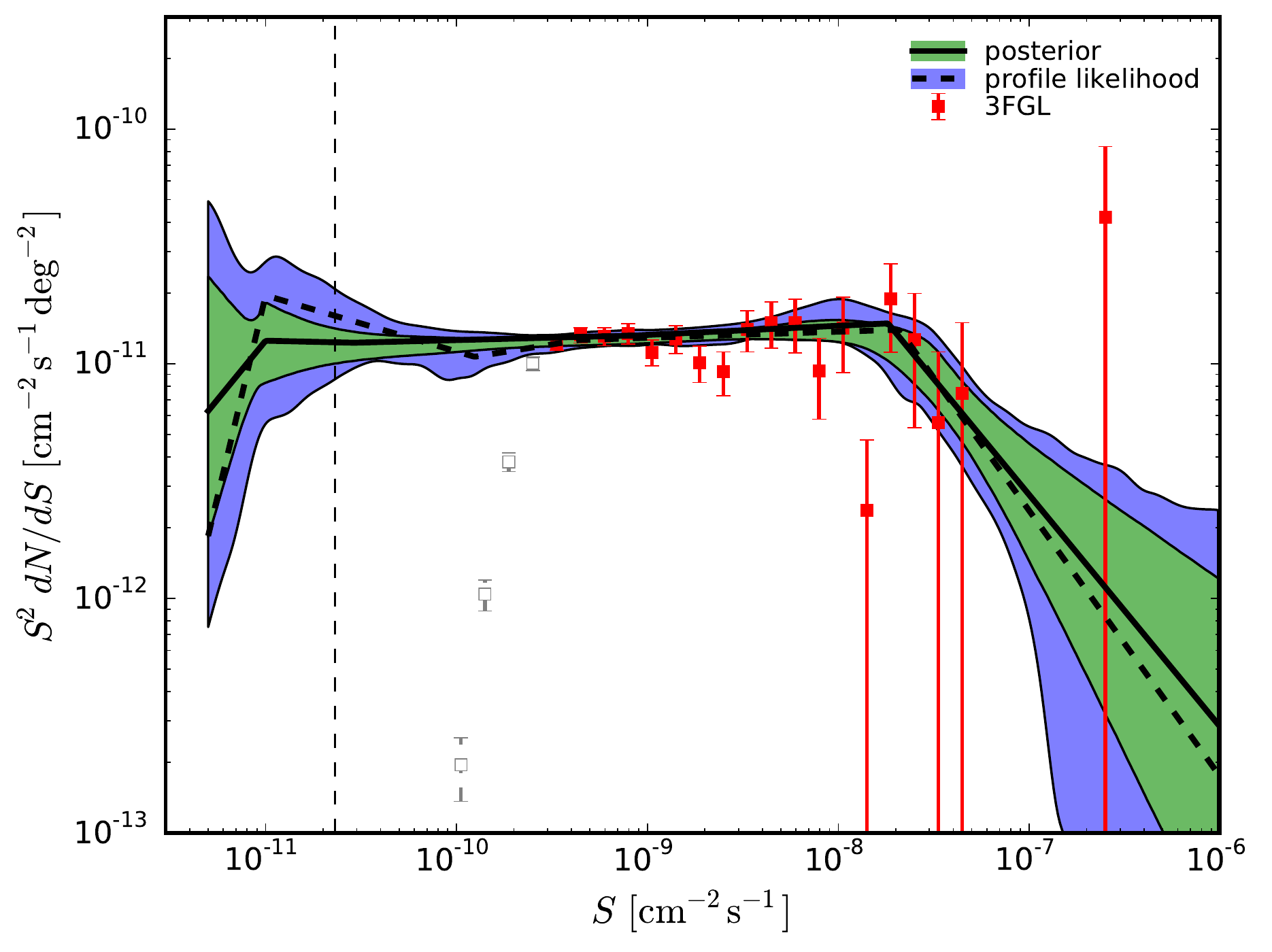}
  \label{sfig:s2dnds_Nb3}
}
\subfigure[$N(>S)$, hybrid, $N^\shy_\mathrm{b}=3$]{%
  \includegraphics[width=0.49\textwidth]{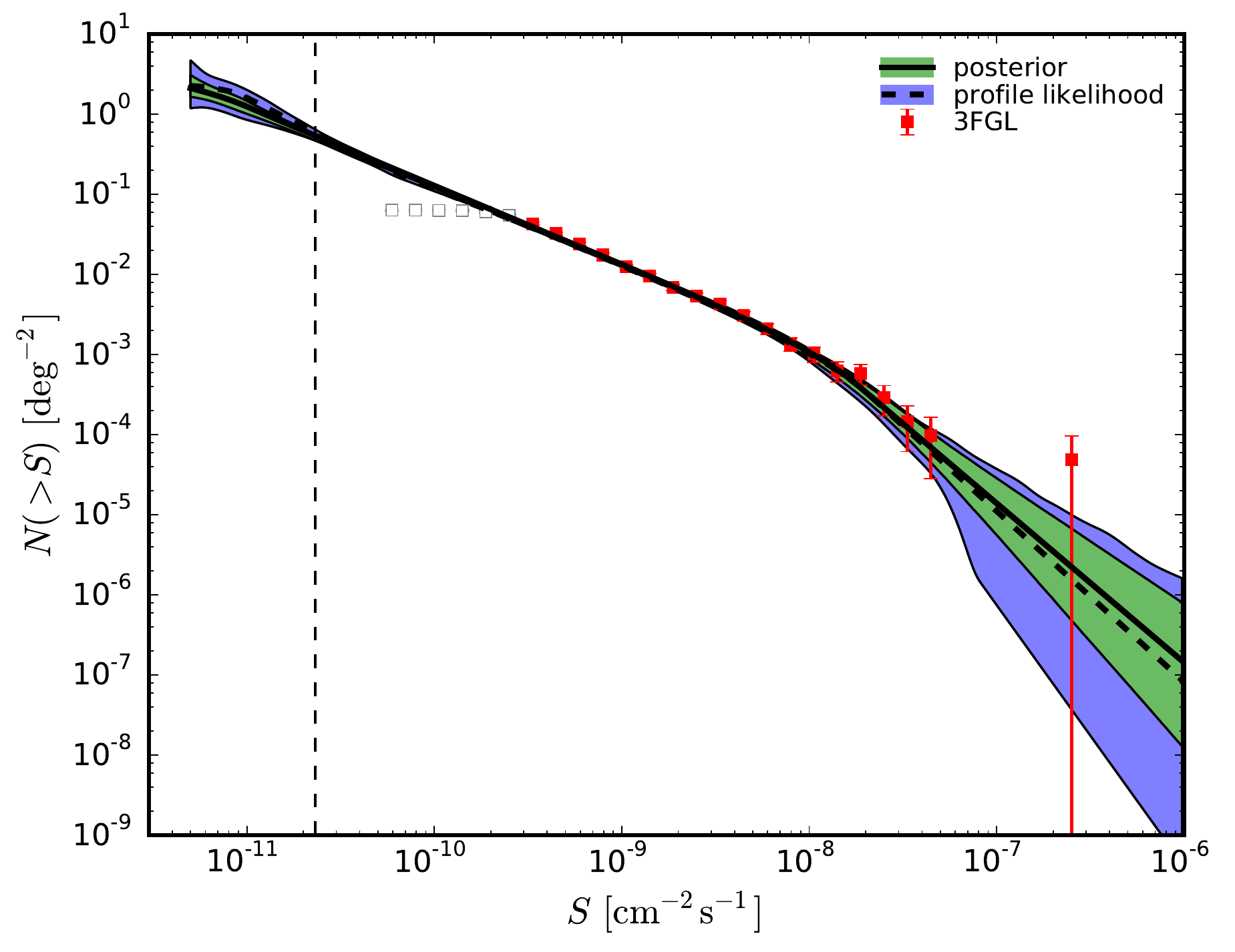}
  \label{sfig:SNS_Nb3}
}\\
\caption{Differential source-count distribution \dnds\ (left column)
  and integral source-count distribution $N(>S)$ (right column) as
  obtained from the 6-year \emph{Fermi}-LAT data using the hybrid
  approach. The \dnds\ distribution has been parameterized with an
  MBPL with one, two, and three free breaks (from top to bottom),
  together with a node at the faint end of the distribution. The use
  of line styles and colors resembles
  Figure~\ref{fig:mbpl_fit_3_2}. \label{fig:hybrid_fit_3_2_1}}
\end{centering}
\end{figure*}

\begin{figure*}[t]
\begin{centering}
\subfigure[simple \opdf, Bayesian parameter estimate]{%
  \includegraphics[width=0.98\textwidth]{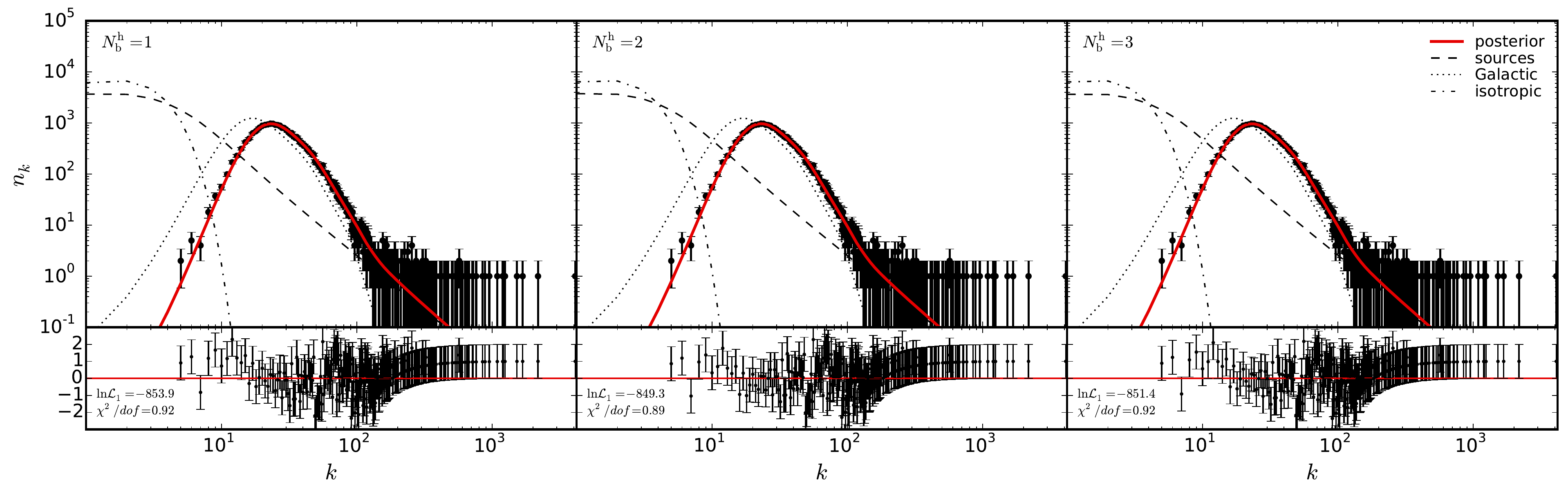}
  \label{sfig:1ppdf_post}
}\\
\subfigure[simple \opdf, maximum likelihood result]{%
  \includegraphics[width=0.98\textwidth]{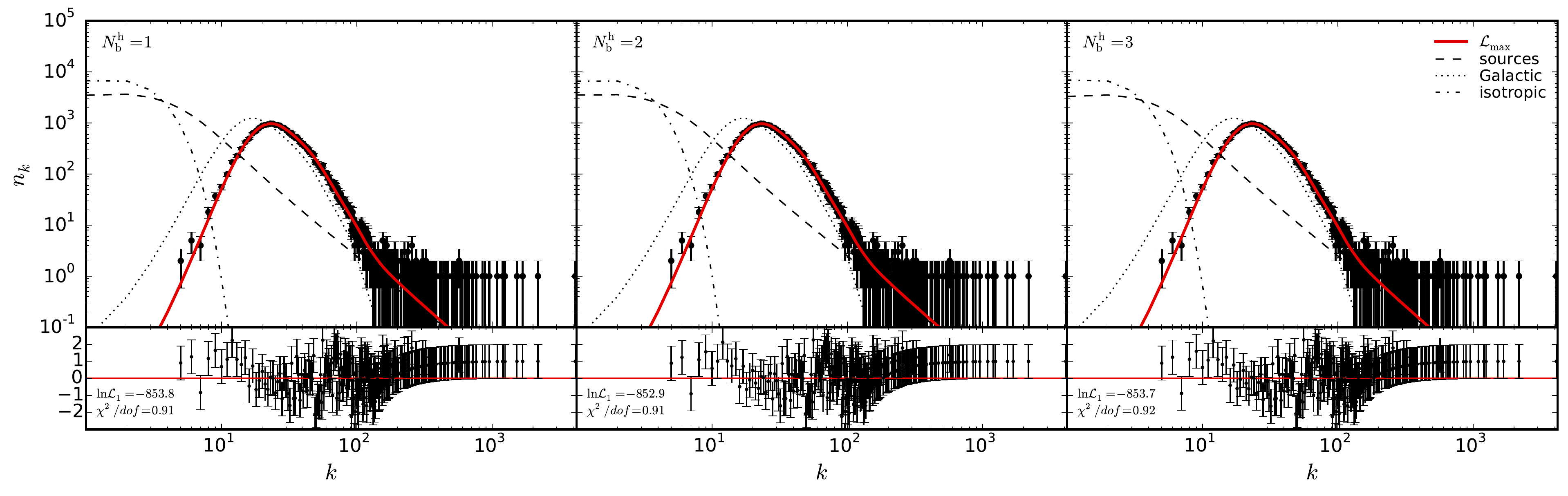}
  \label{sfig:1ppdf_mL}
}
\caption{Pixel-count distribution (black circles) of the
  \emph{Fermi}-LAT 6-year data set compared with the simple
  \opdf\ distributions of the best-fit models (solid red
  lines). Poissonian errors \mbox{$\propto\!\sqrt{n_k}$} have been
  assumed in this figure.  In the top row, the best-fit results
  obtained from the Bayesian posteriors are plotted.  The bottom row
  instead depicts the maximum likelihood results.  The individual
  \opdf\ distributions of the three different contributions are also
  shown, i.e., point sources (dashed black lines), the Galactic
  foreground (dotted black lines), and the isotropic diffuse
  background (dot-dashed black lines). The lower panels of the plots
  show the residuals, given by $(\mathrm{data} -
  \mathrm{model})/\sqrt{\mathrm{data}}$. Error bars represent
  $1\sigma$ uncertainties. \label{fig:phist_3_2_1}}
\end{centering}
\end{figure*}

\begin{figure*}[t]
\begin{centering}
\subfigure[hybrid, $N^\shy_\mathrm{b}=2$, posterior]{%
  \includegraphics[width=1.0\textwidth]{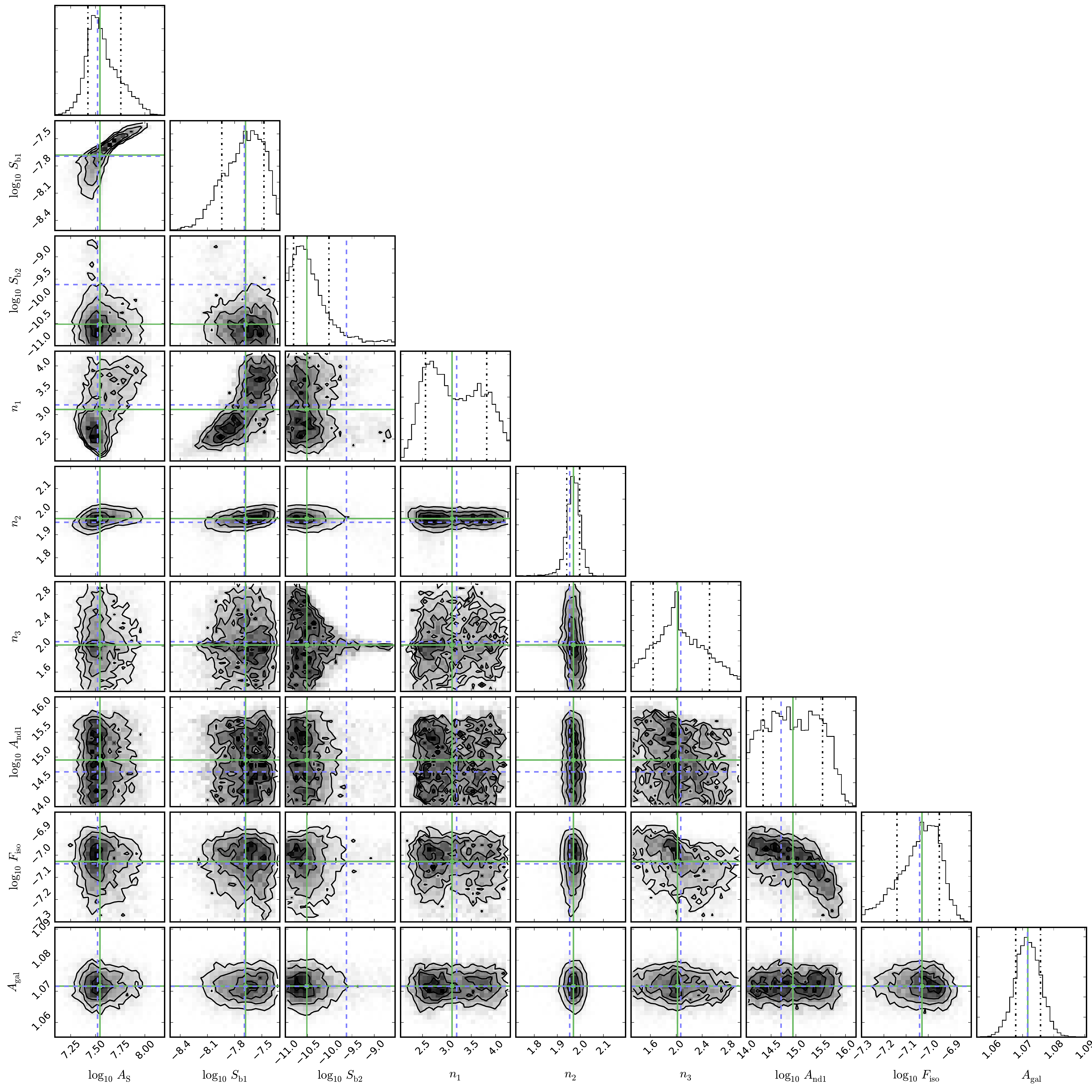}
  \label{sfig:triangle_hybrid_2}
}\\
\subfigure[hybrid, $N^\shy_\mathrm{b}=2$, profile likelihood]{%
  \includegraphics[width=1.0\textwidth]{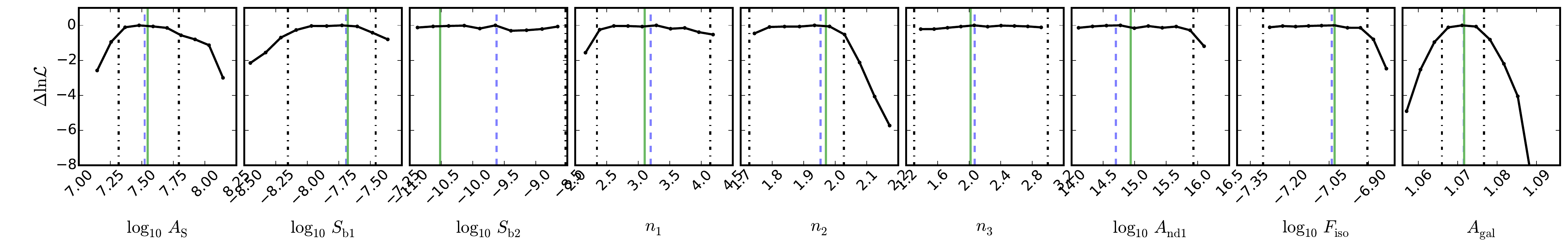}
  \label{sfig:plike_hybrid_2}
}
\caption{(a) Triangle plot of the Bayesian posterior and (b)
  corresponding profile likelihood functions of the sampling
  parameters.  The data have been fit using the hybrid approach with
  two free breaks ($N^\shy_\mathrm{b}=2$) and a node.  The use of line
  styles and colors follows Figure~\ref{fig:triangle_plike_mbpl_3}.
\label{fig:triangle_plike_hybrid_2}}
\end{centering}
\end{figure*}

\subsection{Hybrid Approach}\label{ssec:data_hybrid}
We improved the analysis by applying the hybrid approach
consecutively.  Priors are discussed in Table \ref{tab:priors}: In
particular, the region around the sharp cutoff revealed by the MBPL
approach was parameterized with a node placed at
$A_\mathrm{nd1}=5\times
10^{-12}\,\mathrm{cm}^{-2}\,\mathrm{s}^{-1}$.\footnote{The value
  approximates the faint cutoff positions obtained from the posterior
  of the MBPL fit.}  The lower bound of the prior of the last free
break was set to $\sim\! S_\mathrm{sens}/2$\,. The cutoff was
introduced manually by fixing the index of the power-law component
describing fluxes smaller than $A_\mathrm{nd1}$ to $n_\mathrm{f} =
-10$.

The fit was carried out with \dnds\ parameterizations comprising one,
two, and three free breaks.  Figure~\ref{fig:hybrid_fit_3_2_1} and
Table~\ref{tab:hybrid_fit_3_2_1} summarize the results. The
differential \dnds\ distributions fitting the data best are shown in
the left column of the figure.  In the right column, the corresponding
integral source-count distributions $N(>S)$ are compared to 3FGL
catalog data, providing another reference for investigating the
precision of the fit.

In the bright-source and intermediate regions, the results obtained
with the MBPL approach and with the hybrid approach are consistent
among each other within their uncertainties. As expected, the
determination of the uncertainty bands in the faint-source region
improved in the hybrid fit, given the further degree of freedom
allowed.  In all three scenarios ($N^\shy_\mathrm{b}=1,2,3$), the fits
reproduce well the differential and the integral source-count
distributions from the 3FGL catalog within uncertainties.

Comparing the three \dnds\ models, we find that none are statistically
preferred by the data; see Table~\ref{tab:hybrid_fit_3_2_1}.  The fit
of the model with only a single free break consistently placed the
break in the bright-source region, given that the cutoff in the
faint-source region is effectively accounted for by the node.  As
argued in the previous section, in this case the fit of the
intermediate and faint-source regions of \dnds\ was driven by the high
statistical impact of the relevant brighter sources, yielding a small
uncertainty band also for faint sources (see
Figure~\ref{sfig:s2dnds_Nb1}). To address this issue, we extended the
model with two additional free breaks ($N^\shy_\mathrm{b}=2,3$),
leading to consistent uncertainty bands that were stabilized by the
additonal degrees of freedom added in the intermediate and
faint-source regions (see Figures~\ref{sfig:s2dnds_Nb2} and
\ref{sfig:s2dnds_Nb3}). Because the three-break fit is not
statistically preferred against the two-break fit, we conclude that
two free breaks and a faint node are sufficient to fit the data
properly. A comparison with the maximum likelihood values for the MBPL
fits in Table~\ref{tab:mbpl_fit_3_2} reveals also no statistical
preference for the hybrid result over the MBPL result, confirming that
the data are not sensitive enough to distinguish point sources below
the last node from a purely diffuse isotropic emission.

Figure~\ref{fig:phist_3_2_1} compares the best-fit model
\opdf\ distributions to the actual pixel-count distribution of the
data set.  We plot the results for both the Bayesian posterior and the
maximum likelihood fits.  The residuals $(\mathrm{data} -
\mathrm{model})/\sqrt{\mathrm{data}}$ are shown in addition.  It can
be seen that the pixel-count distribution is reproduced well.  A
comparison with a simple chi-squared statistic, evaluating the
best-fit results using the binned histogram only, leads to reduced
chi-squared values ($\chi^2/\mathrm{dof}$) between 0.89 and 0.92\,.

The triangle plot of the Bayesian posterior and the single-parameter
profile likelihood functions are shown in
Figure~\ref{fig:triangle_plike_hybrid_2} for the \dnds\ fit with two
free breaks and a node.

The stability of the MBPL and hybrid approaches can be further
demonstrated by comparing the respective triangle plots (see
Figures~\ref{sfig:triangle_mbpl_3} and \ref{sfig:triangle_hybrid_2}):
the posteriors of parameters corresponding to each other in both
approaches are substantially equal, with the exception of $n_3$. It
can be seen that the choice of the node in the hybrid approach
stabilized the posterior of $n_3$. We have therefore shown that the
MBPL and hybrid approaches lead to comparable results except in the
faint-source flux region, where the latter improves the determination
of the uncertainty bands.

\subsection{How many breaks?}\label{ssec:}
Both the MBPL approach and the hybrid approach single out a best-fit
\dnds\ distribution that is consistent with a single broken power law
for integral fluxes in the resolved range above $S_\mathrm{sens}
\simeq 2\times 10^{-11}\,\mathrm{cm}^{-2}\,\mathrm{s}^{-1}$. Although
two breaks are preferred to properly fit the \mbox{\textit{entire}}
flux range, the second break found with the MBPL approach in the
faint-source region is consistent with a sensitivity cutoff. Instead,
in the hybrid approach, the second break is needed for a viable
determination of the uncertainty band.

\begin{figure*}[t]
 \epsscale{0.6}
 \plotone{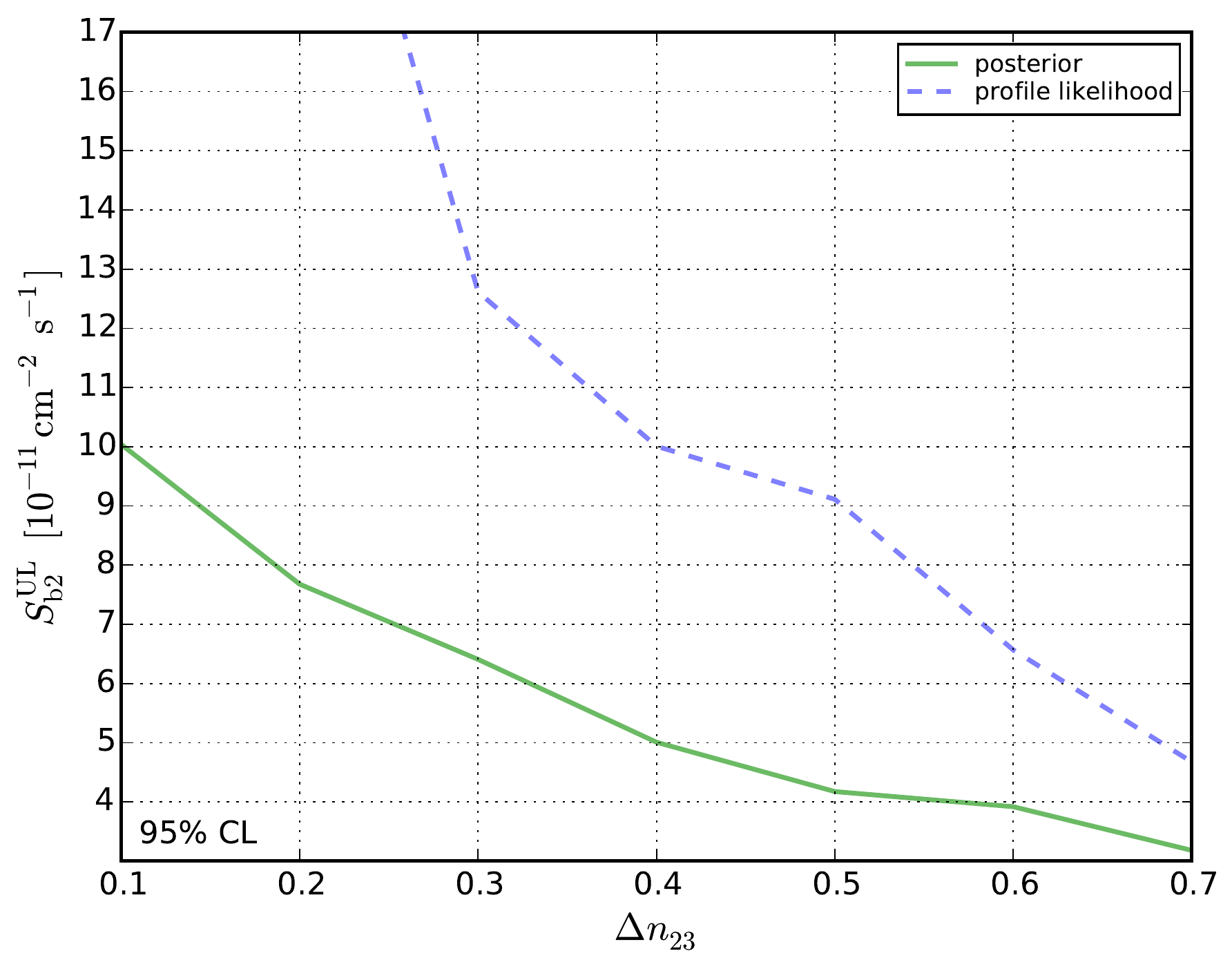}
 \caption{Upper limits on a possible second intrinsic break
   $S_{\mathrm{b}2}$ as a function of $\Delta n_{23}$ at 95\%
   confidence level (see text for details).  The solid green (dashed
   blue) line denotes the upper limits derived from the posterior
   (profile likelihood). \label{fig:Sb2ul}}
\end{figure*}

To further describe the physical \dnds\ distribution at low fluxes, we
therefore derived an upper limit on the position of a possible
intrinsic second break $S_{\mathrm{b}2}$.  The uncertainty band
obtained with the hybrid approach for $N^\shy_\mathrm{b} = 2$ was
used. In general, an intrinsic second break would have been present if
the power-law indices $n_2$ and $n_3$ changed significantly by a given
difference $\left| n_2-n_3 \right| > \Delta n_{23}$\,. We exploited
the full posterior to derive upper limits on $S_{\mathrm{b}2}$ by
assuming given $\Delta n_{23}$ values between 0.1 and 0.7, in steps of
0.1. In detail, the upper limits $S^\mathrm{UL}_{\mathrm{b}2}$ at 95\%
CL were obtained from the marginalized posterior $P( S_{\mathrm{b}2}
|\mathrm{D},H)$, after removing all samples not satisfying the given
$\left| n_2-n_3 \right|$ constraint:
\begin{equation}\label{eq:Sb2ul}
\int_{ \pi_\mathrm{L} (S_{\mathrm{b}2}) }^{S^\mathrm{UL}_{\mathrm{b}2}} 
P_{\left| n_2-n_3 \right| > \Delta n_{23}}(S_{\mathrm{b}2}|\mathrm{D},H)\,\mathrm{d} S_{\mathrm{b}2} = 0.95\,,
\end{equation}
where $\pi_\mathrm{L} (S_{\mathrm{b}2}) =
10^{-11}\,\mathrm{cm}^{-2}\,\mathrm{s}^{-1}$ is the lower bound of the
prior for $S_{\mathrm{b}2}$.  Frequentist upper limits were calculated
from the profile likelihood, constructed from the same posterior as
used in Equation~\eqref{eq:Sb2ul}, by imposing $-2\Delta \ln
\mathcal{L} = 2.71$ for 95\% CL upper limits. The upper limits are
shown in Figure~\ref{fig:Sb2ul}. In consistency with the uncertainty
bands derived in the previous section, $S^\mathrm{UL}_{\mathrm{b}2}$
decreases monotonically as a function of $\Delta n_{23}$, until the
sensitivity limit of the analysis is reached.  Assuming a fiducial
index change of $\Delta n_{23}=0.3$, we find that a possible second
break of \dnds\ is constrained to be below $6.4\times
10^{-11}\,\mathrm{cm}^{-2}\,\mathrm{s}^{-1}$ at 95\% CL.  The
corresponding frequentist upper limit is $1.3\times
10^{-10}\,\mathrm{cm}^{-2}\,\mathrm{s}^{-1}$.

\subsection{Composition of the Gamma-ray Sky}\label{ssec:comp}
\begin{figure}[t]
\epsscale{.65}
\plotone{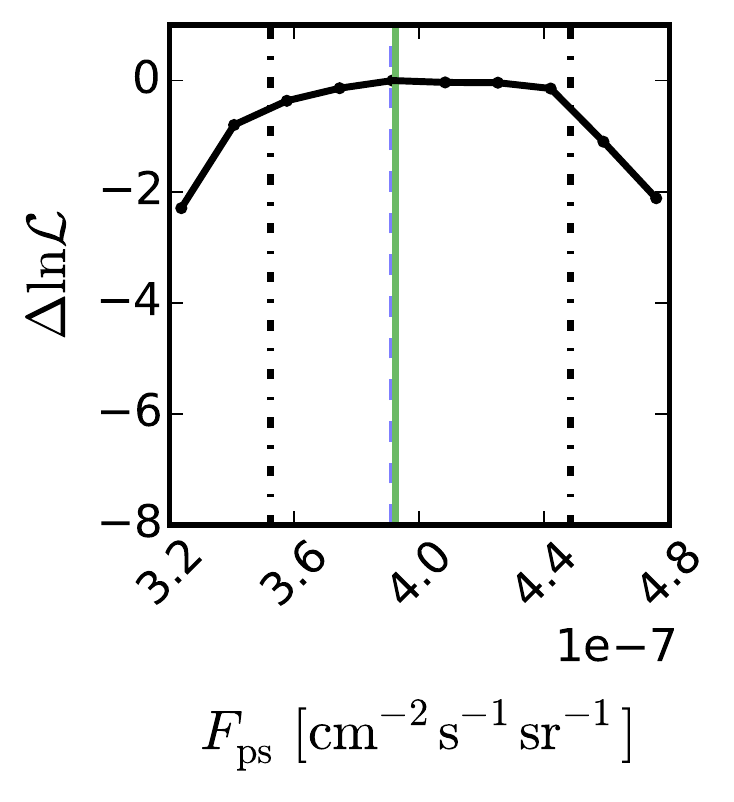}
\caption{Profile likelihood of the integral point-source flux
  $F_\mathrm{ps}$, obtained using the hybrid approach with two free
  breaks and $\kappa=6$. Line styles and colors are as in
  Figure~\ref{sfig:plike_mbpl_3}.
\label{fig:plike_Fps_hp6_2}}
\end{figure}

The method allows decomposing the high-latitude gamma-ray sky
($|b|\geq 30^\circ$) into its individual constituents. The integral
flux $F_\mathrm{ps}$ contributed by point sources was derived by
integrating the posterior samples of $S\,\mathrm{d}N/\mathrm{d}S$ in
the range $[0,S_\mathrm{cut}]$, which effectively corresponds to the
interval $[S_\mathrm{nd1},S_\mathrm{cut}]$ due to the steep cutoff
below the node $S_\mathrm{nd1}$.  Results are presented in
Table~\ref{tab:comp_hybrid_hp6_2}, comparing both Bayesian and
frequentist estimates.  The profile likelihood for $F_\mathrm{ps}$ is
shown in Figure~\ref{fig:plike_Fps_hp6_2}. The integral flux from
point sources is determined as $F_\mathrm{ps} =
3.9^{+0.3}_{-0.2}\times
10^{-7}\,\mathrm{cm}^{-2}\,\mathrm{s}^{-1}\,\mathrm{sr}^{-1}$, thus
with an uncertainty less than 10\%.\footnote{The contribution from the
  interval below the sensitivity estimate,
  $[S_\mathrm{nd1},S_\mathrm{sens}]$, is subdominant, i.e., $(16 \pm
  7)$\% of $F_\mathrm{ps}$.}

The contribution from Galactic foreground emission $F_\mathrm{gal}$
was obtained accordingly by integrating the template (see
Section~\ref{ssec:bckgs}), including the fit results for the
normalization $A_\mathrm{gal}$ (see
Figure~\ref{fig:triangle_plike_hybrid_2}). The isotropic background
emission $F_\mathrm{iso}$ was sampled directly.

\begin{table}[t]
\begin{center}
\caption{Composition of the High-latitude Gamma-ray Sky ($|b|\geq
  30^\circ$); Hybrid Approach, $N^\shy_\mathrm{b}=2$, $\kappa =
  6$ \label{tab:comp_hybrid_hp6_2}}
\begin{tabular}{lcc}
\tableline\tableline
Parameter & \multicolumn{1}{c}{Posterior} & \multicolumn{1}{c}{PL} \\
\tableline
$F_\mathrm{ps}$ &  $3.9^{+0.3}_{-0.2}$ &  $3.9^{+0.6}_{-0.4}$ \\
$F_\mathrm{gal}$ &  $10.95^{+0.04}_{-0.04}$ &  $10.95^{+0.05}_{-0.06}$ \\
$F_\mathrm{iso}$ &  $0.9^{+0.2}_{-0.2}$ &  $0.9^{+0.3}_{-0.4}$ \\
$F_\mathrm{tot}$ &  $15.8^{+0.2}_{-0.1}$ &  $15.7^{+0.3}_{-0.1}$ \\
\tableline
$q_\mathrm{ps}$ &  $0.25^{+0.02}_{-0.02}$ &  $0.25^{+0.03}_{-0.03}$ \\
$q_\mathrm{gal}$ &  $0.693^{+0.007}_{-0.006}$ & $0.697^{+0.015}_{-0.006}$  \\
$q_\mathrm{iso}$ &  $0.06^{+0.01}_{-0.02}$ &  $0.06^{+0.02}_{-0.03}$ \\
\tableline
$F^\mathrm{2FGL}_\mathrm{cat}$ & \multicolumn{2}{c}{$2.097 \pm 0.006$} \\
$F^\mathrm{3FGL}_\mathrm{cat}$ & \multicolumn{2}{c}{$2.494 \pm 0.007$} \\
$F_\mathrm{CR}$ & \multicolumn{2}{c}{$\lesssim 0.7$} \\
\tableline
\end{tabular}
\tablecomments{The table lists Bayesian (posterior) and frequentist
  (PL) estimates for the three flux contributions discussed in the
  text. Fluxes are given in units of
  $10^{-7}\,\mathrm{cm}^{-2}\,\mathrm{s}^{-1}\,\mathrm{sr}^{-1}$.  The
  ratios $q_\mathrm{ps}$, $q_\mathrm{gal}$, and $q_\mathrm{iso}$ refer
  to the total map flux $F_\mathrm{tot} = \sum_i F_i$, which has been
  consistently derived from the posterior and profile likelihood,
  respectively. For comparison, the integral fluxes of all sources
  listed in the 2FGL and 3FGL catalogs for $|b|\geq 30^\circ$ are also
  given. The last row lists an estimate of the flux contributed by
  residual cosmic rays.}
\end{center}
\end{table}

For convenience, individual components can be expressed as fractions
$q$ of the total map flux $F_\mathrm{tot}$\,. The fractions are listed
in Table~\ref{tab:comp_hybrid_hp6_2}.  We found that the high-latitude
gamma-ray emission between 1\,GeV and 10\,GeV is composed of $(25 \pm
2)$\% point-source contributions, $(69.3 \pm 0.7)$\% Galactic
foreground contributions, and $(6 \pm 2)$\% isotropic diffuse
background emission.

Even if not indicated by Figures~\ref{sfig:triangle_mbpl_3} and
\ref{sfig:triangle_hybrid_2}, remaining degeneracies between an
isotropic Galactic component accounted for in the template and the
$F_\mathrm{iso}$ parameter considered in this analysis might be
present.

The flux contribution from point sources can be compared to the flux
of all sources resolved in the 3FGL catalog (for $|b|\geq 30^\circ$;
see Table~\ref{tab:comp_hybrid_hp6_2}). From the difference
$F_\mathrm{ps} - F^\mathrm{3FGL}_\mathrm{cat}$ we conclude that a flux
of $1.4^{+0.3}_{-0.2}\times
10^{-7}\,\mathrm{cm}^{-2}\,\mathrm{s}^{-1}\,\mathrm{sr}^{-1}$ between
1\,GeV and 10\,GeV can be attributed to originate from so far
unresolved point sources.  With regard to the IGRB flux measured by
\citet{2015ApJ...799...86A}, we could therefore clarify between 42\%
and 56\% of its origin between 1\,GeV and 10\,GeV.\footnote{ The IGRB
  obtained by \citet{2015ApJ...799...86A} in the 1\,GeV to 10\,GeV
  energy band is between $\sim\!3.2 \times
  10^{-7}\,\mathrm{cm}^{-2}\,\mathrm{s}^{-1}\,\mathrm{sr}^{-1}$ and
  $\sim\!4.3 \times
  10^{-7}\,\mathrm{cm}^{-2}\,\mathrm{s}^{-1}\,\mathrm{sr}^{-1}$,
  including systematic uncertainties of the Galactic foreground
  modeling. Note that this measurement refers to the 2FGL catalog,
  which has been used for subtracting resolved sources from the
  EGB. We therefore attribute a flux of $1.8^{+0.3}_{-0.2}\times
  10^{-7}\,\mathrm{cm}^{-2}\,\mathrm{s}^{-1}\,\mathrm{sr}^{-1}$ to
  unresolved point sources in this IGRB measurement (using
  $F^\mathrm{2FGL}_\mathrm{cat}$ as quoted in
  Table~\ref{tab:comp_hybrid_hp6_2}).}

\paragraph{Residual Cosmic Rays}
The sum of the values $F_\mathrm{iso}=(0.9\pm 0.2)\times
10^{-7}\,\mathrm{cm}^{-2}\,\mathrm{s}^{-1}\,\mathrm{sr}^{-1}$ and
$F_\mathrm{ps}=3.9^{+0.3}_{-0.2}\times
10^{-7}\,\mathrm{cm}^{-2}\,\mathrm{s}^{-1}\,\mathrm{sr}^{-1}$ listed
in Table~\ref{tab:comp_hybrid_hp6_2} can be compared with the EGB
derived in \cite{2015ApJ...799...86A}.  In the energy range between
1\,GeV and 10\,GeV this amounts to values between $4.7 \times
10^{-7}\,\mathrm{cm}^{-2}\,\mathrm{s}^{-1}\,\mathrm{sr}^{-1}$ and $6.4
\times 10^{-7}\,\mathrm{cm}^{-2}\,\mathrm{s}^{-1}\,\mathrm{sr}^{-1}$,
including systematics in the Galactic diffuse modeling; these values
compare well with the total $F_\mathrm{iso}+F_\mathrm{ps}$ found here.

\begin{figure*}[t]
\epsscale{0.6}
\plotone{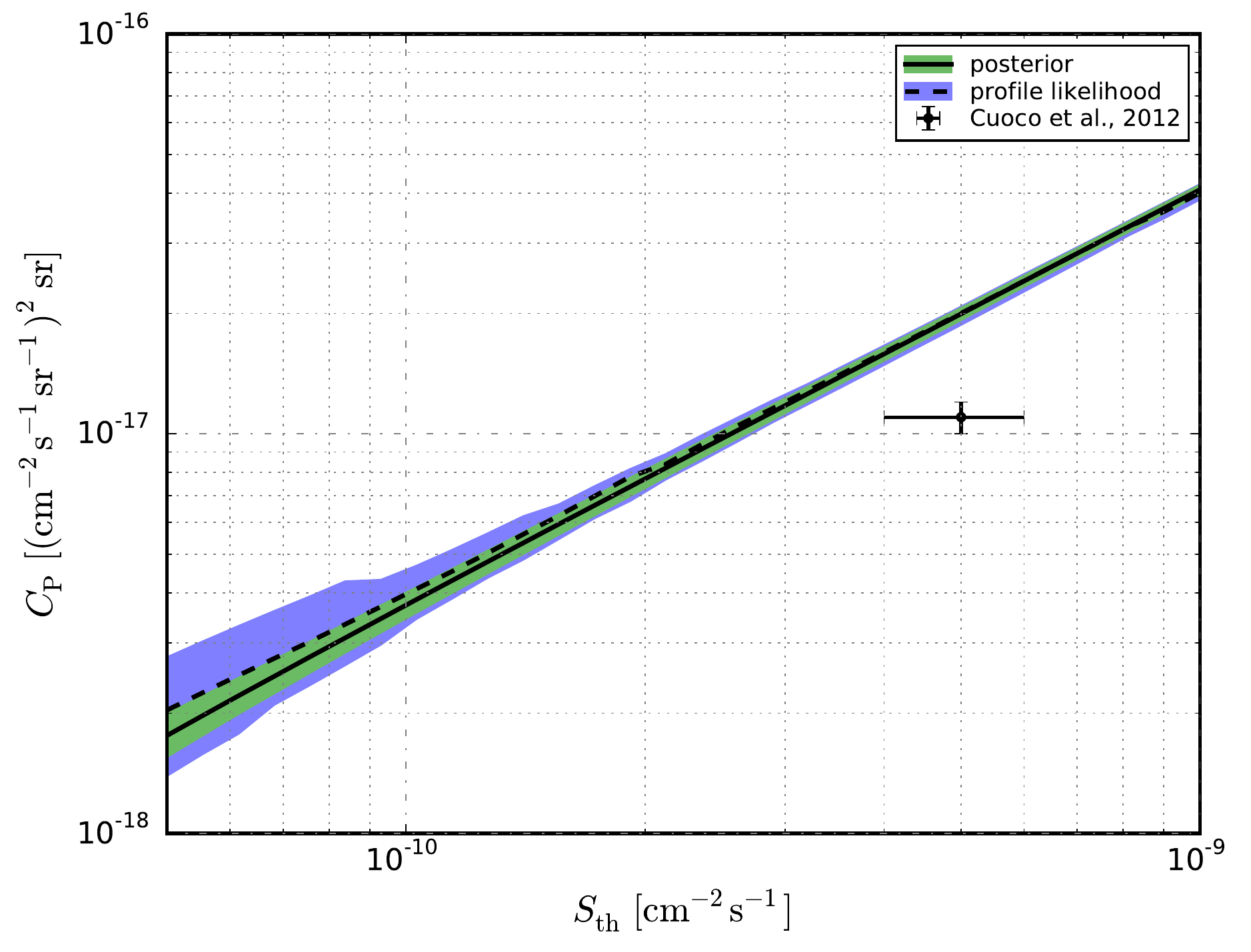}
\caption{Predicted angular power $C_\mathrm{P}$ as a function of the
  point-source detection threshold $S_\mathrm{th}$, derived from the
  \dnds\ distribution measured in this work (hybrid approach,
  $N^\shy_\mathrm{b}=2$).  The solid (dashed) black line and the
  shaded green (blue) band denote best fit and corresponding
  uncertainty derived from the posterior (profile likelihood). The
  data point refers to a measurement by \citet{Cuoco:2012yf} and
  \citet{Ackermann:2012uf}.
\label{fig:Cp}}
\end{figure*}

However, the truly diffuse isotropic background emission
$F_\mathrm{iso}$ incorporates residual cosmic rays (CRs) not rejected
by analysis cuts \citep[see][]{2015ApJ...799...86A}, while for the EGB
derived in \cite{2015ApJ...799...86A} the CR contamination has been
accounted for and subtracted.  The level of residual CR contamination
in the \texttt{P7REP\_CLEAN} selection used in this work has been
estimated to be between 15\% and 20\% of the measured IGRB flux above
1\,GeV \citep[see Figure~28 in][]{2012ApJS..203....4A}, thus amounting
to about $5$-$7\times
10^{-8}\,\mathrm{cm}^{-2}\,\mathrm{s}^{-1}\,\mathrm{sr}^{-1}$.

\section{Anisotropy}\label{sec:anisotropy}
Complementary to the \opdf, the anisotropy (or autocorrelation) probes
unresolved point sources \citep[][]{Ackermann:2012uf,
  Cuoco:2012yf,DiMauro:2014wha,2014JCAP...01..049R}.  The two
observables can thus be compared.  The anisotropy in a given energy
band can be calculated from the \dnds\ distribution by
\begin{equation}
C_\mathrm{P} =  \int_0^{S_{\rm th}} \mathrm{d}S\,S^2 \frac{\mathrm{d}N}{\mathrm{d}S}\,,
\end{equation}
where $S_{\rm th}$ is the flux threshold of detected point sources,
assumed to be `sharp' and independent of the photon spectral index of
the sources. Indeed, the previous assumption is a good approximation
for the 1\,GeV to 10\,GeV energy band \citep[][]{Cuoco:2012yf}.  We
thus calculated the predicted anisotropy from the \dnds\ distribution
measured in this work (hybrid approach, $N^\shy_\mathrm{b}=2$) as a
function of the threshold flux $S_{\rm th}$.  Results are shown in
Figure~\ref{fig:Cp}.  To derive the uncertainty band of
$C_\mathrm{P}$, we sampled the \dnds\ from the posterior and
calculated $C_\mathrm{P}$ from each sampling point of the
\dnds\ parameter space.  The uncertainty on $C_\mathrm{P}$ was then
derived using both the Bayesian and the frequentist approaches; see
Sections~\ref{sssec:par_est} and \ref{sssec:par_est_freq}.  The
predicted $C_\mathrm{P}$ can be compared to the value $(1.1\pm 0.1)
\times 10^{-17}$\,(cm$^{-2}$\,s$^{-1}$\,sr$^{-1}$)$^2$\,sr measured in
\citet{Cuoco:2012yf} and \citet{Ackermann:2012uf}, using a threshold
of about $4$-$6\times 10^{-10}$\,cm$^{-2}$\,s$^{-1}$ suitable for
sources detected in the 1FGL catalog \citep{2010ApJS..188..405A}.  It
can be seen in Figure~\ref{fig:Cp} that the predicted anisotropy is
slightly higher than the measured value.  This can in part be
explained by the approximation of the threshold as a sharp cutoff, as
well as a possible systematic underestimate of the measured anisotropy
itself \citep{Chang:2013ada}.  In addition, a possible clustering of
point sources at angular scales smaller than the pixel size could in
principle be degenerate with the inferred \dnds\ distribution, leading
to systematically higher anisotropies. The anisotropy of clustering
effects is, however, expected to be rather small as compared to the
$C_\mathrm{P}$ values found here, i.e., $C^\mathrm{cluster}_{\ell >
  200 } \lesssim
10^{-20}\,(\mathrm{cm}^{-2}\,\mathrm{s}^{-1}\,\mathrm{sr}^{-1})^2\,\mathrm{sr}$
for multipoles $\ell$ corresponding to angular scales smaller than the
pixel size \citep[e.g.,][]{2007PhRvD..75f3519A,2015ApJS..221...29C}.
Clustering can thus be neglected in this analysis. For the moment, we
deem the agreement reasonable, and we wait for an updated anisotropy
measurement for a more detailed comparison.

\section{SYSTEMATICS}\label{sec:systematics}
The following section is dedicated to systematic and modeling
uncertainties of the analysis framework. In particular, we extensively
investigated possible uncertainties due to the chosen pixel size
(Section~\ref{ssec:hp7}), statistical effects imposed by bright point
sources (Section~\ref{ssec:ps_mask}), and the Galactic foreground
modeling (Section~\ref{ssec:GFsyst}).

\subsection{Pixel Size}\label{ssec:hp7}
\begin{figure}[t]
\epsscale{1.15}
\plotone{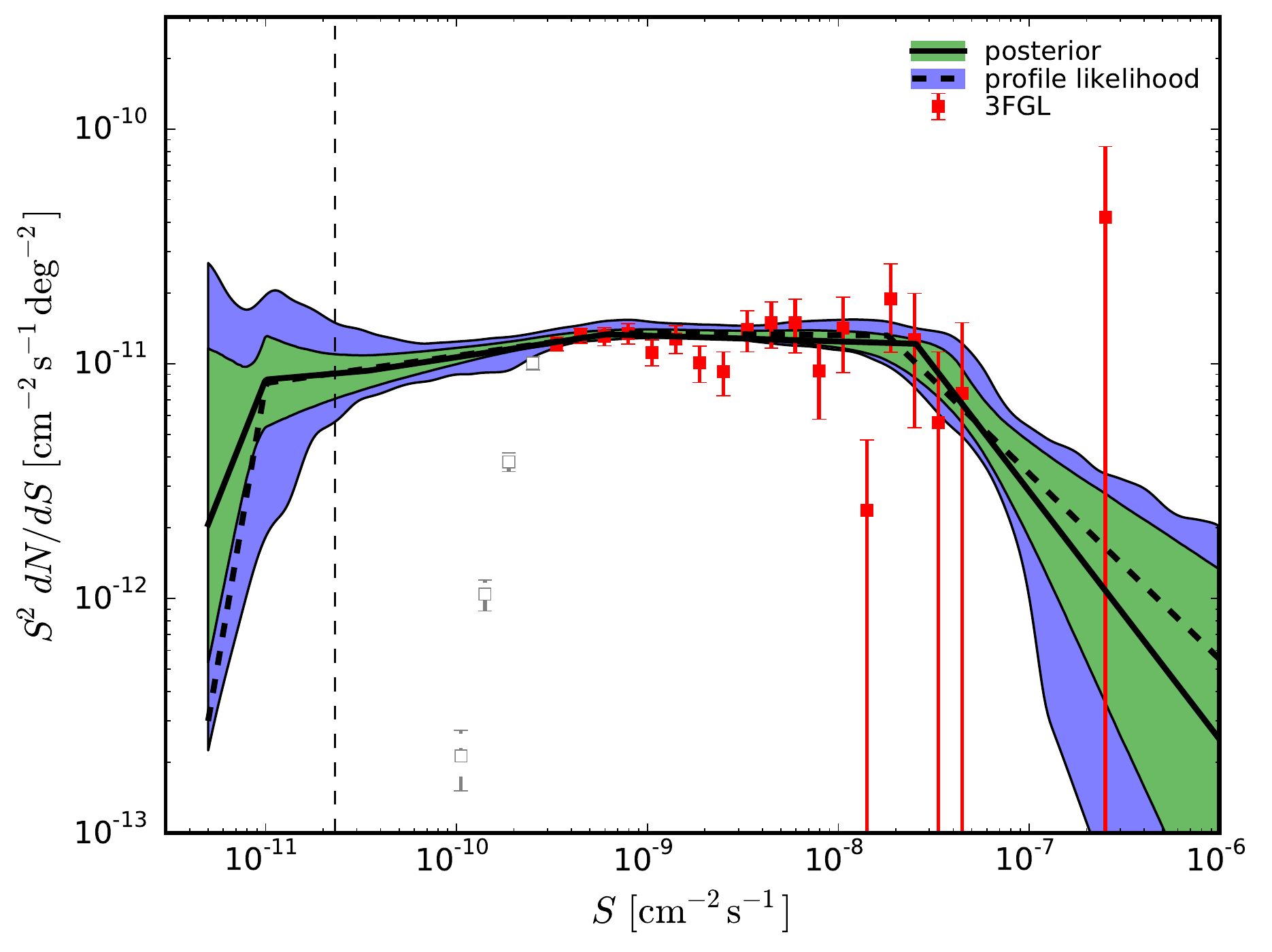}
\caption{Differential source-count distribution \dnds\ obtained with
  the hybrid approach for a HEALPix grid of order $\kappa=7$. The
  hybrid approach was carried out allowing three free breaks and a
  node. Line styles and colors are as in
  Figure~\ref{fig:mbpl_fit_3_2}.
\label{fig:hybrid_fit_hp7_3}}
\end{figure}

The results discussed in Section~\ref{sec:application} were
cross-checked using smaller pixels, i.e., HEALPix order $\kappa=7$,
slightly oversampling the effective PSF (see
Section~\ref{sec:Fermi_data}).  All results were stable against the
resolution change, given the corresponding uncertainty bands.
However, it was found that the enhanced PSF smoothing increased the
uncertainty in determining the first break. An example is given in
Figure~\ref{fig:hybrid_fit_hp7_3}, showing the \dnds\ distribution
obtained with the hybrid approach considering three free breaks and a
node. It is demonstrated in Section~\ref{ssec:ps_mask} that the
increased uncertainty in the bright-source region in turn led to a
small bias in determining the indices $n_2$ and $n_3$.

\begin{table}[t]
\begin{center}
\caption{Hybrid approach, $N^\shy_\mathrm{b} = 3$,
  $\kappa=7$;\\ units are as in
  Tables~\ref{tab:hybrid_fit_3_2_1} and \ref{tab:comp_hybrid_hp6_2}.
\label{tab:hybrid_fit_hp7_3}}
\begin{tabular}{lcc}
\tableline\tableline
Parameter & \multicolumn{1}{c}{Posterior} & \multicolumn{1}{c}{PL} \\
\tableline
$A_\mathrm{gal}$ &  $1.076^{+0.004}_{-0.004}$ &  $1.074^{+0.007}_{-0.004}$ \\
$F_\mathrm{ps}$ &  $3.6^{+0.2}_{-0.2}$ &  $3.4^{+0.5}_{-0.2}$ \\
$F_\mathrm{iso}$ &  $1.3^{+0.1}_{-0.2}$ &  $1.4^{+0.3}_{-0.4}$ \\
\tableline
$\ln \mathcal{L}_1({\bf \Theta})$ &  $-667.2$ &  $-667.9$ \\
$\ln \mathcal{L}_2({\bf \Theta})$ &  $-257817.9$ &  $-257812.0$\\
$\ln \mathcal{Z}$ & \multicolumn{2}{c}{$-257825.9 \pm 0.1$} \\
\tableline
\end{tabular}
\end{center}
\end{table}

Table~\ref{tab:hybrid_fit_hp7_3} summarizes fit results that do not
become evident in Figure~\ref{fig:hybrid_fit_hp7_3}. The integral
point-source flux $F_\mathrm{ps}$ slightly decreased with respect to
the value obtained for $\kappa=6$, with a corresponding increase of
the isotropic background emission $F_\mathrm{iso}$, while the sum
$F_\mathrm{ps} + F_\mathrm{iso}$ remained constant within
(single-parameter) statistical uncertainties.  This is consistent with
resolving fewer point sources due to reduced sensitivity, given that
the value of $A_\mathrm{gal}$ stayed almost the same as found for
$\kappa=6$.

\subsection{Point-source Masking}\label{ssec:ps_mask}
The presence of bright point sources and the corresponding shape of
their source-count distribution may influence the overall fit of the
intermediate region and the faint-source region. The strength of a
possible bias may also depend on the pixel size.

The level of systematics caused by bright point sources was
investigated with point-source masks.  To eliminate the influence of
bright sources, we removed all pixels including sources with an
integral flux larger than or equal to $S_\mathrm{mask} \simeq
10^{-8}\,\mathrm{cm}^{-2}\,\mathrm{s}^{-1}$.  The value of
$S_\mathrm{mask}$ was chosen to be slightly below the first break
determined from the overall fit (see
Section~\ref{sec:application}). Source positions and fluxes were
retrieved from the 3FGL catalog.  For each source, all pixels included
in a circle with a radius of $2.5^\circ$ (corresponding to $\sim 6
\sigma_\mathrm{psf}$)\footnote{Given that most source photons are
  emitted at low energies, we remark that the value of $2.5^\circ$
  corresponds to almost $4 \sigma_\mathrm{psf}(1\,\mathrm{GeV})$. The
  68\% containment radius of the PSF at 1\,GeV is
  $\sigma_\mathrm{psf}(1\,\mathrm{GeV}) \simeq 0.67^\circ$.}  around
the cataloged source position were masked in the counts map. We
checked that the mask area was sufficiently large by comparing radii
between $3 \sigma_\mathrm{psf}$ and $7\sigma_\mathrm{psf}$. Remnant
effects became negligible for radii larger than $\sim 5
\sigma_\mathrm{psf}$.

The masked data were fit using the hybrid approach with three free
breaks, in order to retain full sensitivity to a possible break in the
faint-source region. Priors were chosen as listed in
Table~\ref{tab:priors}, with the exception of changing the upper bound
of the prior of the first break to $S_\mathrm{mask}$. The prior of
$n_1$ was changed accordingly to sample the interval $[1.7,2.3]$,
substantially covering the intermediate region. In addition, the flux
normalization constant was fixed to $S_0 = 3\times
10^{-9}\,\mathrm{cm}^{-2}\,\mathrm{s}^{-1}$ and the upper flux cutoff
of \dnds\ was set to $S_\mathrm{cut} \equiv S_\mathrm{mask}$.

\begin{figure*}[t]
\begin{centering}
\subfigure[point-source mask, HEALPix resolution $\kappa = 6$]{%
  \includegraphics[width=0.49\textwidth]{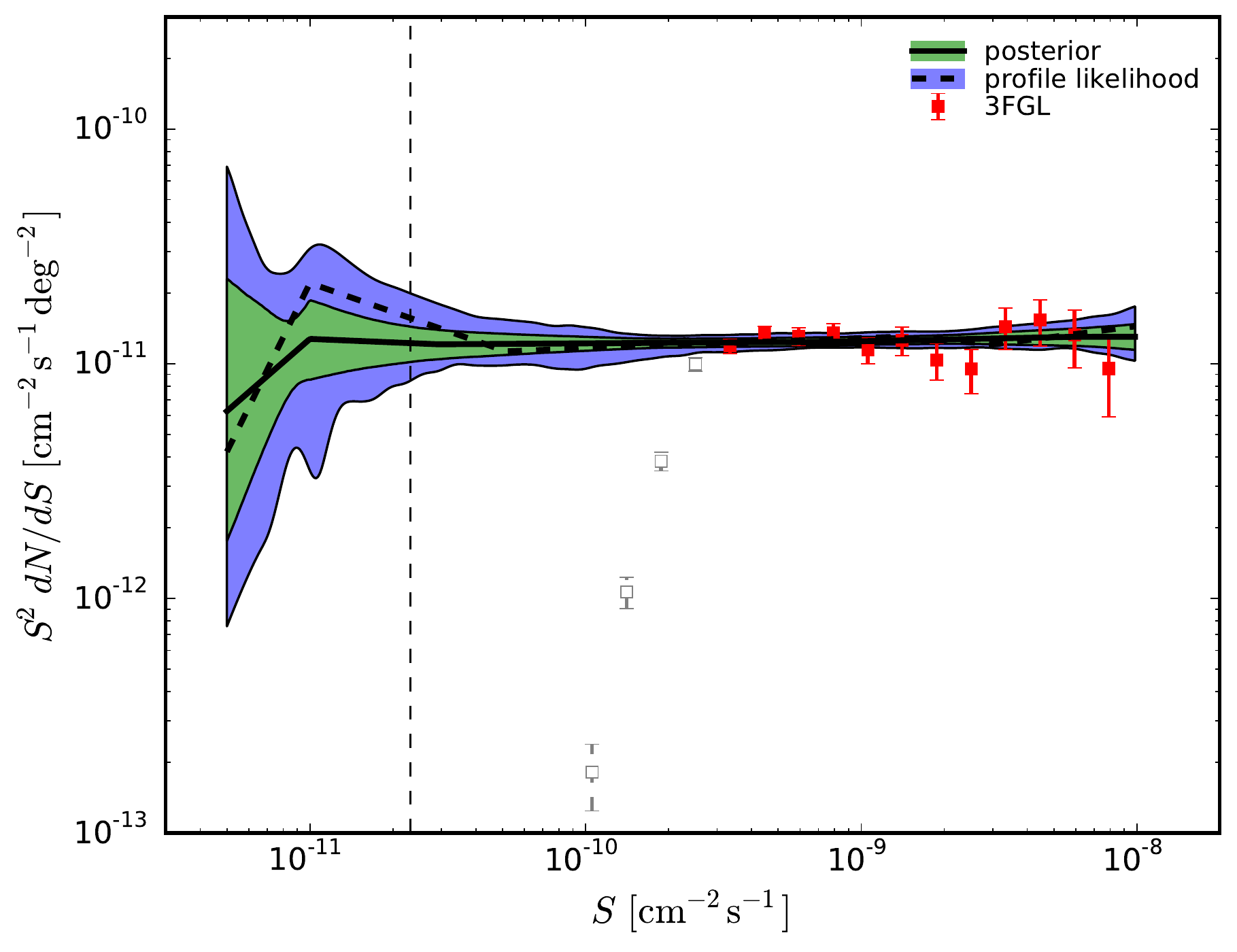}
  \label{sfig:ps_mask_hp6}
}
\subfigure[point-source mask, HEALPix resolution $\kappa = 7$]{%
  \includegraphics[width=0.49\textwidth]{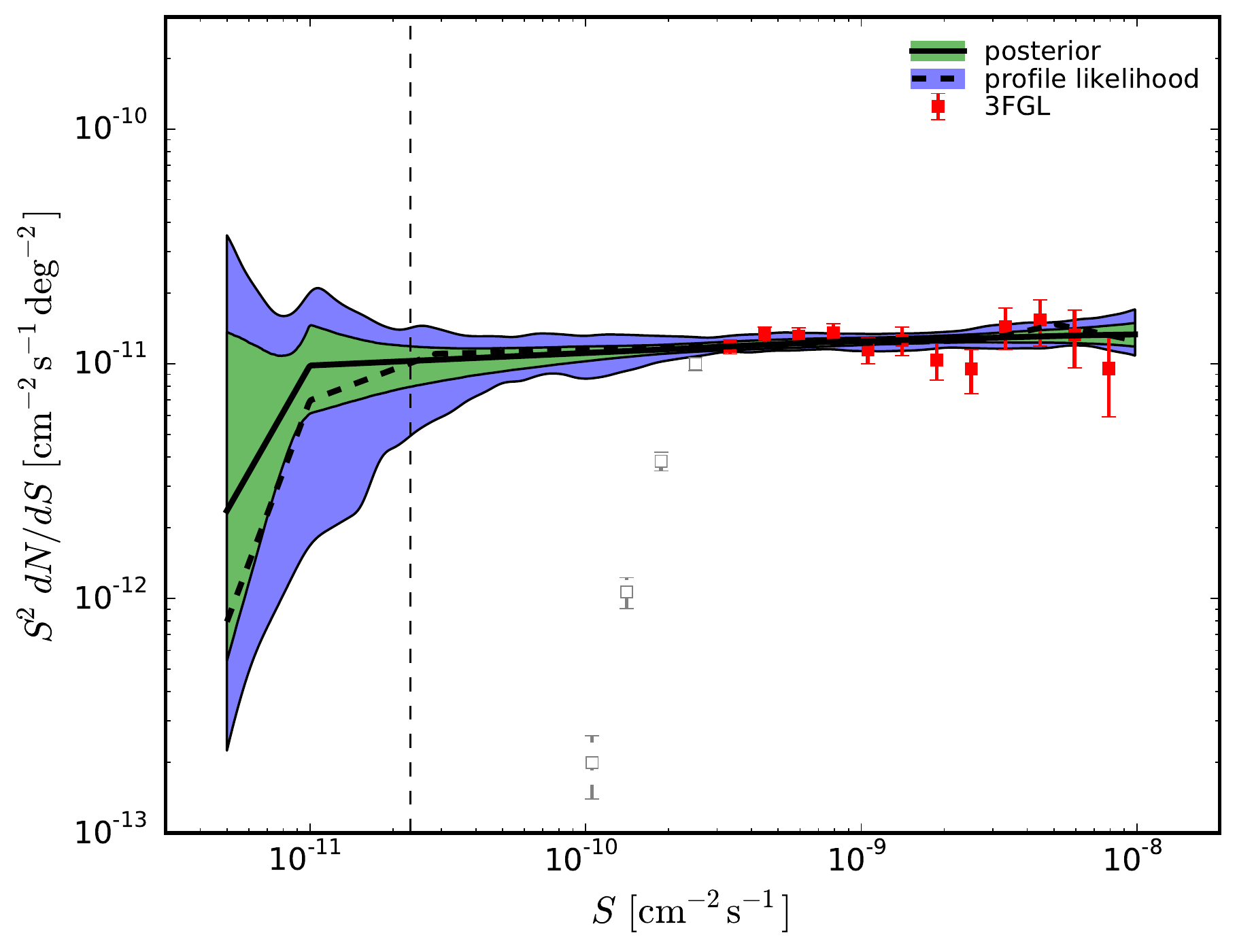}
  \label{sfig:ps_mask_hp7}
}
\caption{Differential source-count distribution \dnds\ obtained with
  the hybrid approach ($N^\shy_\mathrm{b}=3$) for masking bright
  sources with a flux larger than
  $10^{-8}\,\mathrm{cm}^{-2}\,\mathrm{s}^{-1}$.  The figure shows
  results for two pixel sizes, i.e., a HEALPix grid of order (a)
  $\kappa = 6$ and (b) $\kappa=7$. Line styles and colors are as in
  Figure~\ref{fig:mbpl_fit_3_2}.
\label{fig:ps_mask}}
\end{centering}
\end{figure*}

The results are shown in Figure~\ref{fig:ps_mask} for a pixelization
with resolution parameters $\kappa=6$ and $\kappa=7$. It can be seen
that the results are consistent with what was found in
Section~\ref{sec:application}. For $\kappa=7$ we find that the
uncertainty band is slightly down-shifted as compared to $\kappa=6$,
but best-fit results match well within uncertainties.  The value of
$A_\mathrm{gal}$ was determined to be $1.071^{+0.004}_{-0.004}$
($1.072^{+0.005}_{-0.005}$) for $\kappa=6$ and
$1.075^{+0.004}_{-0.004}$ ($1.073^{+0.006}_{-0.004}$) for $\kappa=7$,
using the posterior (profile likelihood). The integral flux of the
isotropic diffuse background emission $F_\mathrm{iso}$ was obtained to
be $0.9^{+0.2}_{-0.2}\,(0.8^{+0.5}_{-0.3}) \times
10^{-7}\,\mathrm{cm}^{-2}\,\mathrm{s}^{-1}$ for $\kappa=6$ and
$1.2^{+0.1}_{-0.2}\,(1.4^{+0.2}_{-0.4}) \times
10^{-7}\,\mathrm{cm}^{-2}\,\mathrm{s}^{-1}$ for $\kappa=7$. The larger
value of $F_\mathrm{iso}$ in the latter case is consistent with the
fact of resolving fewer point sources for $\kappa=7$.

We conclude that systematic effects due to bright point sources are
dominated by statistical uncertainties. Bright point sources do not
affect the determination of the \dnds\ broken power-law indices in the
intermediate and faint-source regions.  For $\kappa=7$, comparing the
analyses of full data (Figure~\ref{fig:hybrid_fit_hp7_3}) and masked
data (Figure~\ref{sfig:ps_mask_hp7}) indicates that systematic effects
slightly increased with enhanced PSF smoothing, but effects on the
indices $n_2$ and $n_3$ remain rather small.

\begin{figure*}[t]
\begin{centering}
\subfigure[]{%
  \includegraphics[width=0.45\textwidth]{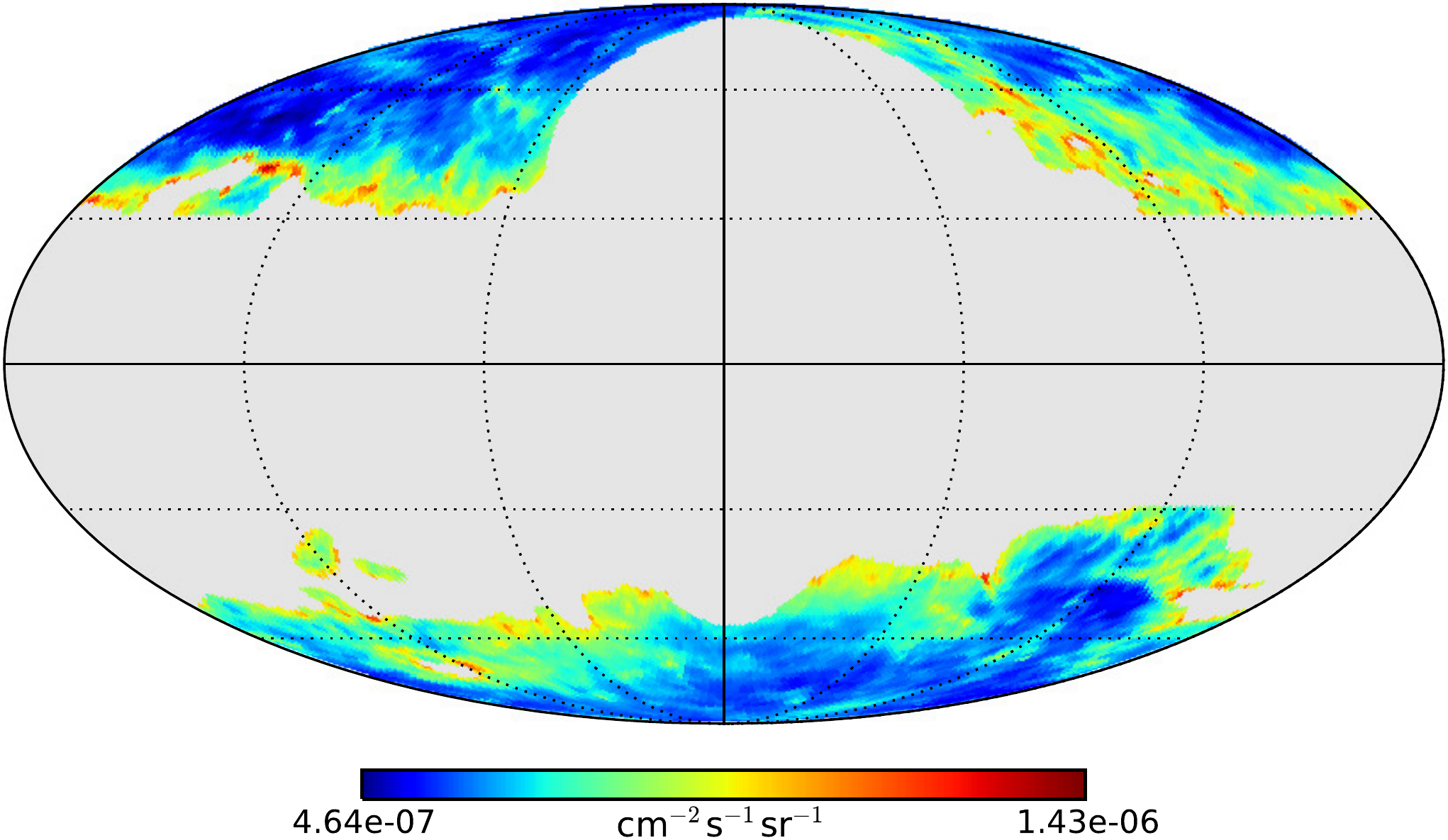}
  \label{sfig:gpll_mask}
}\hspace{1cm}
\subfigure[]{%
  \includegraphics[width=0.45\textwidth]{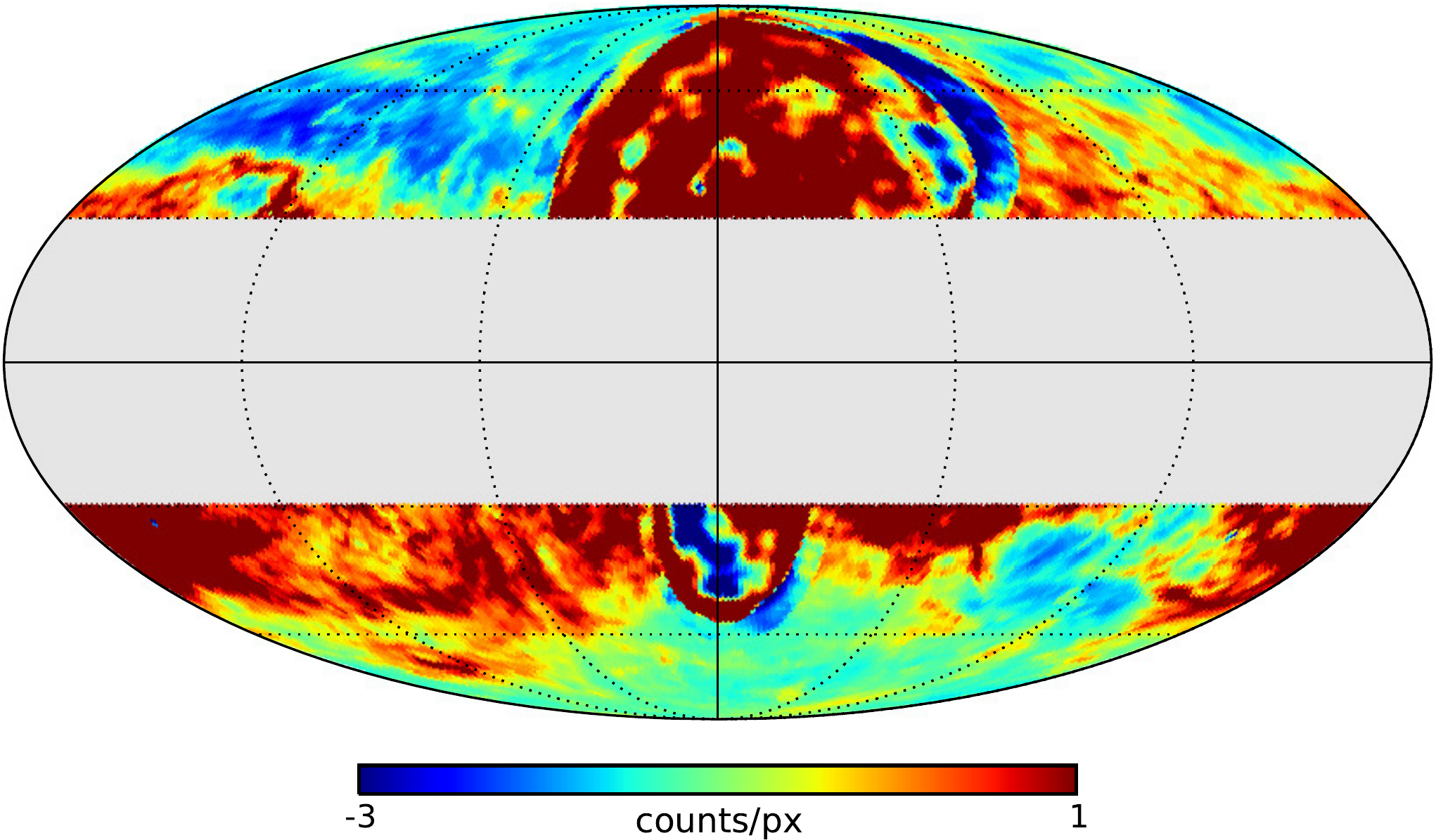}
  \label{sfig:gal_Cdiff_hp6}
}
\caption{(a) Same as Figure~\ref{fig:f1}, but overlaying the GPLL mask
  (in gray). (b) Difference map \texttt{P7}\,$-$\,\texttt{P7REP} of
  the Galactic foreground emission models derived for \texttt{P7} and
  \texttt{P7REP} data between 1\,GeV and 10\,GeV. The difference is
  given in units of counts pixel$^{-1}$, derived with the exposure map
  as discussed in Section~\ref{ssec:bckgs}. Projection and mask are
  the same as in Figure~\ref{fig:f1}.  The color bar has been clipped
  at $-3$ and $1$, in order to visualize small differences.  The
  entire range covered by the data is $(-6.5,12)$. The color mapping
  is linear.\label{fig:GPLL_GFdiff}}
\end{centering}
\end{figure*}

\subsection{Galactic Foreground}\label{ssec:GFsyst}
\begin{figure*}[t]
\begin{centering}
\subfigure[]{%
  \includegraphics[width=0.49\textwidth]{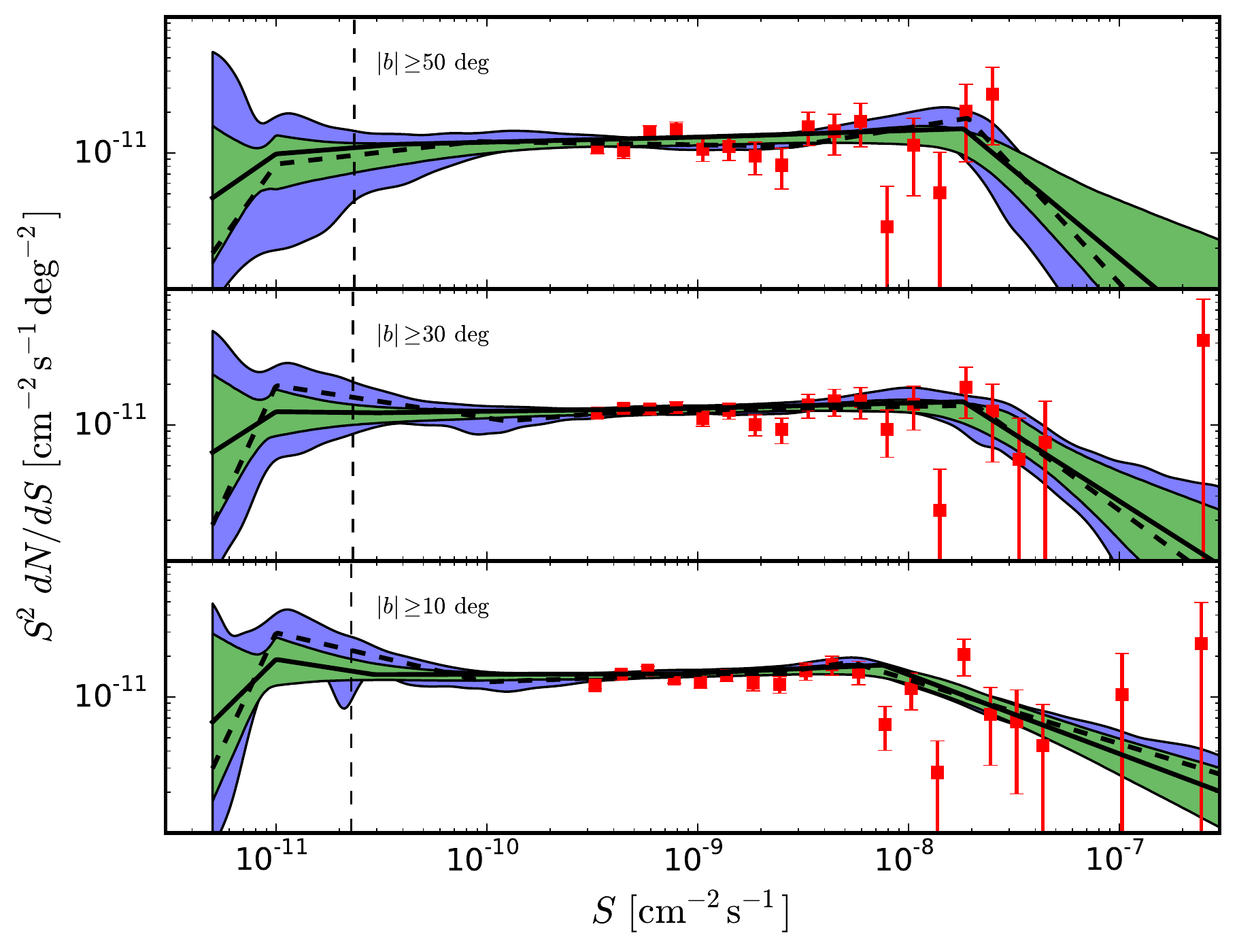}
  \label{sfig:GF_gal_cut}
}
\subfigure[]{%
  \includegraphics[width=0.49\textwidth]{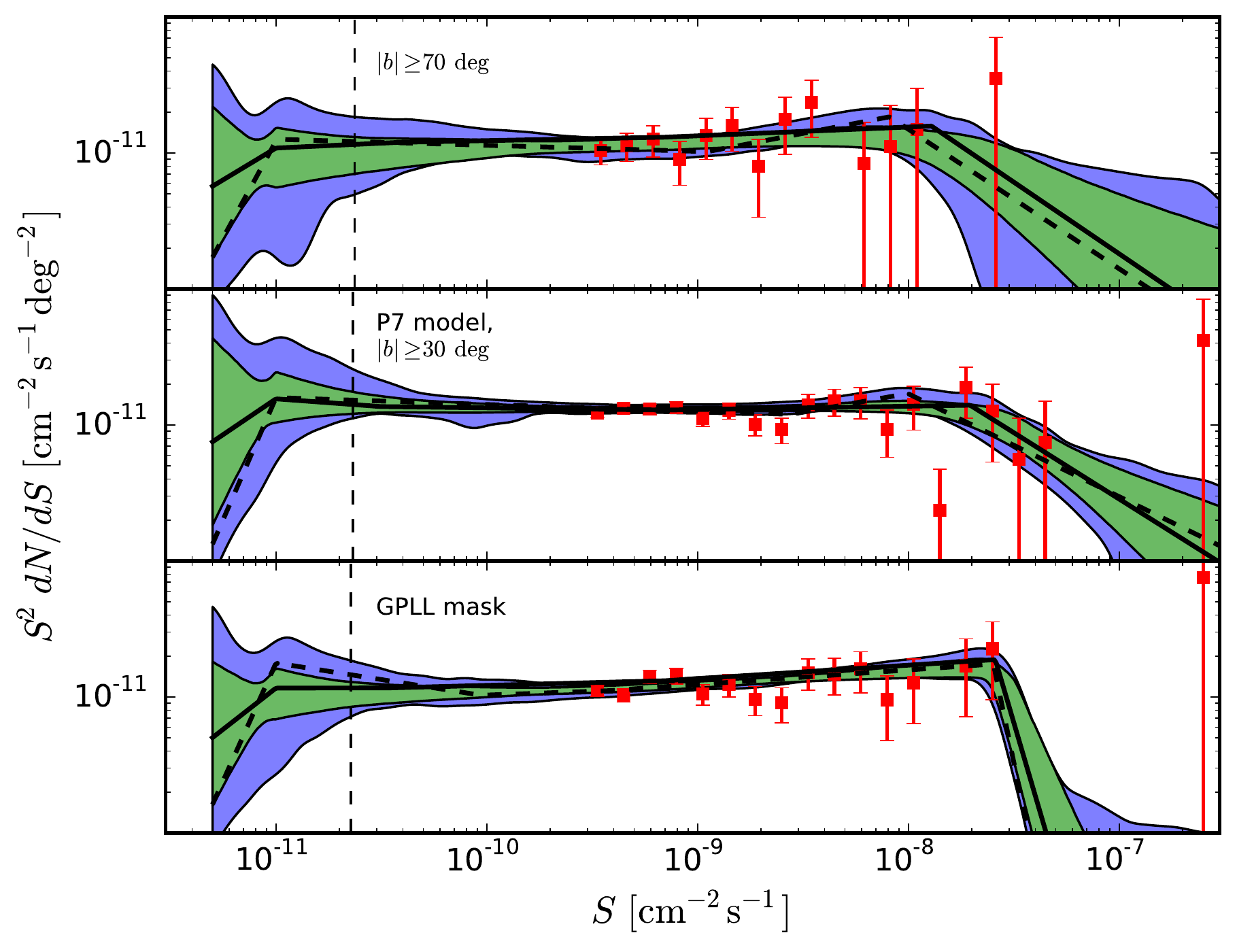}
  \label{sfig:GF_model}
}
\caption{Galactic foreground systematics. The left column shows the
  \dnds\ distributions obtained for different Galactic latitude cuts
  ($10^\circ$, $30^\circ$, and $50^\circ$), using the hybrid approach
  with $N^\shy_\mathrm{b} = 3$ breaks and a node.  Besides the
  $70^\circ$ Galactic latitude cut in the top panel, the right column
  depicts the fit results using the GPLL mask and the \texttt{P7}
  Galactic foreground template model. Line styles and colors are the
  same as in Figure~\ref{fig:mbpl_fit_3_2}.
\label{fig:GFsys}}
\end{centering}
\end{figure*}

We checked our results for systematic uncertainties of the Galactic
foreground model, considering three different approaches:
\begin{itemize}
  \item \textit{Dependence on the Galactic latitude cut.} We selected
    different ROIs, covering regions $|b|\geq b_\mathrm{cut}$.  The
    parameter $b_\mathrm{cut}$ was varied between $10^\circ$ and
    $70^\circ$, in steps of $10^\circ$.
  \item \textit{Extended Galactic plane mask (GPLL mask).} The GPLL
    mask was generated from the Galactic foreground emission model
    discussed in Section~\ref{ssec:bckgs}, by merging mask arrays for
    $|b| < 30^\circ$, a Galactic plane mask removing all pixels above
    a flux threshold\footnote{The Galactic foreground emission model
      was smoothed with a Gaussian kernel of $2^\circ$ before applying
      the threshold.} of
    $10^{-6}\,\mathrm{cm}^{-2}\,\mathrm{s}^{-1}\,\mathrm{sr}^{-1}$,
    and mask arrays for the Fermi bubbles and Galactic Loop\,I
    \citep{Fermi-LAT:2014sfa,2009arXiv0912.3478C,Su:2010qj}.  The GPLL
    mask is shown in Figure~\ref{sfig:gpll_mask}.
  \item \textit{Dependence on the Galactic foreground model.} Given
    systematic uncertainties of the Galactic foreground model in its
    entirety, we incorporated a different foreground model as derived
    for the preceeding \emph{Fermi}-LAT data release \texttt{Pass~7},
    named \texttt{gal\_2yearp7v6\_v0.fits}\footnote{See
      http://fermi.gsfc.nasa.gov/ssc/data/access/lat/
      \\BackgroundModels.html}.  Although mixing different versions of
    data releases and diffuse models is not generally recommended, the
    purpose here is to gauge the effect of a model differing in
    intensity as well as in morphology.  The deviations between the
    two models are shown in Figure~\ref{sfig:gal_Cdiff_hp6} for
    Galactic latitudes greater than $30^\circ$.
\end{itemize}
The hybrid approach was employed for all setups, choosing three free
breaks and a node.  The prior setup resembled the one used in
Section~\ref{sec:application}, but prior ranges were extended in
particular cases to cover the posterior sufficiently well.  The
results of the analyses are summarized in Figure~\ref{fig:GFsys} and
Table~\ref{tab:GFsys}.  We found that all results were stable against
the systematic checks. In addition, it should be noted that the
catalog (3FGL) data points derived for comparison were well reproduced
in all cases.

\begin{deluxetable*}{lcccccc}
\tablecaption{Galactic Latitude Cut and Foreground Systematics\label{tab:GFsys}}
\tablewidth{0pt}
\tablehead{
\colhead{} & \multicolumn{2}{c}{$|b|\geq 10^\circ$} & \multicolumn{2}{c}{$|b|\geq 30^\circ$} & \multicolumn{2}{c}{$|b|\geq 50^\circ$} \\
\colhead{Parameter} & \colhead{Posterior} & \colhead{PL} & \colhead{Posterior} & \colhead{PL}
 & \colhead{Posterior} & \colhead{PL}
}
\startdata
$S_\mathrm{b1}$ &  $0.8^{+0.4}_{-0.3}$ &  $0.5^{+0.4}_{-0.1}$ &  $1.8^{+0.9}_{-1.0}$ &  $2.1^{+1.5}_{-1.5}$ &  $1.8^{+0.8}_{-0.8}$ &  $1.9^{+0.8}_{-1.0}$  \\
$n_1$ &  $2.58^{+0.23}_{-0.14}$ &  $2.47^{+0.33}_{-0.10}$ &  $2.99^{+0.67}_{-0.43}$ &  $3.13^{+0.76}_{-0.76}$ &  $3.29^{+0.60}_{-0.71}$ &  $3.69^{+0.61}_{-0.92}$ \\
$A_\mathrm{gal}$ &  $1.017^{+0.002}_{-0.002}$ &  $1.018^{+0.002}_{-0.002}$ & $1.072^{+0.004}_{-0.004}$ &  $1.070^{+0.006}_{-0.003}$ &  $1.12^{+0.01}_{-0.01}$ &  $1.12^{+0.02}_{-0.02}$ \\
$F_\mathrm{ps}$ & $4.6^{+0.3}_{-0.3}$ & $4.9^{+0.3}_{-0.5}$ &  $3.9^{+0.3}_{-0.2}$ & $3.9^{+0.6}_{-0.3}$  & $3.5^{+0.3}_{-0.2}$ & $3.5^{+0.3}_{-0.5}$ \\
$F_\mathrm{gal}$ & $16.97^{+0.03}_{-0.03}$ & $16.97^{+0.04}_{-0.03}$ &  $10.95^{+0.04}_{-0.04}$ & $10.94^{+0.06}_{-0.03}$  & $8.34^{+0.09}_{-0.09}$ & $8.3^{+0.1}_{-0.1}$ \\
$F_\mathrm{iso}$ &  $1.0^{+0.2}_{-0.3}$ &  $0.8^{+0.3}_{-0.3}$ &  $0.9^{+0.2}_{-0.3}$ &  $0.9^{+0.5}_{-0.4}$ &  $0.8^{+0.2}_{-0.2}$ &  $0.9^{+0.2}_{-0.4}$ \\
\tableline
  &  &  &  \\
  & \multicolumn{2}{c}{$|b|\geq 70^\circ$} & \multicolumn{2}{c}{GPLL Mask} & \multicolumn{2}{c}{\texttt{P7} Model} \\
\tableline
$S_\mathrm{b1}$ &  $1.3^{+0.9}_{-0.8}$ &  $0.8^{+12.3}_{-0.5}$ &  $2.6^{+0.5}_{-0.3}$ &  $2.5^{+0.9}_{-0.7}$ &  $2.0^{+0.9}_{-1.3}$ &  $1.0^{+2.5}_{-0.3}$ \\
$n_1$ &  $3.06^{+0.64}_{-0.58}$ &  $3.03^{+1.27}_{-0.85}$ &  $7.28^{+1.56}_{-2.21}$ &  $9.48^{+0.52}_{-4.93}$ &  $2.98^{+0.61}_{-0.44}$ &  $2.76^{+1.39}_{-0.39}$ \\
$A_\mathrm{gal}$ &  $1.16^{+0.03}_{-0.03}$ &  $1.17^{+0.04}_{-0.05}$ &  $1.12^{+0.01}_{-0.01}$ &  $1.12^{+0.01}_{-0.03}$ &  $0.939^{+0.004}_{-0.004}$ &  $0.938^{+0.005}_{-0.004}$ \\
$F_\mathrm{ps}$ & $3.5^{+0.4}_{-0.4}$ & $3.2^{+1.1}_{-0.3}$ & $3.6^{+0.2}_{-0.2}$ & $3.6^{+0.5}_{-0.4}$ & $4.3^{+0.5}_{-0.3}$ & $4.0^{+1.1}_{-0.3}$ \\
$F_\mathrm{gal}$ &  $7.6^{+0.2}_{-0.2}$  & $7.6^{+0.3}_{-0.3}$ & $7.9^{+0.1}_{-0.1}$  & $8.0^{+0.1}_{-0.2}$ & $9.60^{+0.04}_{-0.04}$ & $9.59^{+0.05}_{-0.04}$ \\
$F_\mathrm{iso}$ &  $0.3^{+0.2}_{-0.2}$ &  $0.3^{+0.4}_{-0.2}$ &  $0.8^{+0.2}_{-0.2}$ &  $0.6^{+0.4}_{-0.1}$ &  $2.0^{+0.2}_{-0.5}$ &  $2.2^{+0.3}_{-1.0}$ 
\enddata
\tablecomments{Selection of fit parameters obtained for different
  Galactic latitude cuts, the GPLL mask, and the \texttt{P7} Galactic
  foreground model template. The fit was carried out with the hybrid
  approach, using $N^\shy_\mathrm{b}=3$ breaks and a node. The units
  resemble the ones used in Tables~\ref{tab:hybrid_fit_3_2_1} and
  \ref{tab:comp_hybrid_hp6_2}.}
\end{deluxetable*}

In the bright-source region, the error band increases almost
monotonically with increasing Galactic latitude cut, due to the
decreasing number of bright sources present in the ROI. We note that
for the $10^\circ$ cut the index $n_1=2.58^{+0.23}_{-0.14}$ matches
well within uncertainties the index deduced by the Fermi Collaboration
from 1FGL catalog data
\citep[$n_1=2.38^{+0.15}_{-0.14}$,][]{2010ApJ...720..435A} for the
same latitude cut and energy band. The first break position, however,
was found to be a factor of $2$ to $3$ larger than in the 1FGL
analysis. The index below the first break is $n_2 \simeq 2$.

The fits of the faint-source region were stable against changing the
Galactic latitude cut.  The slopes of the corresponding \dnds\ fits
match well within uncertaintites for increasing latitude.
Uncertainties grow for higher Galactic latitude cuts given less
statistics. For lower latitude cuts, Figure~\ref{fig:GFsys} indicates
an upturn for very faint sources, which is, however, not significant.
The stability against the Galactic latitude cut is further
supplemented by the integral point-source flux $F_\mathrm{ps}$ (see
Table~\ref{tab:GFsys}), which remains stable within uncertainties.

Table~\ref{tab:GFsys} shows that the normalization of the Galactic
foreground model, $A_\mathrm{gal}$, increases with the latitude cut by
$\sim$10\% from $10^\circ$ to $50^\circ$, while the integral flux of
the isotropic background emission remains constant ($\sim\!9\times
10^{-8}\,\mathrm{cm}^{-2}\,\mathrm{s}^{-1}\,\mathrm{sr}^{-1}$).~\footnote{Given
  large uncertainties and increasing degeneracies, the $|b|\geq
  70^\circ$ ROI has been excluded from this discussion.}  The increase
of $A_\mathrm{gal}$ thus indicates a gradual mismatch between
foreground model and data.  Likewise, it can also indicate the
presence of a new component not covered by our analysis setup.  We
note that a similar behavior has been found in other analyses,
including the 3FGL catalog \citep[see, e.g., Figure~25
  in][]{2015ApJS..218...23A}.

The stability of the results obtained in this article is supplemented
by comparing with the GPLL mask and the \texttt{Pass 7} foreground
model (\texttt{P7} model).  The GPLL mask in particular removes the
Galactic lobes and Galactic Loop\,I, known as regions potentially
affected by large systematic model uncertainties. Employing the
\texttt{P7} model introduces a different Galactic foreground model in
its entirety. As demonstrated in Figure~\ref{sfig:gal_Cdiff_hp6}, the
differences between the models exhibit a nontrivial morphology. The
pixel distribution of photon-count differences extends to $\sim$3 for
the dominating part of the ROI, i.e., systematics can be expected at
the flux level of the sensitivity estimate $S_\mathrm{sens}$.  The
resulting \dnds\ distributions and the integral point-source fluxes
$F_\mathrm{ps}$ are consistent within uncertainties. It is to be
noted, however, that the integral isotropic background flux
$F_\mathrm{iso}$ increased by a factor of $\sim$2 for the \texttt{P7}
model.  At the same time, $F_\mathrm{gal}$ decreased, maintaining a
stable sum $F_\mathrm{gal} + F_\mathrm{iso}$.  We therefore remark
that modeling uncertainties can cause $F_\mathrm{iso}$ to depend on
the Galactic foreground model.

\begin{figure*}[t]
    \epsscale{0.60}
    \plotone{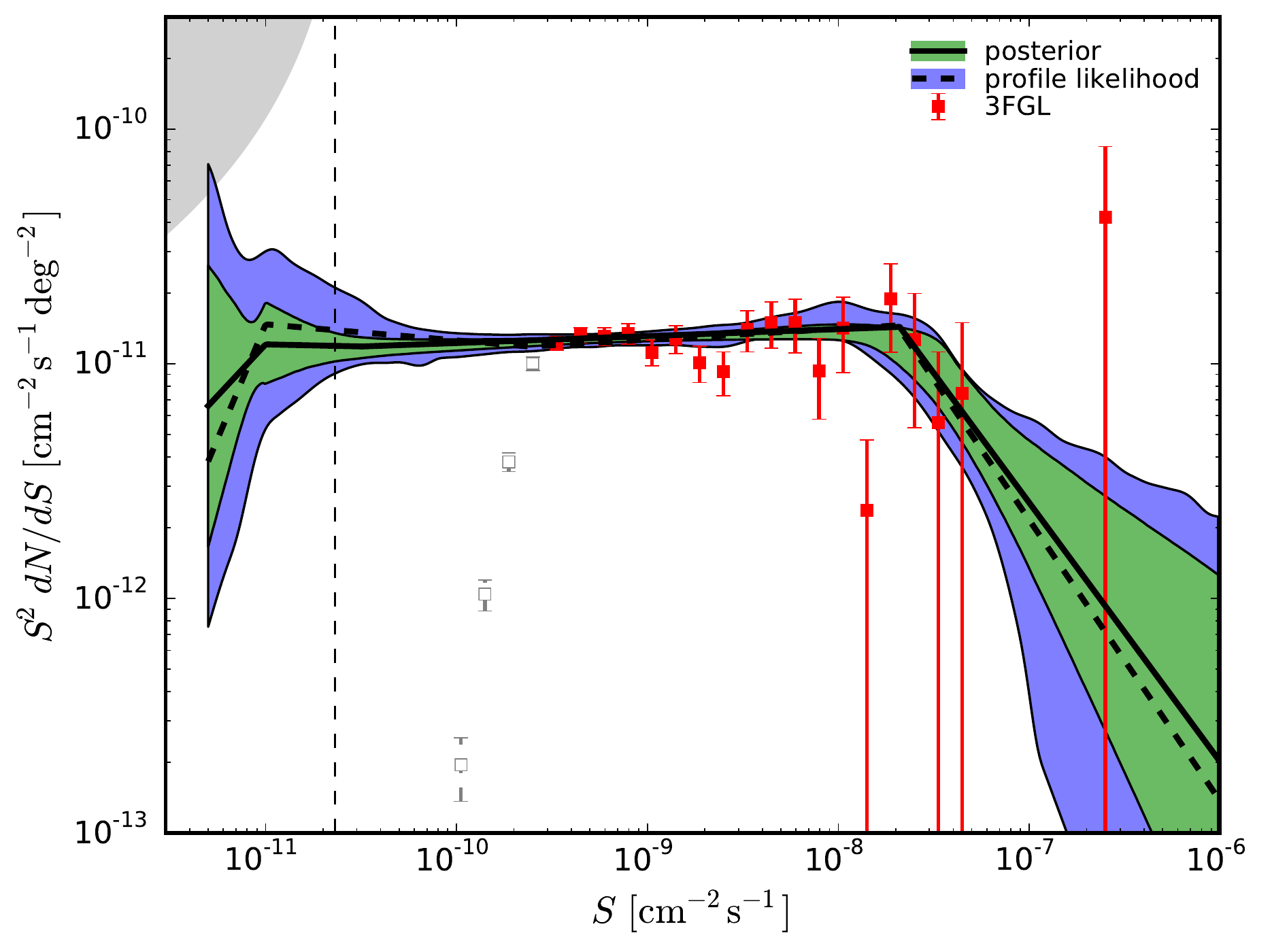}
    \caption{Differential source-count distribution \dnds\ obtained
      from 6-year \emph{Fermi}-LAT data for high Galactic latitudes
      greater than 30$^\circ$. The fit was carried out by employing
      the hybrid approach with two free breaks and a node at the faint
      end of the distribution. Line styles and colors are as in
      Figure~\ref{fig:mbpl_fit_3_2}. The shape of the
      \dnds\ distribution for very faint sources can be further
      constrained by the fact that the sum of the integral
      point-source flux and the Galactic foreground contribution must
      not exceed the total map flux $F_\mathrm{tot}$. Corresponding
      constraints have been derived assuming the \dnds\ distribution
      obtained from the Bayesian posterior down to the best-fit
      position of the last free break, i.e., $3\times
      10^{-11}\,\mathrm{cm}^{-2}\,\mathrm{s}^{-1}$.  Below that value,
      \dnds\ has been extrapolated with a power-law component of
      varying index. At the boundary of the gray-shaded region the
      total point-source flux equals $F_\mathrm{tot} -
      F_\mathrm{gal}$, requiring a break. \label{fig:dnds_final} }
\end{figure*}

\section{CONCLUSIONS}\label{sec:conclusions}
In this article, we have employed the pixel-count distribution
(1-point PDF) of the 6-year photon counts map measured with
\emph{Fermi}-LAT between 1\,GeV and 10\,GeV to decompose the
high-latitude gamma-ray sky.  This statistical analysis method has
allowed us to dissect the gamma-ray sky into three different
components, i.e., point sources, diffuse Galactic foreground emission,
and a contribution from isotropic diffuse background.  The analysis of
the simple pixel-count distribution has been improved by employing a
pixel-dependent approach, in order to fully explore all the available
information and to incorporate the morphological variation of
components such as the Galactic foreground emission.  A summary of the
main results obtained with this analysis follows.

The distribution of point sources \dnds\ has been fit assuming a
multiply broken power law (MBPL approach) with one, two, and three
free breaks. A possible bias in obtaining the correct statistical
uncertainty band for faint-source contributions has been mitigated by
extending the setup with a node, what we called the hybrid approach.
Figure~\ref{fig:dnds_final} summarizes the resulting
\dnds\ distribution at high Galactic latitudes $b$ greater than
30$^\circ$.

We have found that both the MBPL approach and the hybrid approach
single out a best-fit source-count distribution for $|b|\geq 30^\circ$
that is consistent with a single broken power law for integral fluxes
$S$ in the resolved range.  Although two-break models are preferred to
properly fit the \textit{entire} flux range covered by the data, the
second break found in the MBPL approach in the faint-source region is
consistent with a sensitivity cutoff. Instead, in the hybrid approach,
the second break is needed for a viable determination of the
uncertainty band.  The MBPL and hybrid approaches have led to
comparable results except in the faint-source flux region, where the
latter improved the uncertainty band.  For bright sources with an
integral flux above the first break at $2.1^{+1.0}_{-1.3} \times
10^{-8}\,\mathrm{cm}^{-2}\,\mathrm{s}^{-1}$ the \dnds\ distribution
follows a power law with index $n_1=3.1^{+0.7}_{-0.5}$.  Below the
first break, the index characterizing the intermediate region and the
faint-source region of \dnds\ hardens to
$n_2=1.97^{+0.03}_{-0.03}$. It is determined with exceptionally high
precision ($\sim$2\%) thanks to the high statistics of sources
populating that region.  The fit is consistent with the distribution
of individually resolved sources listed in the 3FGL catalog.  We have
measured \dnds\ down to an integral flux of $\sim\! 2\times
10^{-11}\,\mathrm{cm}^{-2}\,\mathrm{s}^{-1}$, improving beyond the
3FGL catalog detection limit by about one order of magnitude.  \\ To
further constrain the physical \dnds\ distribution at low fluxes, we
have derived an upper limit on a possible intrinsic second break from
the uncertainty band obtained with the hybrid approach. We have found
that a possible second break of \dnds\ is constrained to be below
$6.4\times 10^{-11}\,\mathrm{cm}^{-2}\,\mathrm{s}^{-1}$ at 95\% CL,
assuming a change of $\Delta n \geq 0.3$ for the power-law indices
below and above that break.

We have checked our results against a number of possible systematic
and modeling uncertainties of the analysis framework. Likewise, the
behavior of \dnds\ has been investigated as a function of the Galactic
latitude cut. We have considered Galactic latitude cuts in the
interval between $10^\circ$ and $70^\circ$. We have found that the
faint-source and the intermediate regions of \dnds\ are not altered,
while the uncertainty band in the bright end becomes larger due to the
decreasing number of bright sources in the ROI.  At the same time,
fitting the overall normalization of the Galactic foreground template
has revealed that it significantly increases with higher latitude
cuts.  This indicates a possible gradual mismatch between the Galactic
foreground model and the data at high latitudes, or a missing
component not accounted for in our analysis setup. Note, however, that
this increase does not affect the obtained \dnds\ distribution, which
is instead stable.

We have found that the high-latitude gamma-ray sky above $30^\circ$ is
composed of $(25 \pm 2)$\% point sources, $(69.3 \pm 0.7)$\% Galactic
foreground, and $(6 \pm 2)$\% isotropic diffuse background emission.
Both the integral point-source component and the sum of the Galactic
foreground and diffuse isotropic background components were stable
against Galactic latitude cuts and changes of the Galactic foreground
modeling. The choice of the Galactic foreground can, however, affect
the integral value of the diffuse isotropic background component
itself.

With respect to the recent IGRB measurement by
\citet{2015ApJ...799...86A}, this analysis allowed us to clarify
between 42\% and 56\% of its origin between 1\,GeV and 10\,GeV by
attributing it to unresolved point sources.

\acknowledgments
We kindly acknowledge valuable discussions with Luca Latronico and
Marco Regis, and the valuable support by the \textit{Fermi} LAT
Collaboration internal referee Dmitry Malyshev and the anonymous
journal referee in improving the manuscript.

We are grateful for the support of the {\sl Servizio Calcolo e Reti}
of the Istituto Nazionale di Fisica Nucleare, Sezione di Torino, and
of its coordinator Stefano Bagnasco.

This work is supported by the research grant {\sl Theoretical
  Astroparticle Physics} number 2012CPPYP7 under the program PRIN 2012
funded by the Ministero dell'Istruzione, Universit\`a e della Ricerca
(MIUR), by the research grants {\sl TAsP (Theoretical Astroparticle
  Physics)} and {\sl Fermi} funded by the Istituto Nazionale di Fisica
Nucleare (INFN), and by the {\sl Strategic Research Grant: Origin and
  Detection of Galactic and Extragalactic Cosmic Rays} funded by
Torino University and Compagnia di San Paolo.  This research was
partially supported by a grant from the GIF, the German-Israeli
Foundation for Scientific Research and Development.

Some of the results in this paper have been derived using the HEALPix
\citep{2005ApJ...622..759G} package. This analysis made use of the
\mbox{PyMultiNest} package; see \citet{2014A&A...564A.125B} for
details.

The \textit{Fermi} LAT Collaboration acknowledges generous ongoing
support from a number of agencies and institutes that have supported
both the development and the operation of the LAT as well as
scientific data analysis.  These include the National Aeronautics and
Space Administration and the Department of Energy in the United
States, the Commissariat \`a l'Energie Atomique and the Centre
National de la Recherche Scientifique / Institut National de Physique
Nucl\'eaire et de Physique des Particules in France, the Agenzia
Spaziale Italiana and the Istituto Nazionale di Fisica Nucleare in
Italy, the Ministry of Education, Culture, Sports, Science and
Technology (MEXT), High Energy Accelerator Research Organization (KEK)
and Japan Aerospace Exploration Agency (JAXA) in Japan, and the
K.~A.~Wallenberg Foundation, the Swedish Research Council and the
Swedish National Space Board in Sweden.

Additional support for science analysis during the operations phase is
gratefully acknowledged from the Istituto Nazionale di Astrofisica in
Italy and the Centre National d'\'Etudes Spatiales in France.

\newpage

\appendix

\section{DERIVATION OF THE \opdf\ FORMULAE FROM POISSON PROCESSES}\label{app:genfunc_poisson}
The general representation of the generating function
$\mathcal{P}^\mpd (t)$ for photon-count maps can be derived from a
superposition of Poisson processes. In the following, we consider a
population of point sources following a source-count distribution
function \dnds.  In a generic pixel $p$, covering the solid angle
$\Omega_\mathrm{pix}$, we expect an average number of point sources
$\mu = \Omega_\mathrm{pix}\,\Delta S\,\mathrm{d}N/\mathrm{d}S$ in the
flux interval $[S,S+\Delta S]$.\footnote{For clarity, we omit the
  pixel index $\mpd$ in the following.}  The number, $n$, of sources
of this kind in pixel $p$ follows a Poisson distribution,
\begin{equation}\label{app:eq_p1}
\frac{\mu^n}{n!} e^{-\mu} .
\end{equation}
Given $n$ sources in the pixel, the average number of gamma-ray counts
contributed by sources is $n\,\mathcal{C}(\overline{S})$ (see
Equation~\eqref{eq:counts}), where $\overline{S}$ denotes the average
flux of the interval $[S,S+\Delta S]$. In general, the number of
counts, $m$, contributed by these sources also follows a Poisson
distribution,
\begin{equation}\label{app:eq_p2}
\frac{ (n\,\mathcal{C})^m}{m!} e^{-n\,\mathcal{C}} .
\end{equation}
Taking into account the distribution in $n$, the probability
distribution function $p_m$ of counts $m$ in the given pixel can be
obtained from marginalizing over the product of the two distributions
\eqref{app:eq_p1} and \eqref{app:eq_p2}:
\begin{equation}\label{app:pm}
 p_m = \sum_n \frac{\mu^n}{n!} e^{-\mu} \, \frac{ (n\,\mathcal{C})^m}{m!} e^{-n\,\mathcal{C}} .
\end{equation}
This distribution is more conveniently expressed in terms of a
generating function, simplifying to
\begin{equation}\label{app:gen_func}
\sum_m  p_m \, t^m  = \exp \left[  \mu \left(e^{\mathcal{C}(t-1)} -1\right) \right] .
\end{equation}
Equation~\eqref{app:gen_func} is only valid for sources of a given
flux interval $[S,S+\Delta S]$.  To get the final distribution
function of $m$ we need to integrate over the full distribution of
$S$, i.e., the source-count distribution \dnds. The generating
function for the final distribution of $m$ is given by the product of
all individual generating functions~\eqref{app:gen_func}, i.e.,
\begin{equation}
\prod_{\overline{S}} \exp \left[  \mu \left(e^{\mathcal{C} (t-1)} -1\right) \right] =
\exp \left[ \sum_{\overline{S}} \mu \left(e^{\mathcal{C} (t-1)} -1\right) \right] .
\end{equation}
Using the definition of $\mu$ in the limit $\Delta S \rightarrow
\mathrm{d}S$ and rewriting in terms of $x_m$ as defined in
Equation~\eqref{eq:xm} eventually gives the representation of the
generating function quoted in Equation~\eqref{eq:gfgen1}, i.e.,
\begin{equation}
\exp \left[ \sum_{m=1}^{\infty} x_m \left( t^m -1 \right) \right] .
\end{equation}

\section{DERIVATION OF \dnds\ FOR CATALOGED SOURCES}\label{app:dnds_cat}
This section describes our approach of deriving the source-count
distribution \dnds\ (uncorrected for detection efficiency) from the
3FGL catalog. The \dnds\ distribution was derived self-consistently
for each ROI considered in the article.  We first selected all 3FGL
sources contained in a given ROI.  For each source we adopted the
best-fit spectral model (power law, log-parabola, power law with
exponential or super-exponential cutoff) indicated in the catalog,
using the reported best-fit parameters.  The source photon flux in the
energy range of interest was calculated by integrating this spectrum.
The \dnds\ was built as a histogram from the above-mentioned flux
collection, using appropriate binning and normalizing it to the solid
angle covered by the ROI.

\section{NODE-BASED APPROACH}\label{app:node_based}
The node-based approach as introduced in
Section~\ref{sssec:fit_approach} serves as an independent cross-check
for the complementary approach of keeping the positions of breaks as
free fit parameters. We applied the node-based approach to the
$|b|\geq 30^\circ$ data between 1\,GeV and 10\,GeV. The choice of the
node positions was driven by two criteria, i.e., (a) to reasonably
approximate the bright-source and intermediate regions covered by
catalog data, and (b) to approximate possible features in the
faint-source region without overfitting the data. We therefore chose
seven nodes: $5\times 10^{-7}$, $10^{-8}$, $10^{-9}$, $3\times
10^{-10}$, $3\times 10^{-11}$, $10^{-11}$, and $5\times
10^{-12}\,\mathrm{cm}^{-2}\,\mathrm{s}^{-1}$.  Remaining parameters
and priors were chosen in the same way as discussed in
Section~\ref{sssec:priors} for the hybrid approach.

\begin{figure*}[t]
\epsscale{0.60}
\plotone{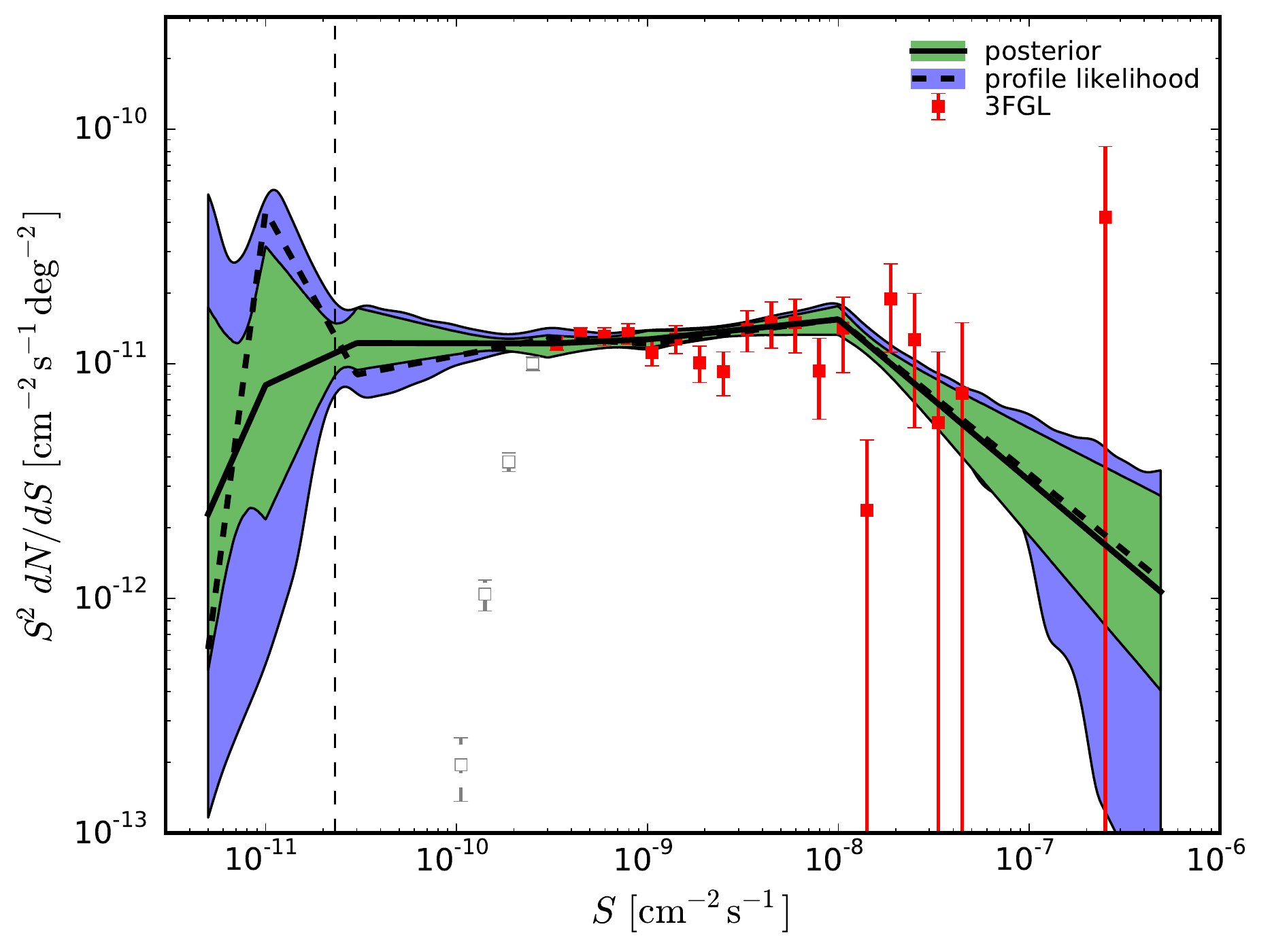}
\caption{Differential source-count distribution \dnds\ obtained with
  the node-based approach for $|b|\geq 30^\circ$.  The
  \dnds\ parameterization is based on a choice of seven nodes (see
  text for details). Line styles and colors are as in
  Figure~\ref{fig:mbpl_fit_3_2}.\label{fig:nodes}}
\end{figure*}

The \dnds\ fit employing the node-based approach is shown in
Figure~\ref{fig:nodes}.  The fit matches well the results found in
Section~\ref{sec:application} within statistical uncertainties.

\section{MONTE CARLO SIMULATIONS}\label{app:sims}
The analysis method and the techniques of fitting the pixel-count
distribution were validated with Monte Carlo simulations. We used the
\texttt{gtobssim} utility of the Fermi Science Tools package to
simulate realistic mock maps including a point-source contribution,
the Galactic foreground, and a diffuse isotropic background
component. Mock maps were analyzed with the same analysis chain as
used for the real data.

\subsection{Setup}\label{sapp:sims_setup}
Mock data were simulated for a time period of 5\,years, using
\texttt{P7REP} instrumental response functions and the
\emph{Fermi}-LAT spacecraft file corresponding to the real data
set. Data selection resembled the procedure applied for real
data. Accordingly, an energy range between 1\,GeV and 10\,GeV was
chosen, and the effective PSF was derived in compliance with the
simulated data set.

To demonstrate the applicability of the analysis and to investigate
the sensitivity, we simulated realizations of four different toy
source-count distributions, tagged A1, A2, B, and C.  In all four
cases, \dnds\ was modeled with a broken power law, where $n_1$ denotes
the index above the break and $n_2$ the index below the break: (A1) no
break, with $n_1 \equiv n_2 = 2.0$, (A2) break at
$10^{-10}\,\mathrm{cm}^{-2}\,\mathrm{s}^{-1}$, with $n_1=2.0$,
$n_2=1.6$, (B) break at $10^{-10}\,\mathrm{cm}^{-2}\,\mathrm{s}^{-1}$,
with $n_1=2.3$, $n_2=1.6$, and (C) break at
$10^{-10}\,\mathrm{cm}^{-2}\,\mathrm{s}^{-1}$, with $n_1=1.6$,
$n_2=2.5$.  In particular, model A1 approximates what was found in the
real data (see Section~\ref{sec:application}).  Model A2 was chosen to
investigate the sensitivity of the analysis in the faint-source
region, while models B and C impose two extreme scenarios.

Point-source fluxes were simulated according to the given
\dnds\ model, and positions were distributed isotropically across the
sky. Realized sources were passed to \texttt{gtobssim} individually.
The flux range covered by the \dnds\ distributions was limited to the
interval $[10^{-12},10^{-8}]\,\mathrm{cm}^{-2}\,\mathrm{s}^{-1}$.  The
lower bound of this interval was chosen to be sufficiently small to
investigate the sensitivity limit. At the same time, the upper bound
ensures a setup that is reasonably simple to study, while resembling
the real data in all flux regions except the bright-source
region. Flux spectra of individual point sources were modeled with
power laws with a fixed power-law index of $\Gamma = 2.0$. In
addition, models A1 and A2 were simulated incorporating a distribution
of point-source spectral indices. We assumed a Gaussian distribution
centered on $\overline{\Gamma}=2.4$, with a half-width
$\sigma_\mathrm{\Gamma} = 0.2$.

The Galactic foreground was modeled using the template discussed in
Section~\ref{ssec:bckgs}.  The isotropic background emission was
modeled with respect to the analysis cuts. The model is given by the
corresponding analysis template
\texttt{iso\_clean\_front\_v05.txt}\,\footnote{See
  http://fermi.gsfc.nasa.gov/ssc/data/access/lat/BackgroundModels.html}.
The simulated background emission between 1\,GeV and 10\,GeV was
normalized to an integral flux of $\sim\!3\times
10^{-7}\,\mathrm{cm}^{-2}\,\mathrm{s}^{-1}\,\mathrm{sr}^{-1}$ in the
case of the fixed-index simulations and to $\sim\!1.5\times
10^{-7}\,\mathrm{cm}^{-2}\,\mathrm{s}^{-1}\,\mathrm{sr}^{-1}$
otherwise.  To investigate a possible bias caused by a distribution of
spectral indices, model A2 was simulated without any backgrounds
(source-only), increasing sensitivity.

\subsection{Results}\label{sapp:sims_results}
The mock data were analyzed applying the procedure established in
Section~\ref{sec:application}. The MBPL approach was conducted
allowing three free breaks.  Priors were adjusted to cover the
intermediate and faint-source regions appropriately. The hybrid
approach was carried out choosing two free breaks and a node at
$5\times 10^{-12}\,\mathrm{cm}^{-2}\,\mathrm{s}^{-1}$
($10^{-12}\,\mathrm{cm}^{-2}\,\mathrm{s}^{-1}$) in the case of
simulations with a fixed (variable) point-source spectral index.  The
node was placed at the faint cutoff deduced from the MBPL fit.

\begin{figure*}[t]
\begin{centering}
\subfigure[MBPL]{%
  \includegraphics[width=0.49\textwidth]{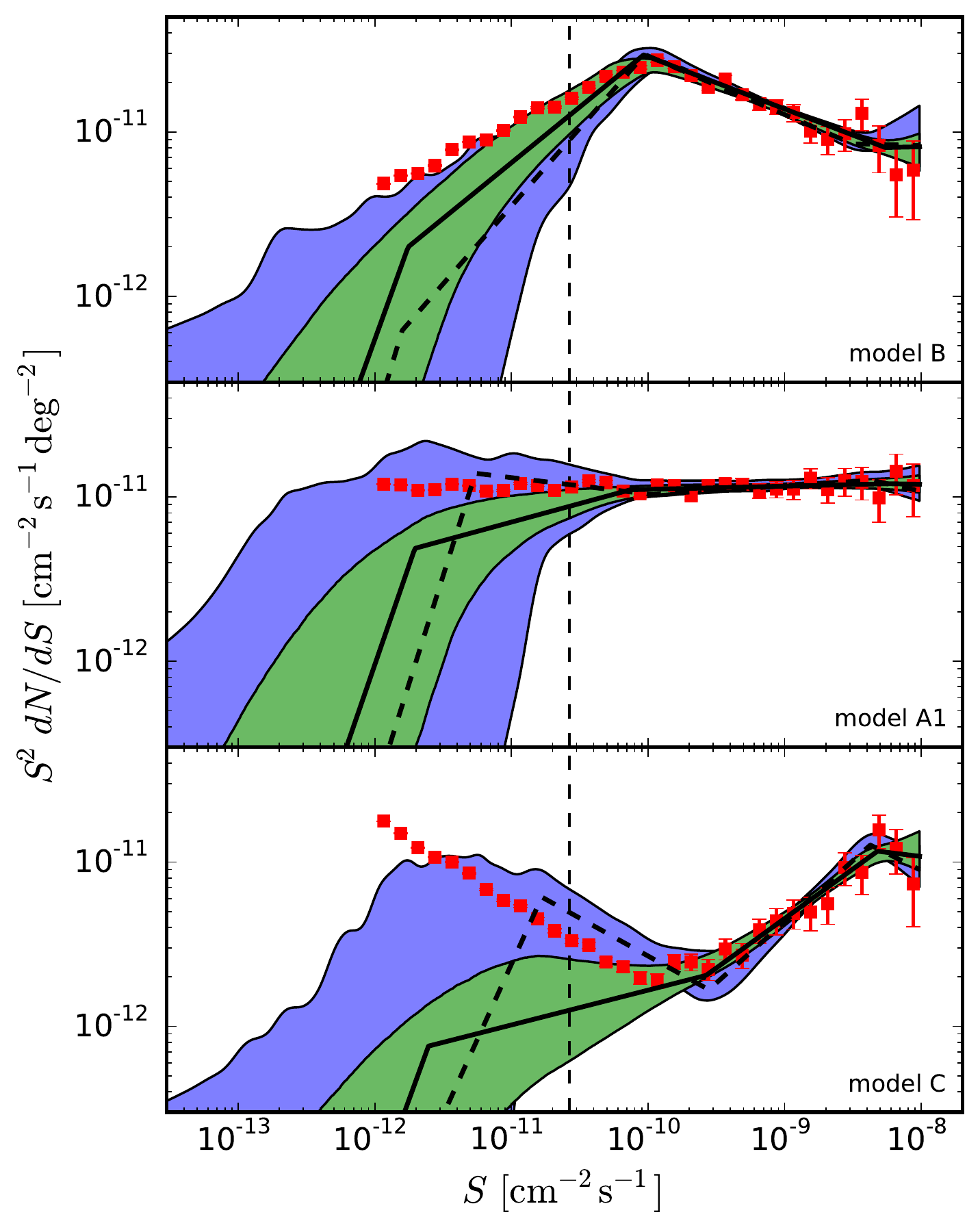}
  \label{sfig:sim_mbpl}
}
\subfigure[Hybrid]{%
  \includegraphics[width=0.49\textwidth]{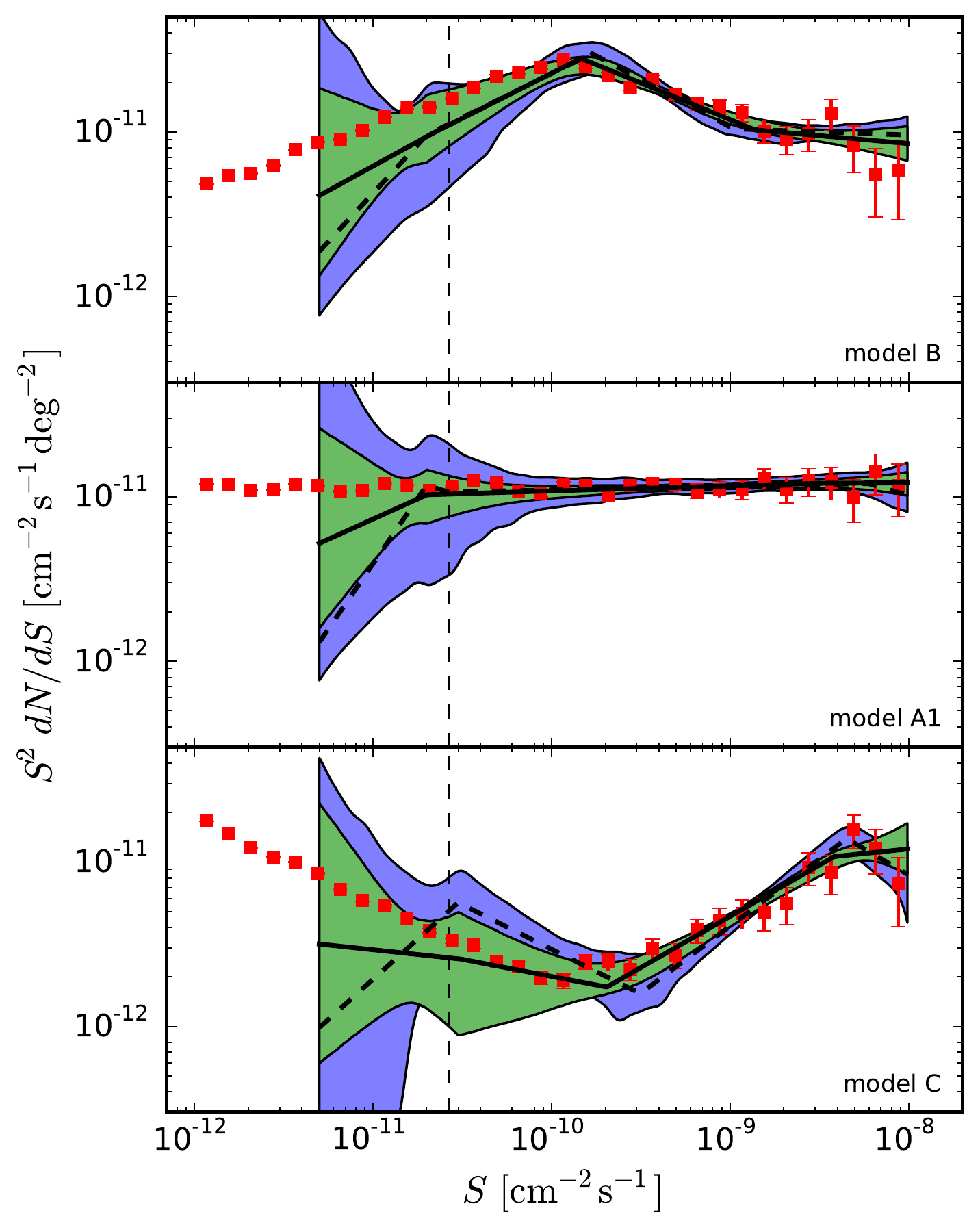}
  \label{sfig:sim_hbd}
}
\caption{Differential source-count distributions \dnds\ obtained from
  the simulated data sets with the MBPL approach (left column) and the
  hybrid approach (right column). Here point sources have been
  simulated with a fixed spectral index. The red data points show the
  actual realization of the simulated model \dnds. Poissonian errors
  $\propto \sqrt{N}$ have been assumed. The solid black line depicts
  the best-fit \dnds\ derived from the Bayesian posterior; the
  corresponding statistical uncertainty is shown by the green
  band. The dashed black line and the blue band show the same
  quantities as derived from the profile likelihood. The vertical
  dashed line depicts the corresponding sensitivity estimate
  $S_\mathrm{sens}$ as discussed in
  Section~\ref{sec:analysis_routine}.
\label{fig:sim}}
\end{centering}
\end{figure*}

\begin{figure*}[t]
\begin{centering}
\subfigure[MBPL]{%
  \includegraphics[width=0.49\textwidth]{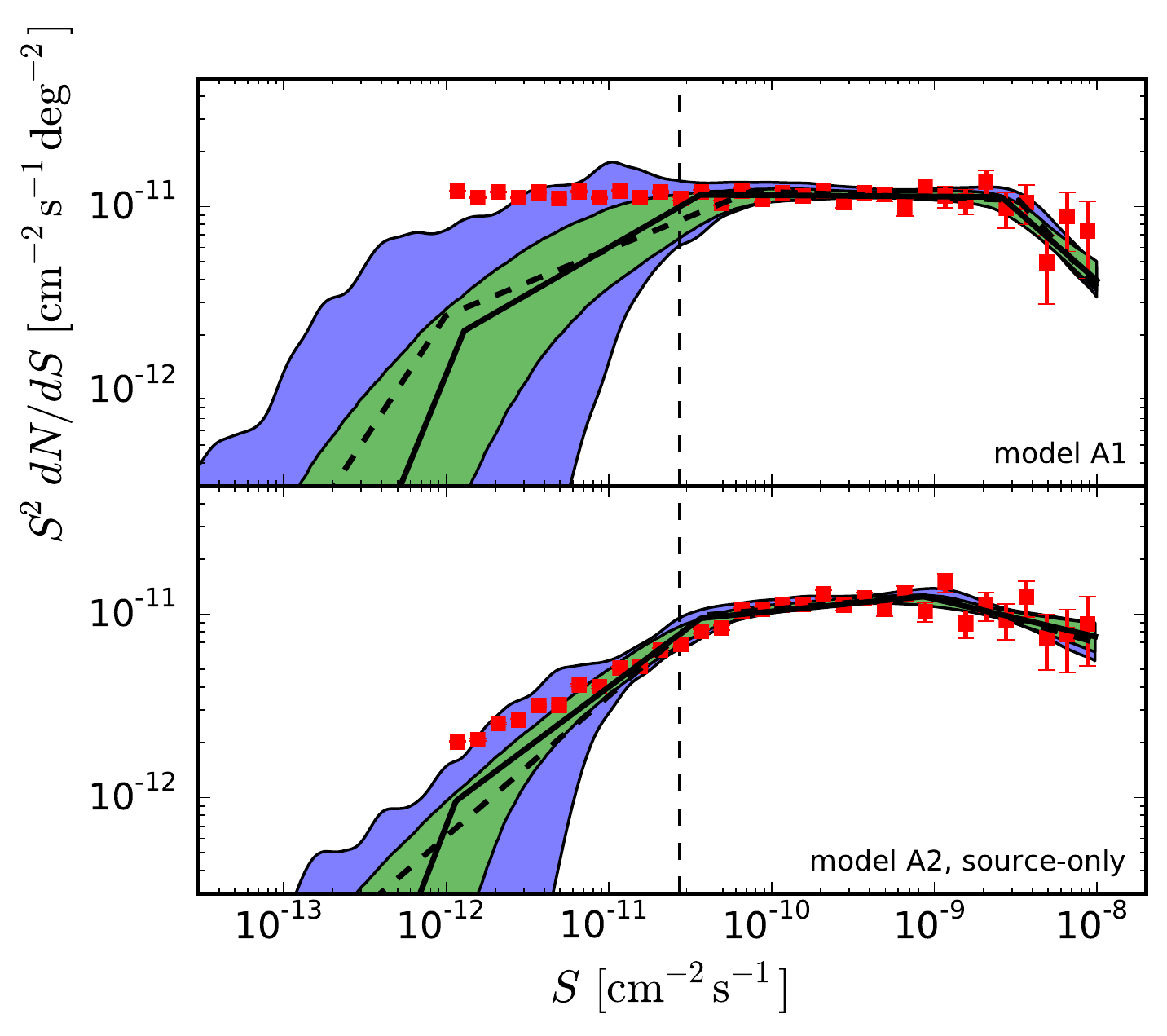}
  \label{sfig:sim_mbpl_sp}
}
\subfigure[Hybrid]{%
  \includegraphics[width=0.49\textwidth]{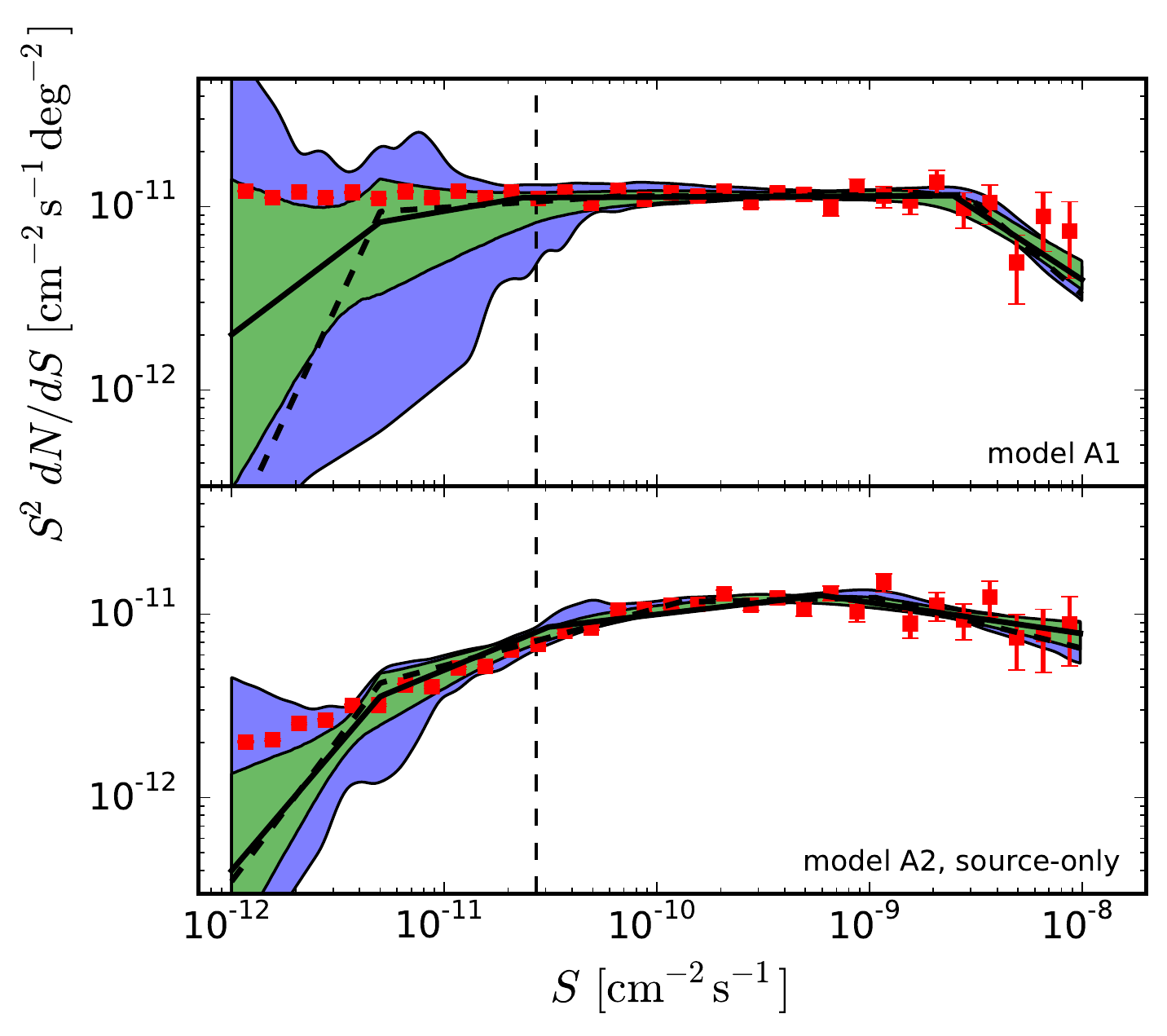}
  \label{sfig:sim_hbd_sp}
}
\caption{Same as Figure~\ref{fig:sim}, but for two realizations of
  simulations including a point-source spectral index distribution
  (see text for details). The analysis has been carried out assuming a
  fixed spectral index of 2.4\,. \label{fig:sim_spread}}
\end{centering}
\end{figure*}

The results of the analyses are depicted in Figure~\ref{fig:sim} for
the fixed-index simulations and in Figure~\ref{fig:sim_spread} for the
simulations including the spectral-index distribution.
Figures~\ref{sfig:sim_mbpl} and \ref{sfig:sim_mbpl_sp} demonstrate
that the MBPL approach recovered well the simulated
\dnds\ distributions (red data points) in the intermediate and
faint-source regions. It can also be seen that the \dnds\ fit follows
statistical fluctuations around the model within allowed degrees of
freedom. The position of the break, corresponding to parameter
$S_\mathrm{b2}$ of the model fit, is well constrained and in good
agreement with the simulated input.  However, uncertainty bands are
biased for very faint sources; in particular, for model~C a
sensitivity cutoff before the faint end of the simulated source
distribution was found. The mismatch increases for the results
obtained from the Bayesian posterior, while the profile likelihood fit
is comparably more accurate. This behavior becomes most pronounced for
model~C.

The bias of the fit in the faint-source region can be significantly
reduced with the hybrid approach; see Figures~\ref{sfig:sim_hbd} and
\ref{sfig:sim_hbd_sp}.  The hybrid approach resolved the sampling
issues affecting the Bayesian posterior.  The data points are well
covered by the derived uncertainty bands.

Possible systematics caused by a distribution of point-source spectral
indices are addressed by Figure~\ref{fig:sim_spread}.  The data sets
with \dnds\ realizations of models A1 and A2, each simulated
incorporating the Gaussian distribution of spectral indices, were
analyzed with the same analysis chain as used for real data, i.e.,
assuming a constant spectral index of
2.4\,. Figure~\ref{fig:sim_spread} shows that no evidence for a
systematic effect on the \dnds\ fit was found for $S \gtrsim
S_\mathrm{sens}$. Below the sensitivity limit $S_\mathrm{sens}$, the
uncertainty bands shift slightly downward in comparison to model A1 in
Figure~\ref{fig:sim}.  The high statistics of the source-only
simulation of model A2 indeed increased the sensitivity (see bottom
row of Figure~\ref{fig:sim_spread}), as expected. We found that the
break was recovered well, again indicating no important systematic
effect.

The Galactic foreground normalization parameter $A_\mathrm{gal}$ was
found to be $\sim$1.05 in all considered scenarios, with no evidence
for a dependence on the Galactic latitude cut. For the realization of
model A1 for fixed spectral indices, for instance, the value of
$A_\mathrm{gal}$ obtained from the posterior was $1.050 \pm 0.002$,
$1.055 \pm 0.005$, and $1.066 \pm 0.014$ for Galactic latitude cuts of
$10^\circ$, $30^\circ$, and $50^\circ$, respectively.  Profile
likelihood parameter estimates were similar, with slightly larger
uncertainties.  The overall effect of obtaining $A_\mathrm{gal}$
larger than 1 can be attributed to remaining degeneracies between the
Galactic foreground model and the diffuse isotropic background
component. However, a slight dependence on the Galactic latitude cut
cannot be excluded within statistical uncertainties.

In conclusion, all toy distributions were well reproduced with the
hybrid approach within statistical uncertainties. The mock data
indicate that the actual sensitivity depends on the source-count
distribution and the background components, matching our expectation
(see Section~\ref{sec:analysis_routine}).  One can nevertheless
conclude from the two extreme scenarios (models~B and C) that the
sensitivity estimate $S_\mathrm{sens}$ constitutes a conservative
benchmark for the energy band between 1\,GeV and 10\,GeV.

\clearpage
\bibliographystyle{apj}
\bibliography{msdnds}

\begin{thebibliography}{}
\expandafter\ifx\csname natexlab\endcsname\relax\def\natexlab#1{#1}\fi

\bibitem[{{Abdo} {et~al.}(2010{\natexlab{a}}){Abdo}, {Ackermann}, {Ajello},
  {et~al.}}]{2010ApJS..188..405A}
{Abdo}, A.~A., {Ackermann}, M., {Ajello}, M., {et~al.} 2010{\natexlab{a}},
  \apjs, 188, 405

\bibitem[{{Abdo} {et~al.}(2010{\natexlab{b}}){Abdo}, {Ackermann}, {Ajello},
  {et~al.}}]{2010ApJ...720..435A}
---. 2010{\natexlab{b}}, \apj, 720, 435

\bibitem[{{Acero} {et~al.}(2015){Acero}, {Ackermann}, {Ajello}, {Albert},
  {et~al.}}]{2015ApJS..218...23A}
{Acero}, F., {Ackermann}, M., {Ajello}, M., {Albert}, A., {et~al.} 2015, \apjs,
  218, 23

\bibitem[{{Ackermann} {et~al.}(2012){Ackermann}, {Ajello}, {Albert},
  {Allafort}, {et~al.}}]{2012ApJS..203....4A}
{Ackermann}, M., {Ajello}, M., {Albert}, A., {Allafort}, A., {et~al.} 2012,
  \apjs, 203, 4

\bibitem[{{Ackermann} {et~al.}(2015{\natexlab{a}}){Ackermann}, {Ajello},
  {Albert}, {et~al.}}]{2015arXiv151100693T}
{Ackermann}, M., {Ajello}, M., {Albert}, A., {et~al.} 2015{\natexlab{a}},
  arXiv:1511.00693

\bibitem[{{Ackermann} {et~al.}(2015{\natexlab{b}}){Ackermann}, {Ajello},
  {Albert}, {et~al.}}]{2015ApJ...799...86A}
---. 2015{\natexlab{b}}, \apj, 799, 86

\bibitem[{Ackermann {et~al.}(2012{\natexlab{a}})}]{Ackermann:2012uf}
Ackermann, M., {et~al.} 2012{\natexlab{a}}, Phys. Rev., D85, 083007

\bibitem[{Ackermann {et~al.}(2012{\natexlab{b}})}]{Ackermann:2012vca}
---. 2012{\natexlab{b}}, Astrophys. J., 755, 164

\bibitem[{Ackermann {et~al.}(2014)}]{Fermi-LAT:2014sfa}
---. 2014, Astrophys.J., 793, 64

\bibitem[{{Ackermann} {et~al.}(2016){Ackermann}, {Ajello}, {Atwood}, {Baldini},
  {Ballet}, {Barbiellini}, {Bastieri}, {Becerra Gonzalez}, {Bellazzini},
  {Bissaldi}, {Blandford}, {Bloom}, {Bonino}, {Bottacini}, {Brandt}, {Bregeon},
  {Bruel}, {Buehler}, {Buson}, {Caliandro}, {Cameron}, {Caputo}, {Caragiulo},
  {Caraveo}, {Cavazzuti}, {Cecchi}, {Charles}, {Chekhtman}, {Cheung}, {Chiang},
  {Chiaro}, {Ciprini}, {Cohen}, {Cohen-Tanugi}, {Cominsky}, {Conrad}, {Cuoco},
  {Cutini}, {D'Ammando}, {de Angelis}, {de Palma}, {Desiante}, {Di Mauro}, {Di
  Venere}, {Dom{\'{\i}}nguez}, {Drell}, {Favuzzi}, {Fegan}, {Ferrara}, {Focke},
  {Fortin}, {Franckowiak}, {Fukazawa}, {Funk}, {Furniss}, {Fusco}, {Gargano},
  {Gasparrini}, {Giglietto}, {Giommi}, {Giordano}, {Giroletti}, {Glanzman},
  {Godfrey}, {Grenier}, {Grondin}, {Guillemot}, {Guiriec}, {Harding}, {Hays},
  {Hewitt}, {Hill}, {Horan}, {Iafrate}, {Hartmann}, {Jogler},
  {J{\'o}hannesson}, {Johnson}, {Kamae}, {Kataoka}, {Kn{\"o}dlseder}, {Kuss},
  {La Mura}, {Larsson}, {Latronico}, {Lemoine-Goumard}, {Li}, {Li}, {Longo},
  {Loparco}, {Lott}, {Lovellette}, {Lubrano}, {Madejski}, {Maldera},
  {Manfreda}, {Mayer}, {Mazziotta}, {Michelson}, {Mirabal}, {Mitthumsiri},
  {Mizuno}, {Moiseev}, {Monzani}, {Morselli}, {Moskalenko}, {Murgia}, {Nuss},
  {Ohsugi}, {Omodei}, {Orienti}, {Orlando}, {Ormes}, {Paneque}, {Perkins},
  {Pesce-Rollins}, {Petrosian}, {Piron}, {Pivato}, {Porter}, {Rain{\`o}},
  {Rando}, {Razzano}, {Razzaque}, {Reimer}, {Reimer}, {Reposeur}, {Romani},
  {S{\'a}nchez-Conde}, {Saz Parkinson}, {Schmid}, {Schulz}, {Sgr{\`o}},
  {Siskind}, {Spada}, {Spandre}, {Spinelli}, {Suson}, {Tajima}, {Takahashi},
  {Takahashi}, {Takahashi}, {Thayer}, {Thompson}, {Tibaldo}, {Torres}, {Tosti},
  {Troja}, {Vianello}, {Wood}, {Wood}, {Yassine}, {Zaharijas}, \&
  {Zimmer}}]{2016ApJS..222....5A}
{Ackermann}, M., {Ajello}, M., {Atwood}, W.~B., {et~al.} 2016, \apjs, 222, 5

\bibitem[{{Ajello} {et~al.}(2014){Ajello}, {Romani}, {Gasparrini},
  {et~al.}}]{Ajello:2013lka}
{Ajello}, M., {Romani}, R.~W., {Gasparrini}, D., {et~al.} 2014, Astrophys. J.,
  780, 73

\bibitem[{{Ajello} {et~al.}(2012){Ajello}, {Shaw}, {Romani}, {Dermer},
  {Costamante}, {King}, {Max-Moerbeck}, {Readhead}, {Reimer}, {Richards}, \&
  {Stevenson}}]{Ajello:2011zi}
{Ajello}, M., {Shaw}, M.~S., {Romani}, R.~W., {et~al.} 2012, Astrophys. J.,
  751, 108

\bibitem[{{Ajello} {et~al.}(2015){Ajello}, {Gasparrini}, {S{\'a}nchez-Conde},
  {Zaharijas}, {Gustafsson}, {Cohen-Tanugi}, {Dermer}, {Inoue}, {Hartmann},
  {Ackermann}, {Bechtol}, {Franckowiak}, {Reimer}, {Romani}, \&
  {Strong}}]{Ajello:2015mfa}
{Ajello}, M., {Gasparrini}, D., {S{\'a}nchez-Conde}, M., {et~al.} 2015,
  Astrophys. J., 800, L27

\bibitem[{{Ando} {et~al.}(2007){Ando}, {Komatsu}, {Narumoto}, \&
  {Totani}}]{2007PhRvD..75f3519A}
{Ando}, S., {Komatsu}, E., {Narumoto}, T., \& {Totani}, T. 2007, \prd, 75,
  063519

\bibitem[{{Bartels} {et~al.}(2016){Bartels}, {Krishnamurthy}, \&
  {Weniger}}]{2016PhRvL.116e1102B}
{Bartels}, R., {Krishnamurthy}, S., \& {Weniger}, C. 2016, Physical Review
  Letters, 116, 051102

\bibitem[{{Baxter} {et~al.}(2010){Baxter}, {Dodelson}, {Koushiappas}, \&
  {Strigari}}]{2010PhRvD..82l3511B}
{Baxter}, E.~J., {Dodelson}, S., {Koushiappas}, S.~M., \& {Strigari}, L.~E.
  2010, \prd, 82, 123511

\bibitem[{Broderick {et~al.}(2014{\natexlab{a}})Broderick, Pfrommer, Puchwein,
  \& Chang}]{Chang:2013yia}
Broderick, A.~E., Pfrommer, C., Puchwein, E., \& Chang, P. 2014{\natexlab{a}},
  Astrophys. J., 790, 137

\bibitem[{Broderick {et~al.}(2014{\natexlab{b}})Broderick, Pfrommer, Puchwein,
  Smith, \& Chang}]{Chang:2013ada}
Broderick, A.~E., Pfrommer, C., Puchwein, E., Smith, K.~M., \& Chang, P.
  2014{\natexlab{b}}, Astrophys. J., 796, 12

\bibitem[{{Buchner} {et~al.}(2014){Buchner}, {Georgakakis}, {Nandra}, {Hsu},
  {Rangel}, {Brightman}, {Merloni}, {Salvato}, {Donley}, \&
  {Kocevski}}]{2014A&A...564A.125B}
{Buchner}, J., {Georgakakis}, A., {Nandra}, K., {et~al.} 2014, \aap, 564, A125

\bibitem[{Calore {et~al.}(2014)Calore, Di~Mauro, Donato, \&
  Donato}]{Calore:2014oga}
Calore, F., Di~Mauro, M., Donato, F., \& Donato, F. 2014, Astrophys. J., 796, 1

\bibitem[{{Casandjian} {et~al.}(2009){Casandjian}, {Grenier},
  {et~al.}}]{2009arXiv0912.3478C}
{Casandjian}, J., {Grenier}, I., {et~al.} 2009, arXiv:0912.3478

\bibitem[{Cholis {et~al.}(2014)Cholis, Hooper, \& McDermott}]{Cholis:2013ena}
Cholis, I., Hooper, D., \& McDermott, S.~D. 2014, JCAP, 1402, 014

\bibitem[{{Condon}(1974)}]{1974ApJ...188..279C}
{Condon}, J.~J. 1974, \apj, 188, 279

\bibitem[{Cuoco {et~al.}(2012)Cuoco, Komatsu, \& Siegal-Gaskins}]{Cuoco:2012yf}
Cuoco, A., Komatsu, E., \& Siegal-Gaskins, J.~M. 2012, Phys. Rev., D86, 063004

\bibitem[{{Cuoco} {et~al.}(2015){Cuoco}, {Xia}, {Regis}, {Branchini},
  {Fornengo}, \& {Viel}}]{2015ApJS..221...29C}
{Cuoco}, A., {Xia}, J.-Q., {Regis}, M., {et~al.} 2015, \apjs, 221, 29

\bibitem[{Di~Mauro {et~al.}(2014{\natexlab{a}})Di~Mauro, Calore, Donato,
  Ajello, \& Latronico}]{DiMauro:2013xta}
Di~Mauro, M., Calore, F., Donato, F., Ajello, M., \& Latronico, L.
  2014{\natexlab{a}}, Astrophys. J., 780, 161

\bibitem[{Di~Mauro {et~al.}(2014{\natexlab{b}})Di~Mauro, Cuoco, Donato, \&
  Siegal-Gaskins}]{DiMauro:2014wha}
Di~Mauro, M., Cuoco, A., Donato, F., \& Siegal-Gaskins, J.~M.
  2014{\natexlab{b}}, JCAP, 1411, 021

\bibitem[{Di~Mauro \& Donato(2015)}]{DiMauro:2015tfa}
Di~Mauro, M., \& Donato, F. 2015, Phys. Rev., D91, 123001

\bibitem[{Di~Mauro {et~al.}(2014{\natexlab{c}})Di~Mauro, Donato, Lamanna,
  Sanchez, \& Serpico}]{DiMauro:2013zfa}
Di~Mauro, M., Donato, F., Lamanna, G., Sanchez, D.~A., \& Serpico, P.~D.
  2014{\natexlab{c}}, Astrophys. J., 786, 129

\bibitem[{Dodelson {et~al.}(2009)Dodelson, Belikov, Hooper, \&
  Serpico}]{Dodelson:2009ih}
Dodelson, S., Belikov, A.~V., Hooper, D., \& Serpico, P. 2009, Phys. Rev., D80,
  083504

\bibitem[{{Feroz} \& {Hobson}(2008)}]{2008MNRAS.384..449F}
{Feroz}, F., \& {Hobson}, M.~P. 2008, \mnras, 384, 449

\bibitem[{{Feroz} {et~al.}(2009){Feroz}, {Hobson}, \&
  {Bridges}}]{2009MNRAS.398.1601F}
{Feroz}, F., {Hobson}, M.~P., \& {Bridges}, M. 2009, \mnras, 398, 1601

\bibitem[{{Feroz} {et~al.}(2013){Feroz}, {Hobson}, {Cameron}, \&
  {Pettitt}}]{2013arXiv1306.2144F}
{Feroz}, F., {Hobson}, M.~P., {Cameron}, E., \& {Pettitt}, A.~N. 2013, ArXiv
  e-prints, arXiv:1306.2144

\bibitem[{Feyereisen {et~al.}(2015)Feyereisen, Ando, \&
  Lee}]{Feyereisen:2015cea}
Feyereisen, M.~R., Ando, S., \& Lee, S.~K. 2015, JCAP, 1509, 027

\bibitem[{Fields {et~al.}(2010)Fields, Pavlidou, \& Prodanovic}]{Fields:2010bw}
Fields, B.~D., Pavlidou, V., \& Prodanovic, T. 2010, Astrophys. J., 722, L199

\bibitem[{{Fornasa} \& {S{\'a}nchez-Conde}(2015)}]{2015PhR...598....1F}
{Fornasa}, M., \& {S{\'a}nchez-Conde}, M.~A. 2015, \physrep, 598, 1

\bibitem[{{G{\'o}rski} {et~al.}(2005){G{\'o}rski}, {Hivon}, {Banday},
  {Wandelt}, {Hansen}, {Reinecke}, \& {Bartelmann}}]{2005ApJ...622..759G}
{G{\'o}rski}, K.~M., {Hivon}, E., {Banday}, A.~J., {et~al.} 2005, \apj, 622,
  759

\bibitem[{Gregoire \& Knodlseder(2013)}]{Gregoire:2013yta}
Gregoire, T., \& Knodlseder, J. 2013, Astron. Astrophys., 554, A62

\bibitem[{Harding \& Abazajian(2012)}]{Harding:2012gk}
Harding, J.~P., \& Abazajian, K.~N. 2012, JCAP, 1211, 026

\bibitem[{{Hasinger} {et~al.}(1993){Hasinger}, {Burg}, {Giacconi}, {Hartner},
  {Schmidt}, {Trumper}, \& {Zamorani}}]{1993A&A...275....1H}
{Hasinger}, G., {Burg}, R., {Giacconi}, R., {et~al.} 1993, \aap, 275, 1

\bibitem[{Inoue(2011)}]{Inoue:2011bm}
Inoue, Y. 2011, Astrophys. J., 733, 66

\bibitem[{Inoue \& Totani(2009)}]{Inoue:2008pk}
Inoue, Y., \& Totani, T. 2009, Astrophys. J., 702, 523, [Erratum: Astrophys.
  J.728,73(2011)]

\bibitem[{Lacki {et~al.}(2014)Lacki, Horiuchi, \& Beacom}]{Lacki:2012si}
Lacki, B.~C., Horiuchi, S., \& Beacom, J.~F. 2014, Astrophys. J., 786, 40

\bibitem[{Lee {et~al.}(2009)Lee, Ando, \& Kamionkowski}]{Lee:2008fm}
Lee, S.~K., Ando, S., \& Kamionkowski, M. 2009, JCAP, 0907, 007

\bibitem[{{Lee} {et~al.}(2015){Lee}, {Lisanti}, \&
  {Safdi}}]{2015JCAP...05..056L}
{Lee}, S.~K., {Lisanti}, M., \& {Safdi}, B.~R. 2015, \jcap, 5, 56

\bibitem[{{Lee} {et~al.}(2016){Lee}, {Lisanti}, {Safdi}, {Slatyer}, \&
  {Xue}}]{2016PhRvL.116e1103L}
{Lee}, S.~K., {Lisanti}, M., {Safdi}, B.~R., {Slatyer}, T.~R., \& {Xue}, W.
  2016, Physical Review Letters, 116, 051103

\bibitem[{{Malyshev} \& {Hogg}(2011)}]{2011ApJ...738..181M}
{Malyshev}, D., \& {Hogg}, D.~W. 2011, \apj, 738, 181

\bibitem[{{Olive} \& {Particle Data Group}(2014)}]{2014ChPhC..38i0001O}
{Olive}, K.~A., \& {Particle Data Group}. 2014, Chinese Physics C, 38, 090001

\bibitem[{{Ripken} {et~al.}(2014){Ripken}, {Cuoco}, {Zechlin}, {Conrad}, \&
  {Horns}}]{2014JCAP...01..049R}
{Ripken}, J., {Cuoco}, A., {Zechlin}, H.-S., {Conrad}, J., \& {Horns}, D. 2014,
  \jcap, 1, 049

\bibitem[{{Rolke} {et~al.}(2005){Rolke}, {L{\'o}pez}, \&
  {Conrad}}]{2005NIMPA.551..493R}
{Rolke}, W.~A., {L{\'o}pez}, A.~M., \& {Conrad}, J. 2005, Nuclear Instruments
  and Methods in Physics Research A, 551, 493

\bibitem[{{Scheuer}(1957)}]{1957PCPS...53..764S}
{Scheuer}, P.~A.~G. 1957, Proceedings of the Cambridge Philosophical Society,
  53, 764

\bibitem[{{Singal}(2015)}]{2015MNRAS.454..115S}
{Singal}, J. 2015, \mnras, 454, 115

\bibitem[{{So{\l}tan}(2011)}]{2011A&A...532A..19S}
{So{\l}tan}, A.~M. 2011, \aap, 532, A19

\bibitem[{Stecker \& Salamon(1996)}]{Stecker:1996ma}
Stecker, F.~W., \& Salamon, M.~H. 1996, Astrophys. J., 464, 600

\bibitem[{Stecker \& Venters(2011)}]{Stecker:2010di}
Stecker, F.~W., \& Venters, T.~M. 2011, Astrophys. J., 736, 40

\bibitem[{Su {et~al.}(2010)Su, Slatyer, \& Finkbeiner}]{Su:2010qj}
Su, M., Slatyer, T.~R., \& Finkbeiner, D.~P. 2010, Astrophys. J., 724, 1044

\bibitem[{Tamborra {et~al.}(2014)Tamborra, Ando, \& Murase}]{Tamborra:2014xia}
Tamborra, I., Ando, S., \& Murase, K. 2014, JCAP, 1409, 043

\bibitem[{Thompson {et~al.}(2006)Thompson, Quataert, \&
  Waxman}]{Thompson:2006qd}
Thompson, T.~A., Quataert, E., \& Waxman, E. 2006, Astrophys. J., 654, 219

\bibitem[{{Vernstrom} {et~al.}(2015){Vernstrom}, {Norris}, {Scott}, \&
  {Wall}}]{2015MNRAS.447.2243V}
{Vernstrom}, T., {Norris}, R.~P., {Scott}, D., \& {Wall}, J.~V. 2015, \mnras,
  447, 2243

\bibitem[{{Vernstrom} {et~al.}(2014){Vernstrom}, {Scott}, {Wall}, {Condon},
  {Cotton}, {Fomalont}, {Kellermann}, {Miller}, \&
  {Perley}}]{2014MNRAS.440.2791V}
{Vernstrom}, T., {Scott}, D., {Wall}, J.~V., {et~al.} 2014, \mnras, 440, 2791

\end{thebibliography}

\end{document}